\documentclass[12pt]{report}
\usepackage{geometry}
\usepackage{CJKutf8}
\usepackage[parfill]{parskip}

\usepackage{enumitem}
\usepackage[pdftex,dvipsnames,table]{xcolor}
\usepackage{textcomp}

\usepackage{titletoc}
\setlistdepth{2000} 

\usepackage[framemethod=tikz]{mdframed}
\usepackage{multicol}
\usepackage{longtable}

\renewenvironment{figure}[1][]{
  \begin{originalfigure}[#1]
    \begin{mdframed}
    [
    backgroundcolor=black!4,
    topline=true,
    bottomline=true,
    leftline=true,
    rightline=true]
}{
    \end{mdframed}
  \end{originalfigure}
}

\renewenvironment{table}[1][]{
  \begin{originaltable}[#1]
    \begin{mdframed}
    [linecolor=lightgray,
    linewidth=0.7pt,
    backgroundcolor=black!4]
}{
    \end{mdframed}
  \end{originaltable}
}

\usepackage[labelfont=bf]{caption}
\usepackage{tabularx}
\usepackage{booktabs}
\usepackage{multirow}
\usepackage{makecell}
\renewcommand{\arraystretch}{1.1}

\usepackage[export]{adjustbox} 
\usepackage{graphicx}
\graphicspath{{images/}}
\usepackage[section,below]{placeins} 
\usepackage[utf8]{inputenc}

\usepackage{inconsolata} 
\usepackage[sfdefault,light,condensed]{roboto} 

\usepackage{pifont} 

\usepackage{amssymb}
\usepackage{amsmath}
\usepackage{xfrac} 
\usepackage[colorlinks,
linkcolor=MidnightBlue,
citecolor=Aquamarine,
urlcolor=RoyalBlue,
bookmarks=false,
hypertexnames=true,
bookmarksopen=true]{hyperref} 
\usepackage{bookmark}
\usepackage[english]{babel}
\usepackage{csquotes,xpatch}

\usepackage[acronym]{glossaries}
\usepackage{fancyref}

\newcommand*{\fancyreflstlabelprefix}{lst}

\frefformat{main}{\fancyreflstlabelprefix}
    {\MakeUppercase{\freflstname}\fancyrefdefaultspacing#1#2}
\Frefformat{main}{\fancyreflstlabelprefix}
    {\MakeUppercase{\Freflstname}\fancyrefdefaultspacing#1#2}
\Frefformat{main}{\fancyreflstlabelprefix}{Listing~#1}
\frefformat{main}{\fancyreflstlabelprefix}{listing~#1}

\frefformat{vario}{\fancyreflstlabelprefix}
    {\MakeUppercase{\freflstname}\fancyrefdefaultspacing#1#2}
\Frefformat{vario}{\fancyreflstlabelprefix}
    {\MakeUppercase{\Freflstname}\fancyrefdefaultspacing#1#2}
\Frefformat{vario}{\fancyreflstlabelprefix}{Listing~#1 on page #2}
\frefformat{vario}{\fancyreflstlabelprefix}{listing~#1 on page #2}
\makeglossaries
\usepackage[explicit]{titlesec}
\usepackage{setspace}

\usepackage{fancyhdr}
\newenvironment{cabstract} 
    {
        \list{}{%
        \setlength{\leftmargin}{1cm}
        \setlength{\rightmargin}{\leftmargin}%
      }%
      \itshape\item\relax
    }
    {\par\vspace{1cm}}

\newcommand{\colorsection}[1]{
    \colorbox{black!10}{
        \parbox{\dimexpr\textwidth-2\fboxsep}{\thesection \hspace*{0.5em} #1 \vspace{0.01em}}
        }
    }

\titleformat{\section}[hang]
    {\bfseries\fontsize{22pt}{22pt}\selectfont }
    {}
    {-8pt}
    {\colorsection{#1}\vspace*{-1em}}

\titleformat{\subsection}
    {\bfseries\fontsize{16pt}{16pt}\selectfont}
    {\thesubsection}
    {1em}
    {#1}
    [\hrule height 0.8pt]
 
\titleformat{\subsubsection}
    {\bfseries\fontsize{14pt}{14pt}\selectfont}
    {\thesubsubsection}
    {1em}
    {#1}

\usepackage{epigraph}

\usepackage{subfig}
\usepackage[fontsize=14.4]{fontsize}
\usepackage{framed}

\newcommand{\blankpage}{\newpage\null\thispagestyle{empty}\newpage}

\newcommand{\etal}{\textit{et al.}}
\newcommand{\aboveparrule}{\leavevmode\rule{0.9\textwidth}{0.15ex}}

\usepackage{blindtext} 

\begin{document}

\begin{titlepage}
\centering
        \includegraphics[width=0.3\linewidth, valign=c]{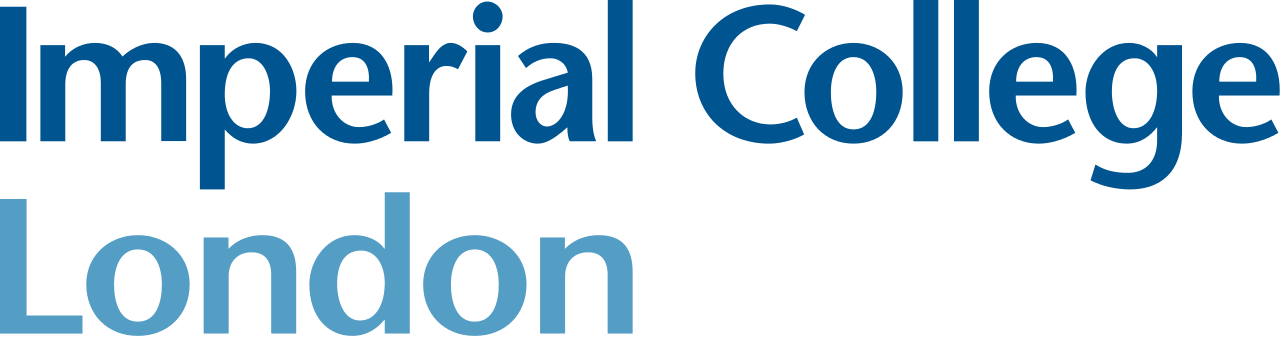}
          \hfill
        \includegraphics[width=0.53\linewidth, valign=c]{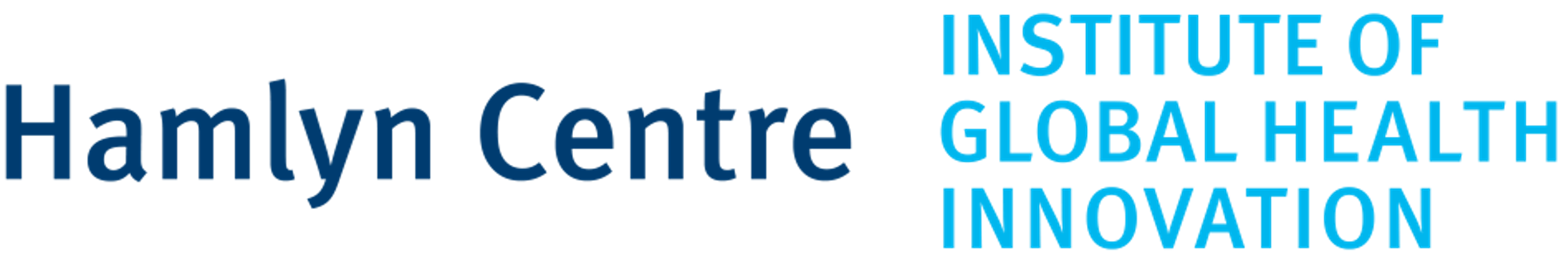}
 
 		\vfill
        
		\rule{\linewidth}{0.5mm}
		\vspace{0.01cm}
		

        {\fontsize{26pt}{26pt}\selectfont\bfseries{Laparoscopic Scene Analysis for Intraoperative Visualisation of Gamma Probe Signals in Minimally Invasive Cancer Surgery}}
        
        \vspace{0.01cm}
		\rule{\linewidth}{0.5mm}
        
        \vspace{0.5cm}
        
        \renewcommand{\arraystretch}{1.4}
            \begin{originaltable}[!h]
            {\fontsize{13pt}{13pt}\selectfont
                \begin{tabularx}{\linewidth}{r l p{2.0cm} r l}
                
                \textbf{Author} & Baoru Huang & & \textbf{Supervisors} & Prof. Daniel S. Elson$^{1, 2}$\\
                & &&& Dr. Stamatia Giannarou$^{1, 2}$\\
        		\end{tabularx}
            }
            \end{originaltable}
		\renewcommand{\arraystretch}{0.8}

        \raggedright
        \vspace{0.5cm}
        
        $^{1}$The Hamlyn Centre for Robotic Surgery, Imperial College London, UK.\\
        $^{2}$Department of Surgery \& Cancer, Imperial College London, UK.
        
        \vfill
        
        \centering

        A thesis submitted in fulfilment of the requirements for the degree of \textit{Philosophiae Doctor}
        
		\vspace{.5cm}
        {\fontsize{14pt}{14pt}\selectfont The Hamlyn Centre for Robotic Surgery, Department of Surgery \& Cancer,\\\vspace{0.2cm}
        \Large Imperial College London}
        
        \vspace{1cm}
        
\end{titlepage}
\clearpage
\thispagestyle{empty}
\vspace*{\fill}

\begin{center}
The Hamlyn Centre for Robotic Surgery,\\
Department of Surgery \& Cancer,\\
Bessemer Building,\\
Imperial College London,\\
180 Queens Gate,\\
South Kensington,\\
SW7 2AZ\\
\vspace{1cm}

Baoru Huang \textcopyright \, 2024\\
\end{center}

\newgeometry{top=3.2cm,bottom=3.2cm,right=3.2cm,left=3.2cm}
	\pagestyle{fancy}
    \fancyhead{} 
    \renewcommand{\chaptermark}[1]{\markboth{#1}{}}
	\fancyhead[R]{\nouppercase{\itshape\leftmark}}
	\fancyhead[L]{\nouppercase{\rightmark}}
    \fancyfoot{} 
    \fancyfoot[R]{\thepage}
    \fancyfoot[C]{\itshape{Gamma Probe for MIS}}
    \fancyfoot[L]{Baoru Huang}
\input{meta/chapterheads}
\onehalfspacing
\thispagestyle{empty}

\vspace*{\fill}

\setlength\epigraphwidth{0.53\textwidth}


\epigraph{\centering \begin{CJK*}{UTF8}{gbsn}便笑人间举子忙。\end{CJK*}}

\vspace*{2cm}

\setlength\epigraphwidth{0.65\textwidth}
\setlength{\epigraphrule}{0pt}


\vspace*{\fill}
\blankpage 
\thispagestyle{empty}
\vspace*{\fill}

\begin{cabstract}
\aboveparrule




In the midst of the chaos and uncertainty of my PhD journey, there was one constant source of comfort and inspiration that filled my heart with inner peace and contentment: my beloved pipa, a Chinese instrument with four strings. Every rainy afternoon in London, as well as every solitary moment at sunrise and sunset, the soothing melody of the pipa accompanied me, as a faithful friend who knew my every joy and sorrow. Throughout the isolation and uncertainty of the pandemic, the pipa was a bridge that connected me to the outside world and carried my longing for home.

As I look towards the future, I know that I will always find solace and tranquility in the gentle strains of the pipa, for striving to maintain a leisurely attitude and pursue peace and happiness. Life is not just about the small and insignificant things that surround me, but also about the thunderous roar of the pipa under my hands, reminding me of the immense power that I hold within myself. Since the day my mother first introduced me to the pipa, it has been my lifelong companion, a dear friend that I cherish beyond measure. May I never be so destitute that I must burn it for warmth, for the audible and silent music that it brings to my life.


\end{cabstract}

\vspace*{\fill}

\thispagestyle{empty}
\chapter*{Copyright}

The copyright of this thesis rests with the author. Unless otherwise indicated, its contents are licensed under a \textcolor{blue}{\href{https://creativecommons.org/licenses/by-nc-nd/4.0/}{Creative Commons Attribution-Non Commercial-No Derivatives 4.0 International Licence}} (CC BY-NC-ND).

Under this licence, you may copy and redistribute the material in any medium or format on the condition that; you credit the author, do not use it for commercial purposes and do not distribute modified versions of the work.

When reusing or sharing this work, ensure you make the licence terms clear to others by naming the licence and linking to the licence text.

Please seek permission from the copyright holder for uses of this work that are not included in this licence or permitted under UK Copyright Law.

The template of this PhD thesis is from Rob Robinson \cite{robinson2020thesis} and the link of the original template is \url{https://github.com/mlnotebook}.

\clearpage
\chapter*{Declaration}

I hereby declare that this thesis and the work herein detailed, was composed and originated by myself, except where appropriately referenced and
credited.

\bigskip
 

\smallskip

\begin{flushright}
    \begin{tabular}{m{5cm}}
        \\ \hline
        \centering{Baoru Huang} \\
    \end{tabular}
\end{flushright}

\chapter*{Abstract}
\glsresetall

\begin{cabstract}

Cancer remains a significant health challenge worldwide, with a new diagnosis occurring every two minutes in the UK\footnote{\href{https://www.cancerresearchuk.org/health-professional/cancer-statistics-for-the-uk}{https://www.cancerresearchuk.org/health-professional/cancer-statistics-for-the-uk}}. Surgery is one of the main treatment options for cancer. However, surgeons rely on the sense of touch and naked eye with limited use of pre-operative image data to directly guide the excision of cancerous tissues and metastases due to the lack of reliable intraoperative visualisation tools. This leads to increased costs and harm to the patient where the cancer is removed with positive margins, or where other critical structures are unintentionally impacted. There is therefore a pressing need for more reliable and accurate intraoperative visualisation tools for minimally invasive surgery to improve surgical outcomes and enhance patient care. 

A recent miniaturised cancer detection probe (i.e., SENSEI\textsuperscript{\textregistered} developed by Lightpoint Medical Ltd.) leverages the cancer-targeting ability of nuclear agents to more accurately identify cancer intra-operatively using the emitted gamma signal. However, the use of this probe presents a visualisation challenge as the probe is non-imaging and is air-gapped from the tissue, making it challenging for the surgeon to locate the probe-sensing area on the tissue surface. Geometrically, the sensing area is deﬁned as the intersection point between the gamma probe axis and the tissue surface in 3D space but projected onto the 2D laparoscopic image. Hence, in this thesis, tool tracking, pose estimation, and segmentation tools were developed ﬁrst, followed by laparoscope image depth estimation algorithms and 3D reconstruction methods.

The problem of detecting the probe axis-tissue intersection point was then transformed to laser point position inference using a custom laser module. Both the hardware and software design of the proposed solution were illustrated
. The best detection results were achieved using a simple network design, allowing real-time inference of the sensing area, establishing a new benchmark for the surgical vision community.

\end{cabstract}

\blankpage 
\tableofcontents
\listoffigures
\listoftables

\singlespacing
\clearpage




\newacronym{dsc}{DSC}{Dice Similarity Coefficient}

\newacronym{mis}{MIS}{Minimally Invasive Surgery}

\newacronym{ar}{AR}{Augmented Reality}
\newacronym{vr}{VR}{Virtual Reality}

\newacronym{sfm}{SfM} {Structure from Motion}

\newacronym{psma}{PSMA}{
Prostate Specific Membrane Antigen}

\newacronym{gan}{GAN}{Generative Adversarial Networks}

\newacronym{drs}{DRS}{Diffuse Reflectance Spectroscopy}

\newacronym{chess}{ChESS}{‘Chess-board Extraction by Subtraction and Summation’}

\newacronym{ippe}{IPPE}{Infinitesimal Plane-Based Pose Estimation}

\newacronym{slam}{SLAM}{Simultaneous Localization and Mapping}

\newacronym{cnns}{CNNs}{Convolutional Neural Networks}

\newacronym{std}{Std.}{standard deviation}

\newacronym{icp}{ICP}{Iterative Closest Point}

\newacronym{absrel}{Abs Rel}{mean absolute error}

\newacronym{sqrel}{Sq Rel}{squared error}
 
\newacronym{rmse}{RMSE}{root mean squared error}

\newacronym{rmselog}{RMSE log}{root mean squared logarithmic error}

\newacronym{pca}{PCA}{Principal Component Analysis}

\newacronym{vit}{ViT}{Vision Transformer}

\newacronym{mlp}{MLP}{Multi Layer Perception}
\newacronym{lstm}{LSTM}{Long Short Term Memory}
\newacronym{resnet}{ResNet}{Residual Networks}

\newacronym{stn}{STN}{Spatial Transformer Network}

\newacronym{ct}{CT}{Computed Tomography}

\newacronym{mri}{MRI}{Magnetic Resonance Imaging}

\newacronym{pet}{PET}{Positron Emission Tomography}

\newacronym{spect}{SPECT}{Single Photon Emission Computed Tomography}

\newacronym{ai}{AI}{Artificial Intelligence}

\newacronym{ssim}{SSIM}{Structural Similarity Index}

\newacronym{dice}{Dice}{Dice}

\newacronym{iou}{IoU}{Intersection over Union}

\newacronym{sensei}{SENSEI\textsuperscript{\textregistered}}{SENSEI\textsuperscript{\textregistered}}
\newacronym{SENSEI}{SENSEI\textsuperscript{\textregistered}}{SENSEI\textsuperscript{\textregistered}}

\newacronym{psa}{PSA}{Prostate-Specific Antigen}

\newacronym{3gc}{3GC}{3D Geometric Consistency}
\printglossary[type=\acronymtype, style=index]

\onehalfspacing

\chapter{Introduction}
\chaptermark{Introduction}
\glsresetall
\label{chap:1-introduction}

\begin{cabstract}

The recent advances of \acrfull{ai} have revolutionised the field of surgical science, offering unprecedented opportunities for enhanced diagnostics, efficient analysis, and improved patient outcomes. This introductory chapter lays the foundation for a comprehensive exploration of different learning methods for integrating a tethered laparoscopic gamma probe into \acrfull{mis}. The chapter begins by providing an overview of the current landscape of cancer treatment research, highlighting the challenges and opportunities in this dynamic field. Next, I discuss the motivation behind this research that stems from the urgent need to address critical gaps in current imaging practices. Despite significant progress, challenges such as diagnostic accuracy, processing speed, and interpretability persist. I then summarise the primary contribution of this thesis, including the development and evaluation of novel learning algorithms tailored for specific \acrshort{mis} cancer applications. Finally, I provide the outline of the thesis and the publication list during my PhD study.

\end{cabstract}


\section{Overview}
\glsresetall

    

Cancer remains a prevalent and challenging global health issue, necessitating a multidisciplinary approach to effective diagnosis and treatment. Among the various treatment modalities, surgery continues to play a fundamental role. However, complete cancer resection, which is often the primary goal of surgical interventions, presents a formidable challenge. 

Although the mature development of pre-operative imaging systems (such as \acrfull{ct}, \acrfull{mri}, or \acrfull{pet} and \acrfull{spect}) can provide surgeons with an accurate indication of the location of cancer tissue within an organ, the substantial deformation of a patient's abdomen, and the limited visual clarity during an intervention inside the organ, present a persistent challenge to precisely register pre-operative images with the intraoperative environment and accurately locate cancer tissue during surgery.

In open surgery, surgeons rely on their sense of touch to identify potential cancer tissue.  \acrshort{mis} has gained much attention due to its advantages -- including reduced trauma and faster recovery times -- and laparoscopy is increasingly employed, including together with robotic control. However, laparoscopic surgeons lose haptic feedback, making the task of precise cancer resection more difficult. In light of these challenges, my doctoral research is dedicated to bridging the gap between cancer surgery and technological innovation. This is achieved through the integration of a gamma detector and advanced computer vision algorithms.

The central focus of this work lies in the integration of a novel tethered laparoscopic  gamma detector into \acrshort{mis}, complemented by the development and exploration of sophisticated computer vision algorithms for the 3D geometric surgical scene analysis and laparoscopic gamma probe sensing area detection. These innovations are aimed at enhancing the surgeon's visual perception of cancer in the operative field and providing invaluable real-time feedback. One of the most promising future applications of this integration could be the creation of \acrfull{ar} and \acrfull{vr} systems customized to the needs of surgeons, providing a display of the location of cancerous tissues and metastases.

In the following sections, I will discuss and motivate the development of this thesis. I will then present a list of summaries of my contributions during my doctoral studies, emphasising key novelties, implementation, and potential impact on real-world cancer treatment applications. Additionally, a curated list of peer-reviewed publications stemming from this research is included, showcasing the academic significance and practical relevance of the work presented in this thesis.

As I embark on this journey, I have explored the fusion of medical science, technology, and \acrshort{ai} innovations, with the ultimate aim of advancing the capabilities of surgeons in their crucial mission to combat cancer and move towards more precise and personalised medicine.

\section{Motivation}


The landscape of modern surgery is characterized by remarkable advancements as well as persistent challenges. At the forefront of healthcare innovation, precision medicine seeks to customize different treatments for each disease. My research aligns with this trend by enhancing surgical precision, contributing not only to disease-specific care but also to universally applicable solutions. Given the increasing incidence of cancer, there is a growing need for more accurate intraoperative cancer detection and dissection as any missed disease can lead to complications, positive surgical margins, and the need for additional surgeries. On the other hand, excessive dissection can impair organ function and postoperative quality of life. These issues not only compromise patient outcomes, but can also escalate healthcare costs. However, surgeons face difficulties in achieving complete cancer tissue removal, even with cutting-edge preoperative imaging systems, due to the lack of reliable intraoperative localisation methods.

Recently, endoscopic radio-guided cancer resection has emerged, featuring a miniaturised gamma radiation detection probe. This probe has the potential to detect signals emitted by targeted nuclear agents, thereby indicating the location of cancer. However, as the probe may not touch the tissue and the probe is non-imaging, there is a visualisation challenge to use this kind of tool during the surgery. Addressing this, and thereby inferring the gamma probe's detection area -- the intersection point between the probe axis and the tissue -- is invaluable for providing visual cues and feedback to surgeons with positive signals eligible for indicating the presence of cancer or affected lymph nodes. Additionally, these aspects, along with the gamma count, represent fundamental knowledge for the potential integration of \acrshort{ar} and \acrshort{vr} technologies into surgical practice, offering the promise of enhanced visualisation and precision.

Furthermore, in the fields of medical imaging and surgical vision, classical problems such as surgical tool tracking, pose estimation, segmentation, and depth estimation of laparoscopic images -- which are related to the gamma probe sensing area inference -- remain inadequately explored. These technical challenges are widely acknowledged and have far-reaching implications for the field, necessitating more effective and efficient solutions across the surgical community.

While the undeniable popularity of \acrshort{mis} stems from its associated benefits, the challenges within this technology demand solutions to ensure its continued advancement and accessibility. Recognising that synergy between hardware and software is required in the medical field, my research explores the potential of combining these elements to facilitate precision medicine. I believe that this interdisciplinary approach will provide the ability to revolutionise healthcare practice. Furthermore, the computer vision algorithms developed for gamma probe localisation and the hardware platforms constructed for validation can be extended to other medical instruments, such as ultrasound probes and \acrfull{drs}, exhibiting scalability for broader applications thus amplifying the impact. 

In light of these pressing challenges, opportunities, and my passion, this research is poised to harness the potential of gamma signal detection, computer vision algorithms, and the synergy of hardware and software. The goal is not only to address immediate issues in cancer surgery but also to lay the foundation for a transformative era in precision medicine and minimally invasive surgery.


\section{Contributions}

This field encompasses a wide range of disciplines, from surgery and cancer to computer science and engineering. My main focus is on developing innovative techniques and technologies that can be used to solve complex problems in a variety of domains. Despite the aforementioned challenges and difficulties, the work presented in this thesis provides several significant contributions to the research field, which are summarised as follows:

\colorlet{shadecolor}{MidnightBlue!5}
\begin{shaded*}
\textbf{Contribution 1}

I proposed a novel method for tracking and visualising the sensing area of a tethered laparoscopic gamma probe used in surgical procedures. A tracking and pose estimation method for the gamma probe was developed involving attaching a dual-pattern marker to the probe to improve detection accuracy. A visualisation system was presented with tissue surface reconstructed using \acrlong{sfm} (\acrshort{sfm}) and sensing area augmented reality was displayed on the surgical view, allowing surgeons to easily identify the location of radioactive sources within the body. 

\end{shaded*}

\begin{shaded*}
\textbf{Contribution 2}

I am one of the first to propose a new self-supervised depth estimation method based on \acrlong{gan} (\acrshort{gan}). These results are significant and can potentially be applied in different applications such as robotic navigation, 3D registration between pre- and intra-operative organ models, or \acrshort{ar}. I further achieved improved results by presenting a self-supervised depth estimator leveraging 3D geometric structural information hidden in stereo pairs while maintaining monocular inference.

\end{shaded*}

\begin{shaded*}
\textbf{Contribution 3}

I built a portable stereo laparoscope system and a control system integrating a gamma probe, laparoscope, structured lighting system, miniaturised cameras, laser module, shutter, and a rotational stage. This platform allowed sequential and automatic data collection on phantoms in the lab and on real human tissue in the operating theatre, for not only RGB images but also the depth ground truth calculation. The newly required dataset can further validate the proposed \acrshort{ai} system.

\end{shaded*}

\begin{shaded*}
\textbf{Contribution 4}

I proposed a new framework for using a laparoscopic drop-in gamma detector in minimally invasive cancer surgery, where a laser module mock probe was utilised to provide training guidance and the problem of detecting the probe axis-tissue intersection point was transformed to laser point position inference. Both the hardware and software design of the proposed solution were illustrated and two newly acquired datasets were publicly released. The problem reformulation and datasets release, together with the initial experimental results with real time inference, established a new benchmark for the surgical vision community. 

\end{shaded*}



\section{Thesis Outline}
The thesis outline is provided below:

I first discuss the background of surgical interventions for cancer treatment in Chapter~\ref{chap:2-rw}, highlighting their critical role in improving patient outcomes and then review advancements in precision and reduced invasiveness, specifically focusing on prostate surgery and lymph node detection. This chapter followed by an introduction to the \acrshort{sensei} probe, detailing its significance and functionality in the thesis. Recent probe-tracking techniques and medical image segmentation methods are examined in this chapter, emphasising their importance in enhancing spatial awareness and accuracy during tumour resections.

Chapter~\ref{chap:3-contribution_1} proposes a robust gamma probe tracking system integrated with \acrshort{ar}. It introduces a dual-pattern marker system combining chessboard vertices and circular dots for improved detection accuracy. The system incorporates temporal information to reduce tracking failures and utilises a 3D point cloud for precise localisation. The proposed method is validated against the OptiTrack system, demonstrating accurate pose estimation.

I then introduce two depth estimation algorithms for laparoscopic images in Chapter~\ref{chap:4-contribution_2-Depth2D} and Chapter~\ref{chap:5-contribution_3-Depth3D}.  Chapter~\ref{chap:4-contribution_2-Depth2D} introduces SADepth, a self-supervised depth estimation method based on \acrshort{gan}. SADepth employs an encoder-decoder generator and discriminator to integrate geometry constraints during training. Multi-scale outputs address local minima issues, and adversarial learning improves output quality. Experiments on public datasets show that SADepth significantly outperforms existing unsupervised methods, narrowing the gap with supervised depth estimation in laparoscopic images. Chapter~\ref{chap:5-contribution_3-Depth3D} presents M3Depth, a self-supervised depth estimator that utilises 3D geometric structural information from stereo image pairs. Unlike previous methods focusing solely on 2D consistency, M3Depth incorporates 3D spatial data to enhance depth estimation. It also implements masking techniques to eliminate border region influence, improving image correspondence. Extensive testing on public and newly acquired datasets indicates that M3Depth surpasses previous self-supervised methods, demonstrating strong generalisation across different samples and laparoscopes.

Chapter~\ref{chap:6-contribution_4-DepthSeg} introduces a unified framework for simultaneous depth estimation and surgical tool segmentation in laparoscopic images. The proposed network features an encoder-decoder architecture with two branches dedicated to each task. A novel multi-task loss function enables effective unsupervised depth estimation while utilising semi-ground truth for tool segmentation. Extensive experiments across various datasets validate that this end-to-end network enhances state-of-the-art performance for both tasks, simplifying their deployment in robotic surgery.

Chapter~\ref{chap:7-contribution_-5PtRegress} addresses the challenge of localising radioactive sources on tissue surfaces during endoscopic procedures. Initial attempts using segmentation or geometric methods were ineffective. Instead, a solution leveraging high-dimensional image features and probe position data is proposed. A simple regression network was developed and validated using datasets from a custom-designed portable stereo laparoscope system. Experiments confirm the method's effectiveness, setting a new benchmark for detecting radioactive sensing areas in endoscopic radio-guided cancer detection and resection.
\section{Publications}

This section presents a list of peer-reviewed publications that were completed during my PhD studies. The full references for each publication are also provided for easy access and citation purposes.

\begin{itemize}

\item \textbf{Baoru Huang}, Yicheng Hu, Anh Nguyen, Stamatia Giannarou, Daniel S. Elson. \textit{Detecting Sensing Area of A Laparoscopic Probe in Minimally Invasive Cancer Surgery}. International Conference on Medical Image Computing and Computer-Assisted Intervention (MICCAI), 2023. (\textbf{\textcolor{blue}{Oral presentation, reported by `MICCAI 2023 Daily'}}). \cite{huang2023detecting}

\item \textbf{Baoru Huang}, Jian-Qing Zheng, Anh Nguyen, Chi Xu, Ioannis Gkouzionis, Kunal Vyas, David Tuch, Stamatia Giannarou, Daniel S. Elson. \textit{Self-supervised Depth Estimation in Laparoscopic Image Using 3D Geometric Consistency}. International Conference on Medical Image Computing and Computer-Assisted Intervention (MICCAI), 2022. \cite{huang2022self}

\item \textbf{Baoru Huang}, Jian-Qing Zheng, Anh Nguyen, David Tuch, Kunal Vyas, Stamatia Giannarou, Daniel S. Elson. \textit{Self-supervised generative adversarial network for depth estimation in laparoscopic images}. International Conference on Medical Image Computing and Computer-Assisted Intervention (MICCAI), 2021. \cite{huang2021self}

\item \textbf{Baoru Huang}, Kunal Vyas, David Tuch, Stamatia Giannarou, Daniel S. Elson. \textit{Self-supervised Monocular Laparoscopic Images Depth Estimation Leveraging Interactive Closest Point in 3D to Enable Image-guided Radioguided Surgery}. European Molecular Imaging Meeting (EMIM), 2022. (\textbf{\textcolor{blue}{Best Poster Award}})

\item \textbf{Baoru Huang}, Ya-Yen Tsai, João Cartucho, Kunal Vyas, David Tuch, Stamatia Giannarou, Daniel S. Elson. \textit{Tracking and visualization of the sensing area for a tethered laparoscopic gamma probe}. International Journal of Computer Assisted Radiology and Surgery (IJCARS), 2020, 15(8): 1389-1397. \cite{huang2020tracking}
    
\item \textbf{Baoru Huang}, Anh Nguyen, Siyao Wang, Ziyang Wang, Erik Mayer, David Tuch, Kunal Vyas, Stamatia Giannarou, Daniel S. Elson. \textit{Simultaneous Depth Estimation and Surgical Tool Segmentation in Laparoscopic Images}. IEEE Transactions on Medical Robotics and Bionics (T-MRB), 2022, 4(2): 335-338. \cite{huang2022simultaneous}

\item Zheng, Jian-Qing, Ziyang Wang, \textbf{Baoru Huang}, Ngee Han Lim, Bartłomiej W. Papież. \textit{Residual Aligner-based Network (RAN): Motion-separable structure for coarse-to-fine discontinuous deformable registration.} Medical Image Analysis (MedIA), 2024: 103038. \cite{zheng2024residual}

\item Kaizhong Deng, \textbf{Baoru Huang}, Daniel S Elson. \textit{Deep Imitation Learning for Automated Drop-In Gamma Probe Manipulation}, Hamlyn Symposium, 2023. \cite{deng2023deep}

\item Xu Chi, \textbf{Baoru Huang}, Daniel S. Elson. \textit{Self-supervised monocular depth estimation with 3-D displacement module for laparoscopic images}. IEEE transactions on medical robotics and bionics (T-MRB), 2022, 4(2): 331-334. \cite{xu2022self}

\item João Cartucho, Chiyu Wang, \textbf{Baoru Huang}, Ara Darzi, Stamatia Giannarou. \textit{An enhanced marker pattern that achieves improved accuracy in surgical tool tracking}. Computer Methods in Biomechanics and Biomedical Engineering: Imaging \& Visualization, 2022, 10(4): 400-408. \cite{cartucho2022enhanced}
    
\end{itemize} 


\setcounter{chapter}{1}
\chapter{Background}
\chaptermark{Background}
\glsresetall
\label{chap:2-rw}

\begin{cabstract}
This chapter first discusses contemporary surgical interventions for cancer treatment, emphasizing the crucial role they play in improving patient outcomes. I summarize the ongoing advancements that have enhanced precision and reduced invasiveness for prostate surgery and lymph node detection. Next, I provide technical details of the \acrshort{sensei} probe which is the state-of-the-art tool that is used throughout this thesis. I review recent probe-tracking techniques and medical image analysis methods, highlighting their pivotal roles in modern cancer surgery. Probe tracking systems offer improved spatial awareness, enabling more accurate and effective tumor resections. 
The latter part of this chapter discusses the segmentation in medical images as segmenting is essential for enhancing surgical navigation and ensuring the safety and precision of procedures. Finally, I discuss depth estimation in medical images. This investigates techniques for accurately estimating depth information from 2D medical images, enhancing the visualization of complex anatomical structures and tumor boundaries. This improved depth perception aids surgeons in making more informed decisions during cancer surgery.


\end{cabstract}


\section{Surgery for Cancer}
\label{rw_cancel_treament}
Surgery is one of the most effective and oldest forms of cancer treatment and is used in several ways for cancer treatment ~\cite{devita2012two}. By offering the opportunity to halt the progression of various cancer forms, surgery plays a pivotal role not only in diagnosing and staging cancer but also in supporting the overall cancer treatment process. There are three different approaches to cancer surgery, including \textit{i)} open, \textit{ii)} laparoscopic, and \textit{iii)} flexible endoscopic surgery. 




\textbf{Open Surgery.} Open surgery involves a surgical procedure where the surgeon creates a single (frequently sizable) incision, and on occasion, multiple incisions ~\cite{velanovich2000laparoscopic}. This approach is often employed in various medical cases, such as in breast surgery, where an additional incision might be necessary in the armpit region to facilitate the removal of lymph nodes. The primary aim of open surgery is to access the affected area directly, enabling the surgeon to perform intricate interventions with precision and care. Though open surgery has been widely used and proven to be effective, advances in medical technology and techniques have led to the development of minimally invasive procedures, which often involve smaller incisions and shorter recovery times. Despite this, open surgery remains a crucial option in complex cases, where a more extensive approach is required to address the patient's medical needs effectively.

\textbf{Laparoscopic Surgery.} Laparoscopic surgery, also known as \acrshort{mis}, has become an essential and increasingly utilised technique in the management of cancer~\cite{sauerland2010laparoscopic}. This advanced surgical approach offers numerous benefits for cancer patients compared to traditional open surgery. In laparoscopic surgery for cancer, the surgeon makes several small incisions in the patient's abdominal area or affected region instead of creating a single large incision. Through these small incisions, a laparoscope -- a thin tube equipped with a light source and camera -- is inserted. This laparoscope provides a magnified view of the internal organs and structures, allowing the surgical team to precisely identify and assess the cancerous tissue.

One of the primary advantages of laparoscopic surgery for cancer is its reduced invasiveness. Smaller incisions lead to less trauma to surrounding tissues, resulting in less postoperative pain and discomfort for the patient. Additionally, the reduced trauma contributes to faster healing and a shorter recovery period, enabling patients to return to their normal activities more swiftly, thus reducing overall costs.

For certain types of cancer, such as colorectal cancer, laparoscopic surgery has become the standard of care for early-stage cases and selected advanced cases~\cite{basunbul2022recent}. The removal of cancerous growths in the colon or rectum through laparoscopic techniques has shown equivalent oncological outcomes compared to open surgery while providing the patient with a better overall surgical experience. Laparoscopic surgery for cancer is not limited to colorectal cancer; it is also used in the management of gynaecologic cancers, such as ovarian and uterine cancers, as well as certain types of liver, kidney, and pancreatic cancers~\cite{antonakis2014laparoscopic}. However, the applicability of laparoscopic techniques in cancer surgery may vary depending on factors such as the tumour size, location, and stage, as well as the surgeon's expertise and experience.

As with any surgical procedure, laparoscopic surgery for cancer carries some risks, such as bleeding, infection, or injury to nearby organs~\cite{kodera2010laparoscopic}. However, overall, this approach has proven to be safe and effective in appropriately selected cases, offering cancer patients a valuable alternative to traditional open surgery. As medical technology continues to advance, laparoscopic techniques are likely to play an even more significant role in the comprehensive and personalised care of cancer patients in the future.

\textbf{Endoscopic Surgery.} In some cases, cancer tissue can be removed without making any cuts through the skin via endoscopic surgery \cite{serra2015transanal}. Endoscopic surgery involves the use of an endoscope -- a flexible tube equipped with a light source and camera -- which is inserted into the body through natural openings or small incisions. The endoscope allows the surgical team to visualise the internal structures and tumour site in real time on a monitor, providing a magnified and detailed view.

Similar to laparoscopic surgery, one of the primary benefits of endoscopic surgery is its minimally invasive nature. The use of small incisions or natural body openings reduces trauma to surrounding tissues, muscles, and organs, resulting in less postoperative pain, decreased scarring, and faster recovery times for the patient. Additionally, the risk of infection and other surgical complications is generally lower compared to open surgery.

Endoscopic surgery can be employed in various types of cancer, depending on the tumour's location and accessibility~\cite{ribeiro2015endoscopic}. For example, it is commonly used in the management of gastrointestinal cancers, such as oesophageal, stomach, and colorectal cancers. In these cases, the endoscope is passed through the mouth or anus to reach the affected area, allowing the surgeon to remove or treat cancerous tissues with precision. Endoscopic procedures are also used in the treatment of certain gynaecological cancers, such as early-stage cervical cancer~\cite{lee2014natural}. The endoscope is introduced through the vagina to access and remove cancerous growth without the need for external incisions.


\section{Prostate Surgery and Lymph Node Detection}
Cancer stands as a prominent public health challenge in the UK. A new diagnosis of this disease occurs every two minutes (Cancer Research UK~\cite{CancerUK}). The total cost for preventable cancer cases diagnosed in the UK in 2023 is estimated at £78 billion, representing 3.5\% of the annual GDP. This assessment considers five key cost categories: individual, healthcare, social care, family and carer, and productivity. Notably, the largest cost contributors for the year 2023 are productivity loss, accounting for £40 billion, and individual costs, amounting to £30 billion. These figures are influenced by factors such as the impact on quality of life and the unpaid productivity lost due to mortality (Frontier Economics~\cite{CostCancer}). With predictions that almost half of the population will be diagnosed with cancer within their lifetime, the health and economic implications for the UK are considerable (NHS~\cite{NHSCancer}).

Among the many kinds of cancers, prostate cancer is the most common cancer in men with 47,700 new cases and 11,500 deaths each year (Cancer Research UK~\cite{CancerUK}), representing an enormous economic burden to the healthcare system. Surgery is one of the main curative treatment options for the disease, however, despite substantial advances in pre-operative imaging (i.e. \acrshort{ct}, \acrshort{mri}, \acrshort{pet}/\acrshort{spect}) to aid diagnosis, surgeons still rely on the sense of touch and naked eye to detect cancer and disease metastases intraoperatively, because of the lack of reliable intraoperative visualisation tools. 

Minimally-invasive approaches by either manually manipulated laparoscopy or robotic-assisted technology are increasingly used due to their proven benefits and significant advantages as illustrated above, such as reduced bleeding, risk of infection, trauma to the patient's tissues, and convalescence~\cite{Trinh2012PerioperativeSample}. However, the increased application and wide use of such approaches make the opportunity for visual inspection and palpation even more limited. As a consequence, cancer is frequently missed, which leads to an increased need for further chemotherapy, radiotherapy, and surgery, or healthy tissue is needlessly removed, resulting in significant consequences for patient health, organ functions and substantial costs for healthcare systems.


The main challenges of prostate cancer surgery are two-fold: \textit{i)} to fully remove the primary tumour, and \textit{ii)} to remove metastases in the lymphatic system. Surgeons often fail to remove all the cancerous tissue surrounding the primary tumour, as evidenced by a positive surgical margin which was reported in 11-38\% of patients~\cite{spahn2013positive}. The consequences of a positive margin include increased risk of cancer recurrence, additional treatment, decreased disease-specific survival, and additional treatment ~\cite{Bott2002AvoidanceProstatectomy}. The recurrence risk is a major impediment to nerve-sparing surgery, which aims to preserve the nerves responsible for erectile function and urinary continence. The approach results in better erectile function and continence than non-nerve sparing surgery, but involves a risk of exposing cancer at the edge of the resected specimen \cite{michl2016nerve}. 

The removal of cancer spread via the lymphatic system during surgery is a high priority since these metastases increase the risk of recurrence and death~\cite{boorjian2007long}. However, due to the challenges of intraoperative detection of cancer, surgeons usually perform an extended pelvic lymph node dissection and extensively dissect healthy lymphatic tissue~\cite{Mottet2017EAU-ESTRO-SIOGIntent}. The consequences of extensive lymph node dissection for patients and healthcare systems are considerable with significantly increased risk of adverse events (deep venous thrombosis, pulmonary embolism, lymphoceles, blood loss), morbidity (wound complications, respiratory, cardiovascular, and neuromusculoskeletal events), and lengthened hospital stays (almost 2.5-fold higher readmission rates and over 2.5-fold higher re-operation rate). In addition to the costs related to managing post-operative complications, performing an extensive lymph node dissection adds approximately an hour to the operating time, at an additional cost of approximately \$$3,000$ per patient~\cite{CostCancer}.


If a surgeon could accurately detect cancer intraoperatively, a more selective dissection of lymph nodes could be applied, whereby cancer spread is assessed in the sentinel lymph node, which is the first node to become metastatically involved, and thus sparing other lymph nodes in the pelvis. In traditional open surgery, the technique is possible and widely validated using a rigid, hand-held gamma probe \cite{maurer2015prostate}. However, with the growing dominance of minimally invasive laparoscopic and robotic cancer surgery, it is extremely challenging to use such a rigid tool and therefore surgeons have continued to extensively dissect the pelvic lymph nodes. Hence, a more suitable and flexible gamma probe is needed.

In addition to the potential benefits to patients in prostate cancer surgery, the radiotracer-guided cancer detection technology could have applications in other minimally invasive cancer procedures such as lung, colorectal, and hepatobiliary cancer surgery which also face daunting challenges of detecting metastatic lymph nodes.

\section{Imaging Agent and \acrshort{sensei} Probe}
The prostate cancer imaging agent 99mTc-PSMA targets \acrfull{psma}. \acrshort{psma} is a cell-surface protein that shows a significant over-expression on prostatic cancer cells and especially in advanced stage prostate carcinomas with low expression in normal human tissue \cite{silver1997prostate, perner2007prostate}. Upon ligand binding, \acrshort{psma} is internalized via clathrin-coated pits and subsequent endocytosis \cite{rajasekaran2003novel}, resulting in effective transportation of the bound molecule into the cells leading to enhanced tumour uptake and retention and \acrshort{psma} imaging is extensively validated clinically with joint consensus guidelines issued by the US and EU professional societies \cite{fendler201768ga}. 99mTc-PSMA has shown high diagnostic performance of $>\!\!90\%$ sensitivity in prostate cancer patients with $>\!\!2ng/ml$ \acrfull{psa} levels \cite{reinfelder2017first, schmidkonz2018spect}. The agent has reported positive Phase \uppercase\expandafter{\romannumeral2} results \cite{goffin2017phase}. 

Intraoperative use of 99mTc-PSMA in prostate cancer surgery has been applied previously for open surgery with a hand-held gamma probe \cite{maurer2015prostate, rauscher2017value}. In those studies, investigators found a high detection rate of $>\!\!90\%$ for lymph node metastases in recurrent patients. However, most prostate cancer procedures are performed minimally invasively using robotic assistance. Therefore, a laparoscopic trocar compatible probe is required to address the true clinical need.

\begin{figure*}[h]
    \centering
    \includegraphics[height=0.6\linewidth, width=0.99\linewidth]{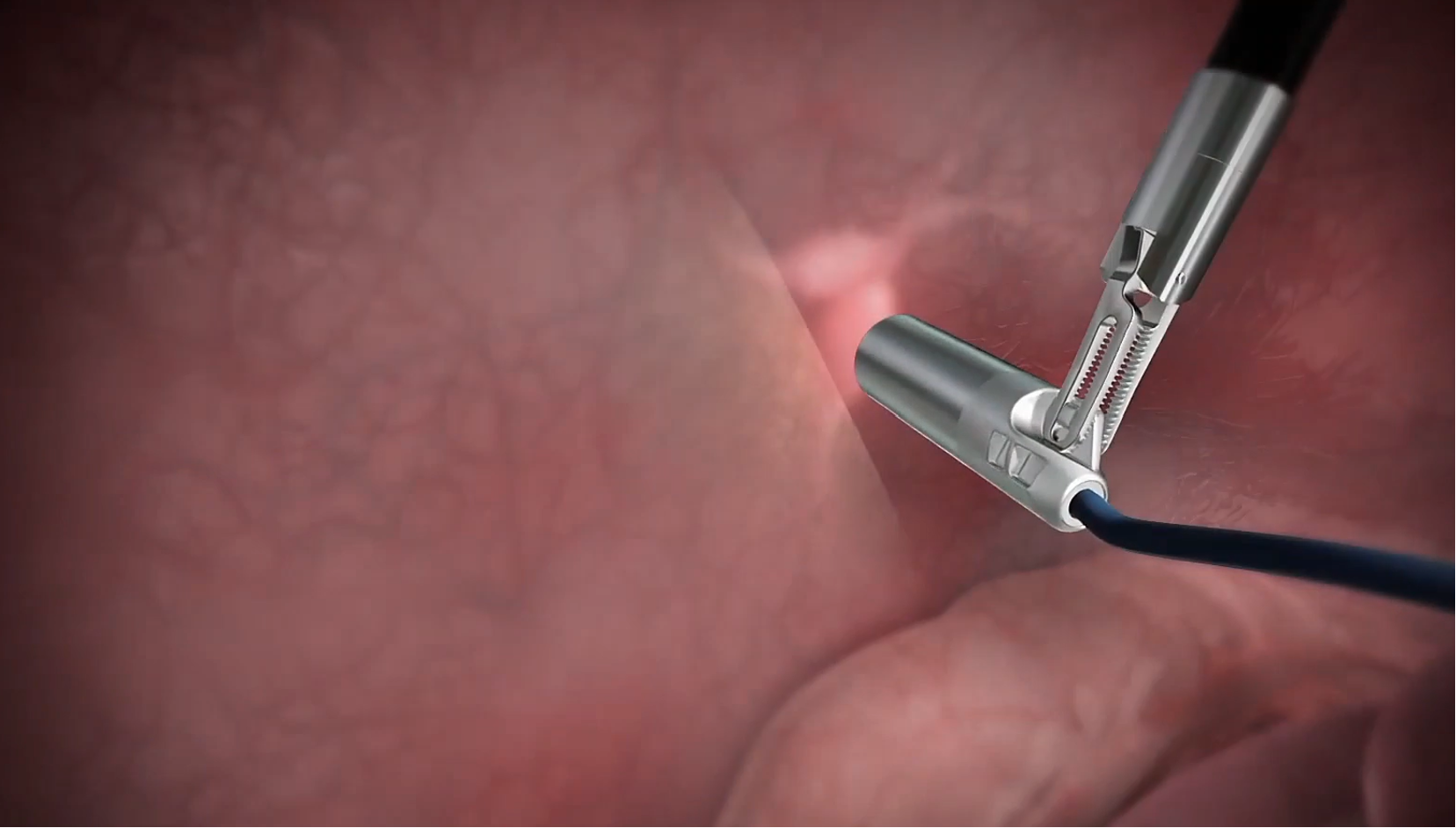}
    \caption{The \acrshort{sensei} probe. Image courtesy of \href{https://lightpointmedical.com/}{Lightpoint Medical Ltd.}}
\label{fig_SENSEI_probe}
\end{figure*}

\href{https://lightpointmedical.com/}{Lightpoint Medical Ltd.} has developed a miniaturised cancer detection probe that is fully compatible with all contemporary surgical techniques, including minimally invasive laparoscopic and robotic surgical systems, and provides the potential for intraoperative detection of cancer in surgery, reducing complications to patients and costs to the healthcare system. The device, \acrshort{sensei}, is a tethered laparoscopic probe, operating on the same detection principles of rigid, hand-held probes. It enables high-sensitivity detection of gamma imaging agents that are administered to the patient prior to surgery and selectively accumulate in cancerous tissue, by providing an audible and numerical read-out (Fig.~\ref{fig_SENSEI_probe}). Compared with the previous hand-held gamma probe, the \acrshort{sensei} probe has been miniaturised to a length of only 30 millimetres to ensure full compatibility with minimally invasive key-hole laparoscopic surgery and robotic systems. In addition, the probe has a novel grip to grasp the probe with a standard tissue grasper, enabling full manoeuvrability within the surgical cavity. Moreover, the tethered probe can remain \textit{in vivo} throughout the procedure, without preventing other instruments from being inserted through the surgical trocar, ensuring full compatibility with the clinical workflow and ease of use. 

However, the use of such a gamma detector presents a visualisation challenge and a need for image-guided intervention and navigation. As the probe may not be in contact with tissue during the surgery and the target site may be obscured from the perspective of the operating laparoscope, making the interpretation of the \acrshort{sensei} data even more difficult. Furthermore, the surgeon has to memorise the detected tissue surface and the previously acquired probe data, which is inefficient and increases the surgeon's workload. Additionally, it may lead to an increased risk of cancer not being fully removed or missing positive lymph nodes. Hence, there is a requirement for a method that can aid the visualisation of the signal and that integrates well with the endoscope camera.

\section{Probe Tracking}
\label{rw_proble_tracking}

Probe tracking in \acrshort{mis} represents an important task that addresses critical challenges inherent in procedures characterised by reduced invasiveness~\cite{mohiuddin2013maximizing}. In \acrshort{mis}, where surgeons operate through small incisions or natural body orifices, maintaining precise control and spatial orientation can be challenging due to limited direct visibility. The integration of probe tracking methods offers a dynamic solution. By accurately tracking the position and orientation of surgical probes or instruments in real-time, surgeons gain invaluable insights into the internal anatomy, enhancing their ability to navigate complex anatomical structures with enhanced precision~\cite{sinha2017three}. This not only improves the overall safety of the procedure by reducing the risk of inadvertent tissue damage but also contributes to increased efficiency and success rates in minimally invasive interventions. As the field of \acrshort{mis} continues to evolve, the role of probe tracking becomes increasingly central in advancing surgical capabilities and improving patient outcomes.

In the past few years, several probe tracking methods have been proposed. Hughes \textit{et al.}~\cite{hughes2013intraoperative} introduced an intraoperative probe tracking method to identify anatomical structures like vessels and tumours. However, a notable drawback of this method is that surgeons still need to interpret the ultrasound images and mentally relate them to the laparoscopic view~\cite{mohareri2015intraoperative}. This task is particularly demanding since the ultrasound images and laparoscopic videos are typically displayed on separate screens. To alleviate the cognitive burden on surgeons, a probe tracking method has been proposed in~\cite{hughes2013intraoperative}. This method involves registering the 2D ultrasound images into the surgical scene, allowing surgeons to simultaneously view relevant details from both the ultrasound and laparoscope on a single display. This integrated approach provides consistent image overlays, improving the surgeon's ability to comprehend critical information during the procedure. In practice, it is crucial to track the ultrasound probe throughout the operation to ensure the accuracy and alignment between the two imaging modalities~\cite{hughes2013intraoperative}.

There are two popular methods to track medical devices, including vision-based methods and tracking-devices-based methods~\cite{zhang2017real}. Due to the nature of laparoscopy, using laparoscopic images directly for instrument tracking is a more practical approach compared to other methods that require additional tracking devices~\cite{feuerstein2007magneto}, which take up valuable space in the surgical field and are also susceptible to inherent limitations such as ferromagnetic interference and line-of-sight issues.

\begin{table}[h]
 \scriptsize
\caption{Visual marker-based tracking methods comparison for the intraoperative probe.}
\def\arraystretch{1.1}
\begin{center}
\begin{tabular}{l l c c c c c c c c }
\toprule
\multicolumn{0}{c}{Method} & 
\multicolumn{0}{c}{Publication}& \multicolumn{0}{c}{Probe Shape} & \multicolumn{0}{c}{Pattern Type} & \multicolumn{0}{c}{Usage} & \multicolumn{0}{c}{AR/VR} & \multicolumn{0}{c}{Real-time} & \multicolumn{0}{c}{Code?}  \\
\midrule
Pratt~\etal~\cite{pratt2012intraoperative} & MICCAI12        & Planar & Chessboard     & Freehand  & No  & Yes & No     \\
Edgcumbe~\etal~\cite{edgcumbe2013calibration} &MICCAIW13 & Planar        & Circular dot  & Robotic & No  & No        & No      \\
Jayarathne~\etal~\cite{jayarathne2013robust}     &MICCAI13       & Cylindrical        & Chessboard & Freehand & No  & No    & No     \\Pratt~\etal~\cite{pratt2015robust}                 &IJCARS15  & Planar        & Chessboard  & Robotic & No  & Yes        & No    \\
Zhang~\etal~\cite{zhang2017real}   &IJCARS17       & Cylindrical       & Hybrid & Both & No      & Yes & No     \\   
\cmidrule{1-8}
Huang~\etal~\cite{huang2020tracking} &IJCARS21                          & Cylindrical                & Hybrid  & Both & Yes & Yes   & Yes \\
\bottomrule
\end{tabular}
\label{tab:trackingproble_comparison}
\end{center}
\end{table}

Table~\ref{tab:trackingproble_comparison} shows a summary of marker-based tracking methods for intraoperative probes. Pratt~\etal~\cite{pratt2012intraoperative} introduced the first tracking system using a planar probe shape. The chessboard pattern is utilised as the vision marker enabling real-time tracking. However, the chessboard pattern design is limited to fast motion and occlusion. The work in~\cite{edgcumbe2013calibration} proposed to use the circular dot as the pattern for robot-assisted probe tracking, but the system is not fast enough for real-time applications. To make the probe tracking more useful in laparoscopic surgery, the work  in~\cite{jayarathne2013robust} proposed to use a cylindrical shape and chessboard pattern. Pratt~\etal~\cite{pratt2015robust} designed a planar probe with the chessboard pattern for robotic tracking. However, as in other previous works that used the chessboard pattern, this method does not handle occlusions well. More recently, Zhang~\etal~\cite{zhang2017real} addressed the occlusion problem by designing a hybrid tracking pattern.

In this thesis, recognising the limitations of previous probe tracking methods, I propose a new solution for real-time probe tracking~\cite{huang2020tracking}. In particular, I design the probe shape as cylindrical with a new hybrid pattern. This new design allows the probe to be tracked in both freehand and robotic usage scenarios. Furthermore, the tracking method is integrated into an \acrshort{ar}/\acrshort{vr} environment. The whole system is lightweight and has real-time performance. Finally, unlike other methods, my source code is made publicly available to the community. Chapter~\ref{chapter_marker_probe_tracking} describes in detail my new method for tracking the intraoperative probe.

\section{Surgical Image Segmentation}
\label{re_surgical_segmentation}

\subsection{Datasets}

\begin{table}[h]
\footnotesize
\centering
\caption{Publicly available medical image segmentation datasets}
\label{tab:segmentation-datasets}
\begin{tabular}{|l|l|c|c|}
\hline

\hline
\textbf{Dataset} & \textbf{Type} & \textbf{\#Images} & \textbf{Annotation} \\ \hline
RMIT~\cite{sznitman2012data} & Retinal Instruments & 1,500 & Manual \\ \hline
InstrumentCrowd~\cite{maier2014can} & Laparoscopic Instruments & 120 & Manual \\ \hline
NeuroSurgicalTools~\cite{bouget2015detecting} & Neurosurgical Microscopes & 2,476 & Manual \\ \hline
EndoVis2015~\cite{endo15} & Endoscopic Tools  & 9,000 & Manual \\ \hline
EndoVis2017~\cite{allan20192017} & Endoscopic Tools & 1,800 & Manual \\ \hline
EndoVis2018~\cite{allan20202018} & Endoscopic Tools & 3,384 & Manual\\ \hline
ROBUST-MIS2019~\cite{ross2020robust} & Endoscopic Tools & 10,040 & Manual \\ \hline
Kvasir-Instrument~\cite{jha2021kvasir} & Gastrointestinal Endoscopy & 590 & Manual\\ \hline
CaDIS/CATARACTS~\cite{grammatikopoulou2021cadis} & Cataract  &  4,670 & Manual \\ \hline
CholecSeg8K~\cite{hong2020cholecseg8k} & Laparoscopic Cholecystectomy & 8,080 & Manual\\ \hline
RoboTool~\cite{garcia2021image} & Surgical Tools & 514 & Automatic \\ \hline

\hline
\end{tabular}
\end{table}

The availability of openly accessible datasets has played a pivotal role in driving advancements in medical image segmentation in recent years, as highlighted in~\cite{rodrigues2022surgical}. One of the early datasets, RMIT~\cite{sznitman2012data} consists of approximately 1,500 images from three sequences during retinal microsurgery. The instrument position and groundtruth have been manually annotated. The InstrumentCrowd~\cite{maier2014can} dataset has data collected from three laparoscopic adrenalectomies and three laparoscopic pancreatic resections. From each surgery, 20 images containing one or several medical instruments are extracted and manually labeled. The NeuroSurgicalTools dataset~\cite{bouget2015detecting} consists of 2,476 images, of which 1,221 are used for training and 1,255 for testing. The images of NeuroSurgicalTools are collected from \textit{in vivo} neurosurgeries. EndoVis2015~\cite{endo15} is a popular dataset with approximately 9,000 \textit{in vivo} images from four laparoscopic colorectal surgeries. More recently, the EndoVis2017~\cite{allan20192017} has approximately 1,800 frames from robotic surgical videos. Each image in this dataset is manually labeled with different tool parts and types. 

In the EndoVis2018~\cite{allan20202018} dataset, the images were collected from 16 robotic nephrectomy procedures recorded using da Vinci Xi systems. The image resolution is 1280$\times$1024 and both the left and right images and stereo camera calibration parameters are provided. However, only the left images are labeled. The ROBUST-MIS2019 dataset~\cite{ross2020robust} has approximately 10,000 images from the video snippet of endoscopic procedures. The Kvasir-Instrument dataset~\cite{jha2021kvasir} consists of 590 annotated images from gastrointestinal endoscopy procedure tools such as snares, balloons, biopsy forceps, etc. The CaDIS and CATARACTS dataset~\cite{grammatikopoulou2021cadis} provided groundtruth for semantic segmentation of cataract surgery videos. The CholecSeg8K dataset~\cite{hong2020cholecseg8k} has 8,080 laparoscopic cholecystectomy images extracted and annotated from 17 video clips of the Cholec80 dataset. Recently, the RoboTool dataset~\cite{garcia2021image} is proposed with 514 images extracted from the videos of robotic surgical procedures. This dataset provides the annotation for binary tool and background segmentation. Table~\ref{tab:segmentation-datasets} shows the summary of medical image segmentation datasets.

\subsection{Segmentation Algorithms}
Surgical tool segmentation is an active research field~\cite{rodrigues2022surgical,huang2024cathaction}. There are three key directions to perform medical image segmentation: \textit{i)} Utilising deep networks for segmentation, \textit{ii)} Attention mechanism for medical image segmentation, and \textit{iii)} Data augmentation for medical image segmentation.

\textbf{Segmentation Network.} Employing different deep networks for medical segmentation is well-investigated in the literature. U-Net~\cite{ronneberger2015u} was one of the first novel methods for medical segmentation. The authors in~\cite{ross2020robust} tackled segmentation tasks using state-of-the-art vision networks, including Mask R-CNN~\cite{he2017mask} and U-Net. The work in~\cite{zisimopoulos2017can} used a VGG backbone~\cite{simonyan2014very} to conduct semantic segmentation. Isensee \textit{et al.}~\cite{isensee2020or} introduced a network with U-Net and residual blocks from ResNet in the encoder for adaptable segmentation map generation. Shvets \textit{et al.}~\cite{shvets2018automatic} used an encoder-decoder architecture similar to U-Net on the EndoVis 2017 dataset. Hasan and Linte~\cite{hasan2019u} modified the U-Net by using VGG16 as an encoder. The authors in~\cite{mohammed2019streoscennet} proposed a multi-encoder and a single decoder for part and binary segmentation tasks on the EndoVis 2017 dataset. Gonzalez \textit{et al.}~\cite{gonzalez2020isinet} introduced the fine-grained segmentation problem on the EndoVis 2018 dataset for instrument segmentation using a Mask R-CNN network. To increase inference time, Andersen \textit{et al.}~\cite{andersen2021real} proposed a mobile U-Net for the segmentation of surgical tools and suture needles. The work in~\cite{islam2019learning} proposed a spatio-temporal learning model with a shared encoder and spatio-temporal decoders for real-time surgical instrument segmentation. More recently, Allan \textit{et al.}~\cite{allan20202018} used the ResNeXt-101 architecture with squeeze-excitation blocks for segmentation. The authors in~\cite{choi2021video} use YOLOv4~\cite{bochkovskiy2020yolov4} for real-time object detection and semantic segmentation of surgical tools in a mastoidectomy surgery dataset.

\textbf{Attention Mechanism.} Following the recent trends in computer vision, the attention mechanism is widely used in medical segmentation. Fox \textit{et al.}~\cite{fox2020pixel} proposed to use the attention mechanism for surgical tool segmentation in ophthalmic cataract surgery with a Mask-RCNN backbone. Further work~\cite{jha2021exploring} introduced a dual decoder attention network for surgical tool segmentation. 
Other authors ~\cite{chen2021semi} proposed a novel approach with a cross-consistency mechanism for microscopic image segmentation using the Deeplab network. Ceron \textit{et al.}~\cite{ceron2021assessing} introduced a network architecture with a cross attention module for real-time instance segmentation of surgical instruments. Ni \textit{et al.}~\cite{ni2019rasnet} introduced a refined network that simultaneously segments and classifies surgical instruments using an attention fusion module. Lee \textit{et al.}~\cite{lee2019segmentation} combined attention with two-phase deep learning for laparoscopic image segmentation on the EndoVis 2017 dataset.  Rocha \textit{et al.}~\cite{da2019self} proposed a two-step algorithm for surgical tool segmentation using kinematic information. Kletz \textit{et al.}~\cite{kletz2019learning} used ResNet50 architecture and a feature pyramid network for instance image segmentation in gynecological surgeries.

\textbf{Data Augmentation.} 
Training deep networks requires a vast amount of data, but this requirement is not always available in the medical field due to privacy concerns. Data augmentation is a popular technique to overcome this limitation. Pissas \textit{et al.}~\cite{Theodoros} introduced the data oversampling strategy to overcome the limitation in medical segmentation. Sahu \textit{et al.}~\cite{sahu2021simulation} used knowledge distillation to learn from annotated simulation data and unlabeled real data. Zhang \textit{et al.}~\cite{zhang2021surgical} proposed a \acrshort{gan}-based method for unpaired image-to-image translation for surgical tool image segmentation. Kanakatte \textit{et al.}~\cite{kanakatte2020surgical} proposed a pixel-wise instance segmentation algorithm for the segmentation and localization of surgical tools using data augmentation and a spatial-temporal deep network.

\section{Depth Estimation in Surgical Imaging}
\label{sec:depthestimation}

\subsection{Dataset}

\begin{table}[h]
\footnotesize
\centering
\caption{Depth estimation datasets in medical imaging}
\label{tab:depth-datasets}
\begin{tabular}{|l|l|c|c|}
\hline

\hline
\textbf{Dataset} & \textbf{Procedure} & \textbf{\#Images} & \textbf{Annotation} \\ \hline
C3VD~\cite{bobrow2023colonoscopy} & Colonoscopy & 10,015 & \acrshort{ct} Data Alignment \\ \hline
EndoMapper~\cite{azagra2022endomapper} & Endoscopy & 59 sequences & Vicalib Tool~\cite{vicalib} \\ \hline
EndoSLAM~\cite{ozyoruk2021endoslam} & Endoscopy  & 64,587 & CV Toolbox~\cite{cvtoolbox} \\ \hline
ColonoscopyDepth~\cite{rau2019implicit} & Colonoscopy & 16,000 & \acrshort{ct} Data Alignment \\ \hline
dVPN~\cite{ye2017self} & Partial Nephrectomy& 48,702 & N/A\\ \hline
LATTE~\cite{huang2022self} & Laparoscopy & 839 & Structured Lighting(SL)\\ \hline
SCARED~\cite{allan2021stereo}  & Endoscopy & 9 sequences & SL and Interpolation \\ \hline

\hline
\end{tabular}
\end{table}

Table~\ref{tab:depth-datasets} shows an overview of recent depth estimation datasets in medical imaging. The C3VD dataset~\cite{bobrow2023colonoscopy} features imagery sourced from a phantom scenario, complete with known depth values. These images were captured employing an Olympus CF-HQ190L endoscope within a synthetic silicone model mimicking the structure of the human colon. The dataset was annotated with precise depth and normal information, achieved through the process of 2D-3D registration. The C3VD dataset comprises a collection of 10,015 images, rendering it an extensive and valuable asset for depth estimation in the medical image domain.

EndoMapper~\cite{azagra2022endomapper} is a challenging dataset for depth estimation due to its inclusion of authentic colonoscopy and gastroscopy procedures conducted within the human body by endoscopists in clinical practice. This dataset exhibits a rich array of real-world textures such as veins, blood, and contaminants, along with various artifacts like blur, and water interference. To ensure data quality, the authors employ a manual inspection process on the selected sequences, wherein occluded and excessively blurred frames are removed.

EndoSLAM~\cite{ozyoruk2021endoslam} is a recent dataset crafted for 6-DoF pose estimation and dense 3D reconstruction. The dataset is recorded using multiple endoscope cameras and ex-vivo porcine gastrointestinal organs from diverse animals, addressing critical requirements for the development of endoscopic \acrshort{slam} methods. To enrich the dataset, synthetically generated data from Unity environment is incorporated, facilitating the exploration of sim-to-real challenges such as domain adaptation and transfer learning. In addition, the authors also proposed an unsupervised approach for depth and pose estimation in endoscopic videos using spatial attention mechanisms and a brightness-aware hybrid loss. 

ColonoscopyDepth~\cite{rau2019implicit} represents the synthetic dataset for depth estimation. This dataset comprises a total of 16,016 RGB images, each image is paired with its respective ground truth depth information. Prior to analysis, the images underwent resizing to dimensions of 256 $\times$ 256 pixels. The dVPN dataset~\cite{ye2017self} was collected from da Vinci partial nephrectomy, with 34,320 pairs of rectified stereo images for training and 14,382 pairs for testing. 

The SCARED dataset \cite{allan2021stereo} was released at the MICCAI Endovis Challenge 2019. As only the ground truth depth map of keyframes was provided (from structured light), the other depth maps were created by reprojection and interpolation of the keyframe depth maps using the kinematic information from the \textit{da Vinci} robot, causing a misalignment between the ground truth and RGB data. Hence, only keyframe ground truth depth maps were used in the test set while the RGB data formed the training set, but with the similar adjacent frames removed. 

To overcome the SCARED dataset misalignment and improve the validation, in this thesis, I collected and proposed an additional laparoscopic image dataset (namely the LATTE dataset~\cite{huang2022self}). This includes RGB laparoscopic images and corresponding ground truth depth maps calculated from a custom-built structured lighting pattern projection. The LATTE dataset provided 739 image pairs for training and 100 pairs for validation and testing for depth estimation in laparoscopic images.

\subsection{Monocular Depth Estimation}

Monocular depth estimation has been intensively studied in different medical procedures such as endoscopic surgery. Visentini \textit{et al.}~\cite{visentini2017deep} used \acrshort{ct} renderings for depth supervision in bronchoscopies. In practice, obtaining \acrshort{ct} scans and ground-truth depth data is time-consuming, which makes self-supervision a necessity. The
common approach is to formulate a depth estimation problem as the minimisation of a photometric reprojection loss at the training stage~\cite{wang2020parallax,godard2019digging}. Recently, several works have been proposed to estimate the depth from monocular endoscopic images. In~\cite{liu2020reconstructing}, the authors incorporated recomputed matched points and camera poses from the \acrshort{sfm} to train a self-supervised
monocular depth estimation network on sinus video. Ozyoruk \textit{et al.}~\cite{ozyoruk2021endoslam} jointly estimated the camera pose and depth map on synthetically generated data. Other works have utilised video-based training to use temporal information for depth estimation~\cite{karaoglu2021adversarial,hwang2021unsupervised}. 

The key challenge in monocular depth estimation is due to the presence of deformations and weak texture. The authors in~\cite{rodriguez2022uncertain} proposed a method to estimate the uncertainty in monocular depth estimation in colonoscopies. The work in~\cite{li2022geometric} used geometric constraints for self-supervised monocular depth estimation, which includes a dual-task consistency loss for laparoscopic images. The authors in~\cite{oda2022depth} estimated the depth from a single-shot monocular endoscope image using domain adaptation and edge-aware loss.

\subsection{Stereo Depth Estimation}
Recently, methods~\cite{godard2019digging,yin2018geonet} that utilise stereo images
as supervision signals have achieved remarkable results in computer vision~\cite{huang2022h}. One of the first self-supervised stereo-matching methods
implemented on a pair of stereo endoscopic images was developed by Ye \textit{et al.} in~\cite{ye2017self}. Following this, many works have explored multi-view integration for depth estimation in medical imaging~\cite{luo2019details,xu2019unsupervised}, or combining tracking and \acrshort{slam} pipelines~\cite{recasens2021endo,ozyoruk2021endoslam}. The authors in~\cite{liu2019dense} used monocular videos and multi-view stereo to provide weak depth supervision. More recently, the authors in~\cite{yang2021dense} proposed a method for dense depth estimation from stereo endoscopy videos using unsupervised optical flow networks.

The specificity of the medical domain leads to a difficulty in data collection, and therefore many works have utilised synthetic data. Consequently, simulation-to-real (sim2real), as a challenge for depth estimation on real dataset has also been extensively explored. For instance, in~\cite{shen2019context} a conditional \acrshort{gan} was proposed for depth recovery while integrating \acrshort{slam} and multi-view inputs. In~\cite{chen2019slam}, the authors trained a network with synthetic images of a simple colon model and then fine-tuned it with domain-randomised photorealistic images rendered from \acrshort{ct} scans. Many recent works also focused on the domain shift between simulated and real colons~\cite{cheng2021depth,rodriguez2022uncertain}.  

\section{Chapter Transitions}
This chapter reviews the literature on common methods for cancer surgery, imaging agents, probe tracking, segmentation, and depth estimation methods in \acrshort{mis}. Segmentation, tracking, and depth estimation play crucial roles in advancing medical imaging. Segmentation, the process of identifying and delineating specific structures or regions within an image, allows for precise localisation of organs, tissues, and abnormalities. Tracking provides real time information on the tools being used during procedures. Depth estimation, on the other hand, provides essential spatial information, allowing for the creation of three-dimensional representations of internal structures. Improving the accuracy of each task and combining the outcomes of these tasks for \acrshort{mis} and \acrshort{sensei} probe applications remain open challenges. In the following chapters, I propose several techniques that aim to address these open questions.


\setcounter{chapter}{2}
\chapter{Marker-Based Tracking for a Tethered Laparoscopic Gamma Probe}
\label{chapter_marker_probe_tracking}

\chaptermark{Probe Tracking}
\glsresetall
\label{chap:3-contribution_1}

\begin{cabstract}

The tethered laparoscopic gamma detector does not clearly indicate where on the tissue surface the activity originates, making localization of pathological sites difficult and increasing the mental workload of surgeons. To solve this problem, a robust real-time gamma probe tracking system integrated with  \acrshort{ar} is proposed.

In this work, a dual-pattern marker was attached to the gamma probe, which combined chessboard vertices and circular dots for higher detection accuracy. Both patterns were detected simultaneously based on blob detection and pixel intensity-based vertices detector, and used to estimate the pose of the probe. Temporal information was incorporated into the framework to reduce tracking failure. Furthermore,  the 3D point cloud generated from \acrshort{sfm} was utilised to find the intersection between the probe axis and the tissue surface. When presented as an augmented image this can provide visual feedback to the surgeons.

The method has been validated with ground truth probe pose data generated using the \href{https://optitrack.com/}{OptiTrack} system. When detecting the orientation of the pose using circular dots and chessboard dots alone, the mean error obtained was $0.05^{\circ}$ and $0.06^{\circ}$, respectively. As for the translation, the mean error for each pattern was 1.78mm and 1.81mm. The detection limits for pitch, roll and yaw were $360^{\circ}$, $360^{\circ}$ and $8^{\circ}\sim82^{\circ}\cup188^{\circ}\sim352^{\circ}$.

The performance evaluation results showed that this dual-pattern marker can provide high detection rates, as well as more accurate pose estimation and a larger workspace than the previously proposed hybrid markers. The \acrshort{ar} would be used to provide visual feedback to the surgeons on the location of the affected lymph nodes or tumour.

\end{cabstract}
This chapter’s research has been previously published in the International Journal of Computer Assisted Radiology and Surgery in 2021~\cite{huang2021self}. 

\section{Introduction}
\label{sec:3-introduction}

The use of the laparoscopic gamma probe presents a visualisation challenge, since the probe may not be in contact with tissue during the surgery, which makes it difficult to detect the location of the sensing area on the tissue surface. Additionally, when scanning a tissue the surgeon needs to memorise the previously acquired probe data. This is inefficient, increases the surgeon's workload and the probability of the cancerous tissue not being entirely removed or positive lymph nodes missed. Therefore, the development of a visualisation tool that shows the surgeon directly where the cancerous tissue is located is of extreme importance.

Pose estimation and tracking is a popular problem in computer vision and surgical imaging~\cite{wen2024foundationpose,nguyen2023language}. To date, many probe tracking methodologies have been proposed. The first \textsl{in vivo} \acrshort{ar} surgical anatomy visualisation system with the probe tracked by an optical tracker was proposed in \cite{kang2014stereoscopic}. A magnetic tracking method was presented in \cite{cheung2010fused} combined with stereoscopic video. However, the introduced additional tracking devices are likely to occupy valuable operating space and bring some intrinsic limitations, such as line-of-sight and ferromagnetic interference. A commonly used approach is through laparoscopic image based optical pattern detection, which searches the pattern attached to a probe. Previous studies used corner detection to detect chessboard pattern attached to instruments \cite{Jayarathne2013RobustCamera} \cite{Edgcumbe2013CalibrationSurgery}. This method was extended in \cite{jayarathne2018robust} by computing the probe pose with randomly distributed fiducial pattern over the curved surface, which allowed the occlusion on fiducials and the outliers to be properly handled. Later, the circular dot pattern was proposed, which relied on a more efficient and robust ‘blob detector' rather than the intersection of edges to estimate the pose of the instrument \cite{Pratt2015RobustNephrectomy}. Zhang \textit{et al.} \cite{Zhang2017Real-timeMarker} proposed a hybrid type, incorporating both aforementioned patterns, chessboard circular dot patterns, which provide more information when ambiguous pose problems occurred. However, for the \acrshort{sensei} probe used in this project, the rotation around its own axis does not affect the detection results since the probe is non-imaging. Therefore, these chessboard vertices are redundant.

In this work, a new dual-pattern cylindrical marker is proposed to leverage strengths of two kinds of markers and  facilitate gamma probe tracking. The dual-pattern marker consists of circular dots and chessboard vertices which are simultaneously detected and tracked. To improve the robustness of the whole system and reduce detection failures, temporal information is employed to complement marker detection. The new marker and tracking framework are assessed using an OptiTrack system from where I collected the ground truth data. The detection rates, pose estimation accuracy, and workspace coverage were calculated and I observed that using the novel dual-pattern marker outperformed the current state-of-the-art. The tissue surface is reconstructed using a \acrshort{sfm} algorithm and the intersection point between the surface and the probe axis is estimated. Using that intersection point, my framework highlights to the surgeon the part of the tissue that is being scanned.

\section{Methodology}
\label{sec:3-methods}

\subsection{Dual-pattern marker design}
\label{sec:3-marker_design}

\begin{figure*}[ht]
    \centering
    {
\subfloat[]
{\includegraphics[height=0.235\textwidth]{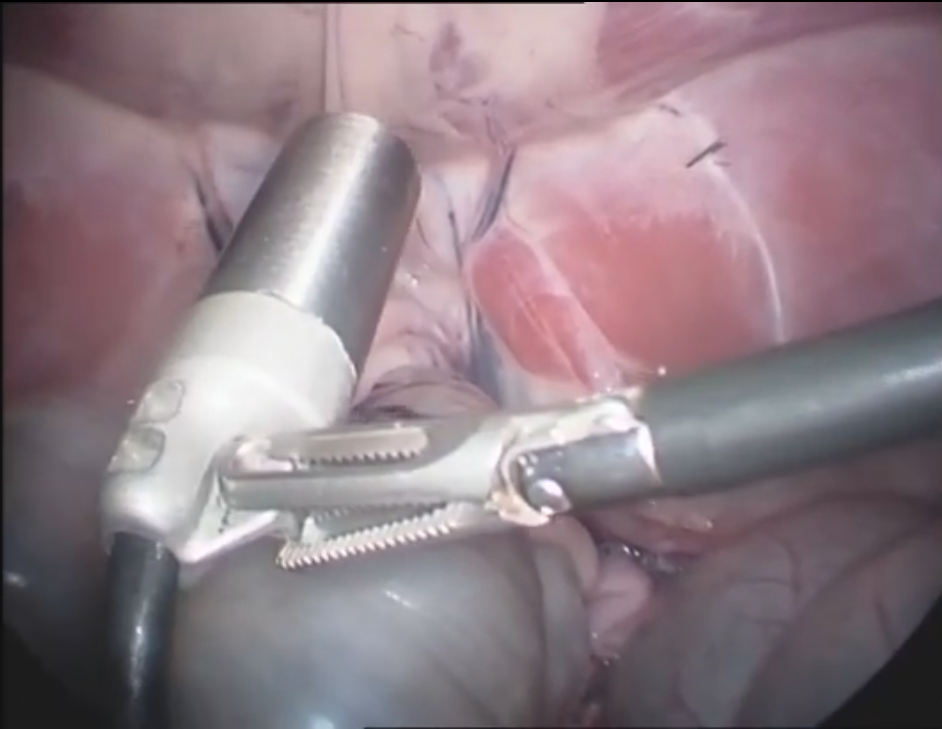}}
\hfill
\subfloat[] {\includegraphics[height=0.235\textwidth]{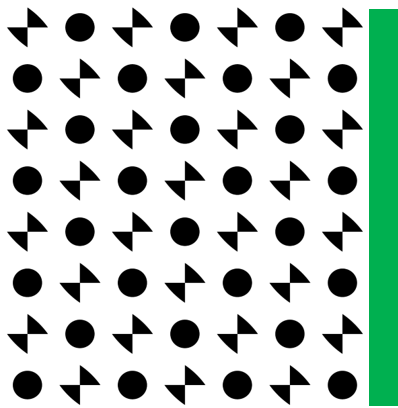}}
\hfill
\subfloat[] {\includegraphics[height=0.255\textwidth]{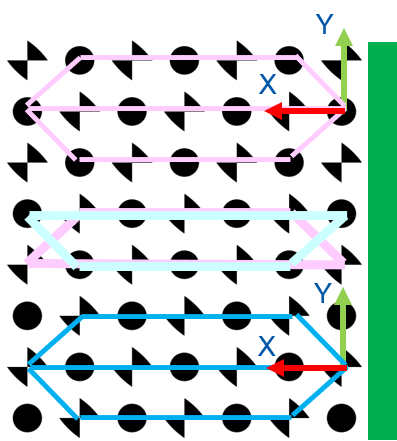}}
\hfill
\subfloat[] {\includegraphics[height=0.235\textwidth]{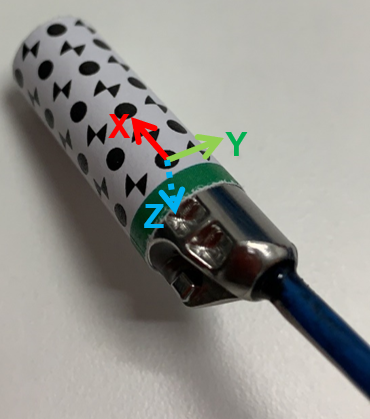}}
\caption{(a) An example of a tethered probe being used in MIS; (b) The gamma probe marker; (c) Example detected circular dots and chessboard vertices; (d) The local coordinates defined on the probe}
\label{fig:marker}}
\end{figure*}

In this work, we proposed a dual-pattern marker (Fig~\ref{fig:marker}(b)) that combines the chessboard vertices and circular dots to estimate the instrument pose. The two patterns were equally spaced and placed circumferentially and appeared alternately. Every two lines of the pattern formed a trapezoidal shape and were considered as a detection unit (Fig~\ref{fig:marker}(c)) for pose estimation and tracking. A green stripe was placed at one end of the marker to resolve ambiguous poses and introduce asymmetry. The marker was attached to the cylindrical instrument such that the overall width matched the circumference and the patterns were aligned with its axis. To ensure that every two lines of the pattern are visible in one video frame, the size of the marker was adjusted correspondingly, considering the near-to-far distance between the probe and the laparoscope.

A local coordinate frame was set at the surface of the probe (Fig~\ref{fig:marker}(d)) and its origin was regarded as the coordinate pivot. When the marker is flattened, the relative position of each feature in the X-Y coordinate frame can be determined from their size and separation. Thus, for a given radius of the probe, the 3D position ($\mathbf{P} = [X, Y, Z]^T$) of each dot and vertex in the 3D local coordinate frame can be determined from their 2D position ($\mathbf{p} = [x, y]^T$).

\subsection{Feature detection}
\begin{figure*}[ht]
\centering
\includegraphics[width=\linewidth]{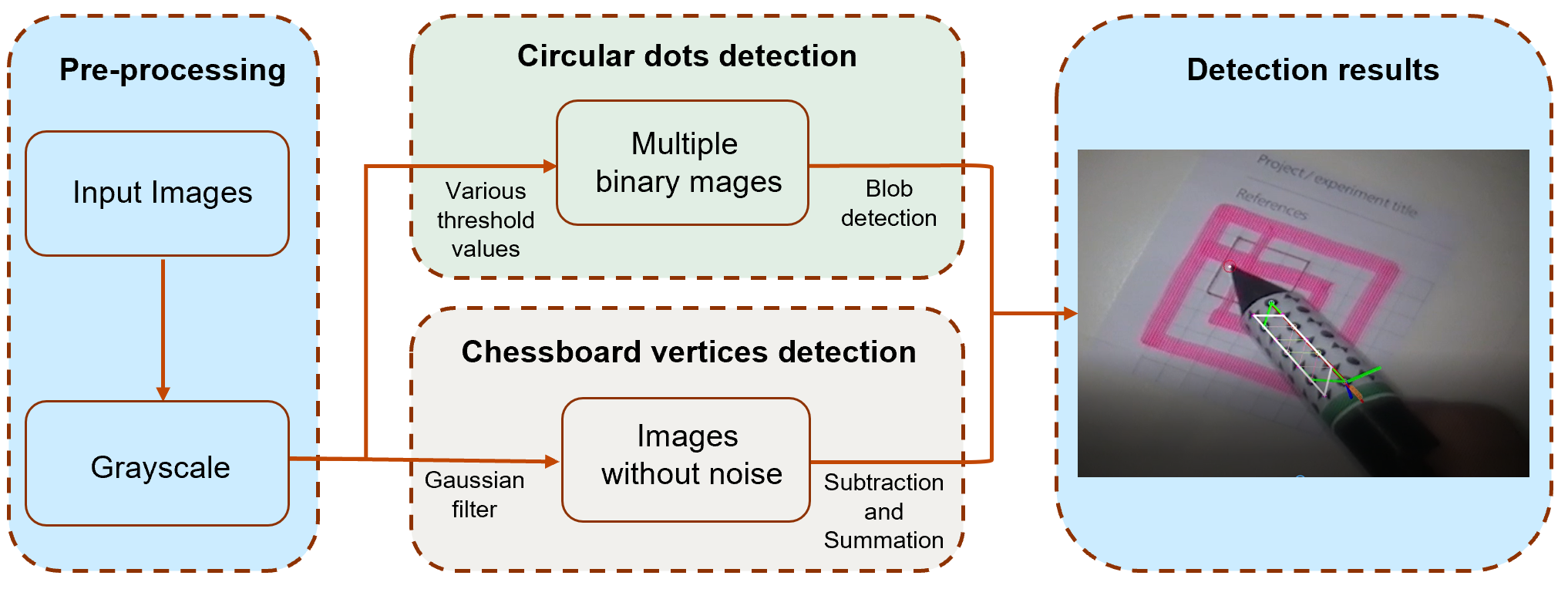}
\caption{Feature detection algorithm workflow}
\label{fig:workflow}
\end{figure*}

The detection of the proposed marker consisted of two parts: blob detection and chessboard vertices detection. The detection algorithm workflow is shown in Fig~\ref{fig:workflow}. For blob detection, a relatively simple algorithm for extracting circular blobs from images in OpenCV was used~\cite{bradski2008learning}. For the chessboard vertices detection (Fig~\ref{fig:workflow}), a Gaussian filter was first applied to the grayscale image to eliminate noise and speckles, and then a chessboard detector~\cite{Bennett2014ChESSFeatures} was applied. To further filter spurious features that give weaker responses, an efficient non-maximum suppression method \cite{Neubeck2006EfficientSuppression} was adopted to retrieve features with the maximum local responses. In addition, the area formed by the intersection of two lines at the centre of the chessboard vertex was relatively easy to be misdetected as a dot. Hence, accurate detection of chessboard vertices would also help to eliminate incorrectly detected circular dots.

\subsection{Marker identification}

The correspondence between the identified markers in the image and model points is necessary to conduct marker pose estimation. First, circular dots and chessboard vertices patterns are clustered based on their vicinity into different feature groups. The group with the largest number of features is used to find the trapeziums for transformation. The four endpoints located at the corners that form two trapeziums are identified from both vertex and dot patterns in this group. The trapezoidal shapes must be convex hulls and lie on the two parallel edges. Once the four vertices were identified, the pattern was transformed into a pattern in the image with the help of the corresponding information. Then, by comparing the transformed pattern and the projected pattern, the identity of each dot and vertex in the projected pattern can be determined as the nearest point to the transformed pattern \cite{Zhang2017Real-timeMarker}.

The addition of the green stripe introduces asymmetry to the markers which helps to identify the orientation of the marker frame. It was placed at the near side of the probe. For each iteration, the RGB image was converted to HSV to separate colour from intensity which made it more robust to the changes in lighting.

\subsection{Marker tracking}
Once all the features that correspond to the model points have been identified, the pose of the probe can be estimated directly by computing a homography. The homography - \textit{i.e.} the transformation that relates the markers and camera - can be estimated through $\mathbf{P}_m = \mathbf{H}\mathbf{P}_r$, where $\mathbf{H}$ presents the homography matrix, $\mathbf{P}_r$ denotes the locations of points on the pattern expressed in a coordinate reference frame, and $\mathbf{P}_m$ denotes the locations of the projected points onto the image plane in the camera frame. During surgery, marker occlusion and invisibility are inevitable due to causes such as strong light reflections and blood staining. If the detection component fails to detect the whole marker and extract its location, the tracking method is used to complement the detection. In this tracking method, the optical flow is computed by the pyramidal affine Lucas Kanade feature tracking algorithm \cite{bouguet2001pyramidal} and temporal information is taken into consideration. By using the optical flow, the current position of the remaining features could be found. Then the position of missing features could also be derived from the correspondence in the reference coordinate frame with the help of a homography. This homography can be estimated with only four pairs of non-collinear feature points, which indicates that it is robust to occlusion.

\subsection{Pose estimation}
Once the position of the model points in the local coordinate frame of the marker and the corresponding projections on the image are found, a framework called \acrfull{ippe} is employed \cite{Collins2014InfinitesimalEstimation}, which is much faster than the current methods based on PnP and is more accurate in most cases. It returns several solutions and the geometric relationship of these solutions is clear. Normally, the correct solution will lead to a smaller re-projection error representing the difference between the tracked results and projections. Hence, in each video frame, the re-projection errors from both circular dots and chessboard vertices are compared and the pose with the smallest error should always be chosen. In this case, two solutions can be derived from each pattern, creating four solutions. If all of them give similar errors close to zero, then there is ambiguity. This situation typically happens when the marker is placed too far from or too close to the camera and the projection of the pattern is close to affine. Some methods are proposed to solve this issue, for instance, \cite{Zhang2017Real-timeMarker} applies points from a different plane to create a large reprojection error for the wrong solution. However, the gamma probe collects gamma data from its tip and the rotation around the probe axis will not influence the detection results of the probe. Hence, the affine problem faced by \cite{Zhang2017Real-timeMarker} can be ignored as long as the re-projection error is sufficiently small.

\subsection{Augmented reality}
The probe signals when the targeted tissue is detected, but it lacks the functionality to provide important visual feedback to the surgeon about the locations. Given the transformation matrix between the laparoscope and the local coordinate frame defined on the probe, the equation of the probe axis can be obtained from the geometrical relationship between the axis and the coordinate pivot. If the equation of the tissue surface is known then the intersection location between the probe axis and the tissue surface can be estimated. To this end, we used a functioning \acrshort{sensei} probe and a prostate phantom with sealed radioactive sources Cobalt-57 hidden inside. The diameter of the Cobalt-57 disk was 25mm and it was placed about 5mm below the tissue surface. The experimental setup is shown in Fig~\ref{fig:AR_gamma} (a) and (c). The \acrshort{sensei} probe was grasped with a laparoscope surgical grasper and the control unit nearby indicated the gamma counts. The laparoscope captured the video of the whole procedure with the image displayed on a monitor. The 3D reconstruction of the prostate phantom surface was conducted using \acrshort{sfm} in MATLAB and a corresponding surface point cloud was generated. The actual scale of this point cloud was calculated with the help of the \acrshort{sensei} probe of the known physical size. By calculating the distance between the points in the point cloud to the probe axis, points with short distances were determined. As the 3D reconstruction by \acrshort{sfm} was quite dense, these points were considered to be the potential intersection points. Besides, the distance between the intersection point and the marker pivot point should be longer than the distance between the probe tip and the marker pivot.

\section{Experimental Setup}

\subsection{Hardware setup}

\begin{figure*}[ht]
\centering
{
\subfloat[]{\includegraphics[height=0.33\textwidth]{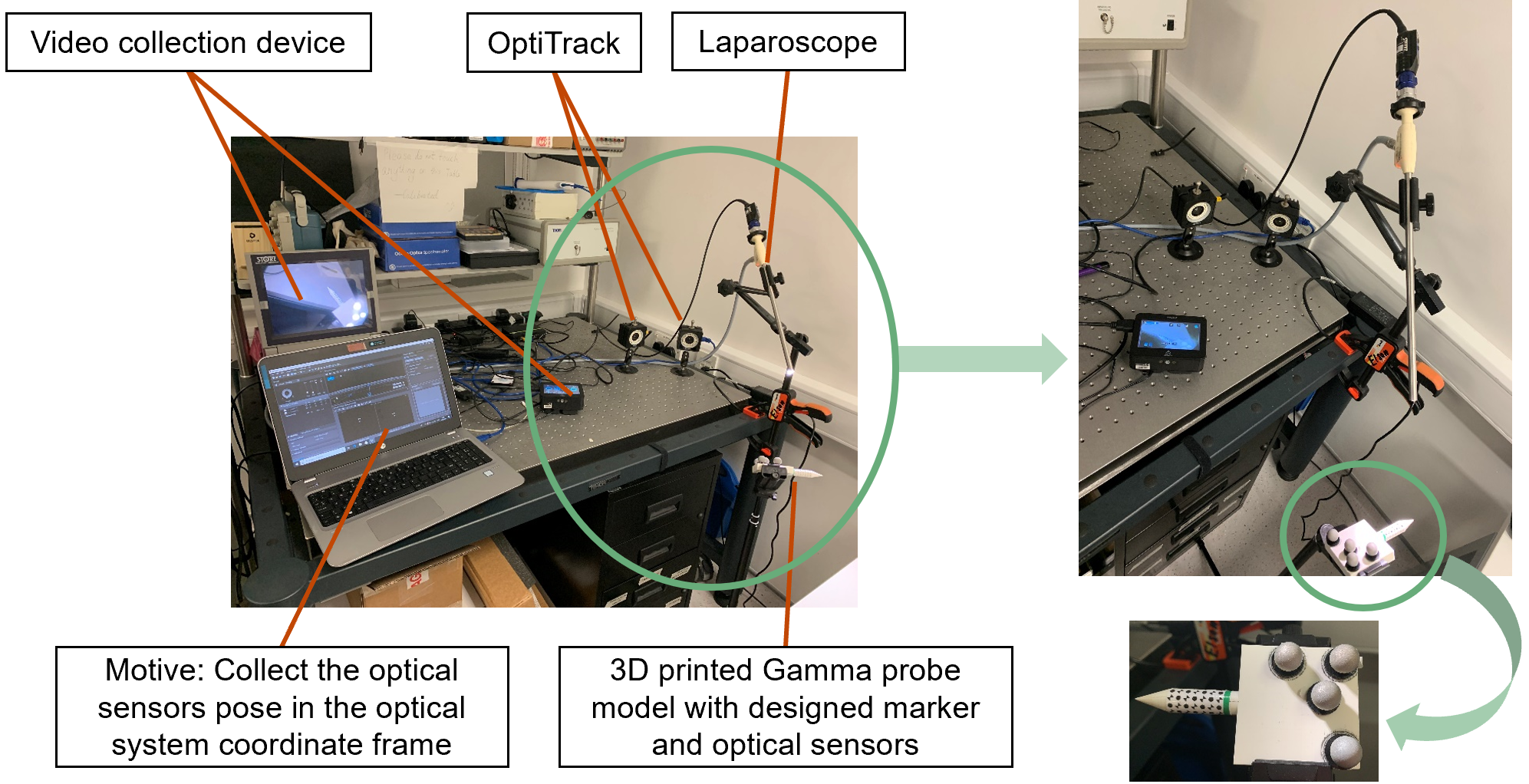}}
\hspace{2mm}
\subfloat[]
{\includegraphics[height=0.33\textwidth]{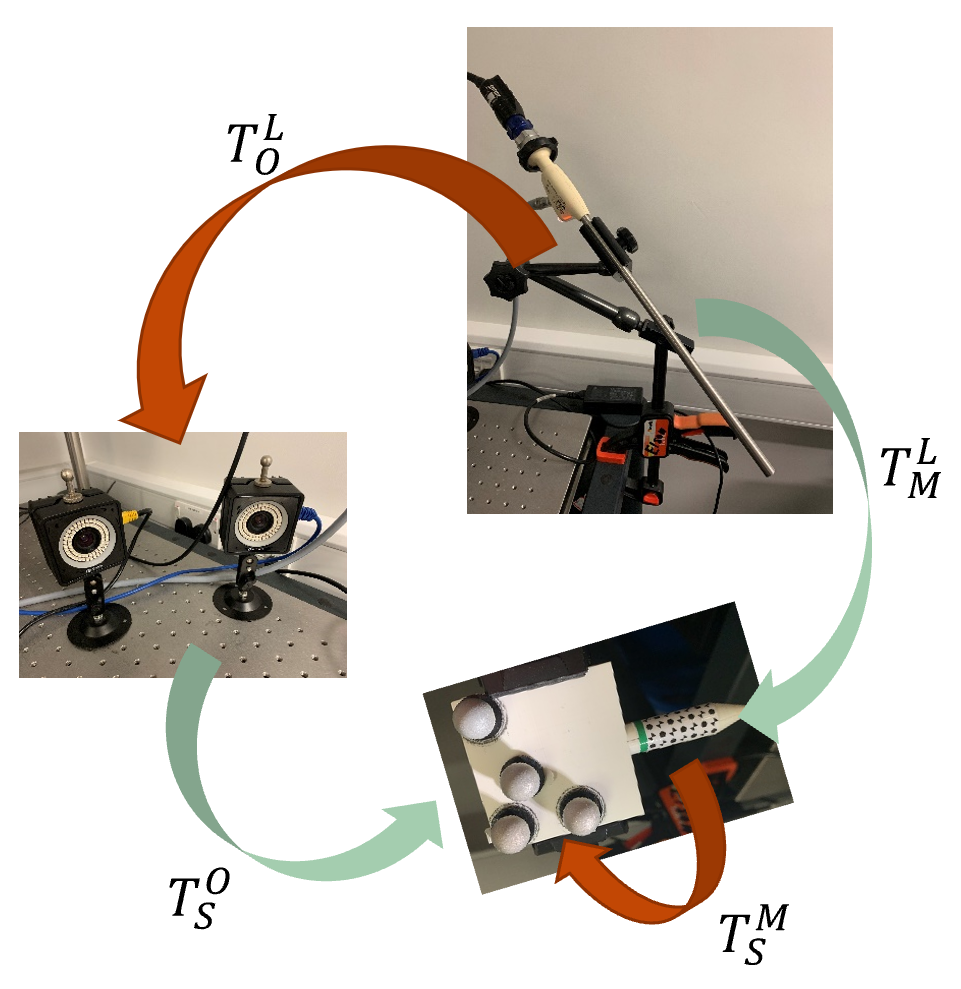}}
\caption{(a) Hardware set-up for experiments; (b) The transformation matrices between laparoscope, OptiTrack system, optical sensors, and designed marker}
\label{fig:regi_hardware}}
\end{figure*}

 Fig~\ref{fig:regi_hardware} (a) shows the experimental setup illustrating a 3D printed model with the same dimensions as the real probe. During the detection procedure, the tip of the probe was positioned 2 to 3cm from the tissue surface. Therefore, a cone with a height of 2cm was added to the front end of the probe model to maintain a fixed distance to the tissue surface for validation. The designed marker was attached to the cylindrical probe and four optical sensors were mounted on a flat plate attached to the model for validation via OptiTrack (NaturalPoint Inc., America). The diameter of the probe was 12mm and it can be placed directly into the patient's abdominal cavity through standard MIS trocars. In this experiment, the probe could be placed in the view field of a standard 10mm diameter monocular calibrated \cite{zhang2000flexible} laparoscope (KARL STORZ SE \& Co. KG, Tuttlingen, Germany). The videos were displayed on a monitor and captured using a Ninja-2 box (Atomos Global Pty Ltd, Australia). The videos were streamed to a computer (2.5 GHz CPU, 8GB RAM) using S-Video to HDMI and HDMI to USB video converters (StarTech.com Ltd, America).    

\subsection{Pose estimation error}


To validate the pose estimation algorithm, the OptiTrack system and its software, Motive, were used to obtain the ground truth and calculate the transformation matrix between the OptiTrack system and the optical sensors \(\mathbf{T}_S^O \). In addition, the marker pose in the laparoscope coordinate frame \( \mathbf{T}_M^L\) can be estimated. However, there were still two unknown registrations: the laparoscope to the OptiTrack system \( \mathbf{T}_O^L\) and optical sensors to the designed marker \( \mathbf{T}_S^M \). As shown in Fig~\ref{fig:regi_hardware} (b), the green arrows indicate parameters that can be directly obtained while the red arrows represent the unknowns. The relationship between these four transformation matrices is given as follows: 
\begin{equation}\label{eq1}
\mathbf{T}_M^L \cdot \mathbf{T}_S^M = \mathbf{T}_O^L \cdot \mathbf{T}_S^O
\end{equation}

This problem can be treated as an $\mathbf{A}\mathbf{X} = \mathbf{Y}\mathbf{B}$ problem and 10 pairs of \( \mathbf{T}_M^L\) and \(\mathbf{T}_S^O \) were required to obtain the \( \mathbf{T}_S^M \) and \( \mathbf{T}_O^L\) \cite{Shah2011ComparingData}. However, the error from the registration accumulates in the final pose estimation error. During experimental validation, the probe was placed at the `typical' position at 100mm from the laparoscope to match a typical surgery. As two different patterns could be detected on the marker, the final transformation matrix used was from the one that led to a smaller re-projection error. For each pattern, 60 video trials were made and 10 of these were for registration to calculate \( \mathbf{T}_S^M \) and \( \mathbf{T}_O^L\) while 50 of these were for pose estimation error calculation. The position of the laparoscope and the two OptiTrack cameras were always fixed. In every video trial, the probe was static but the background of the scene was not static and changed over time. Besides, from trial to trial, the position of the probe was changed. In each trial, the relative pose between the ground truth and the estimated result was calculated as:
\begin{equation}\label{pose error}
\text{Relative\_pose\_matrix} =  (\mathbf{T}_S^M)^{-1} \cdot (\mathbf{T}_M^L)^{-1} \cdot \mathbf{T}_O^L \cdot \mathbf{T}_S^O
\end{equation}

Ideally, the relative pose matrix should be equal to the identity matrix. However, this was not the case due to the error from the registration and pose estimation. The translation error was set as the mean of the fourth column in the matrix. To have a more intuitive understanding of the rotation error, the rotation matrix was converted to an axis-angle.

\begin{figure*}[ht]
\centering
{
\subfloat[]{\includegraphics[height=0.46\textwidth]{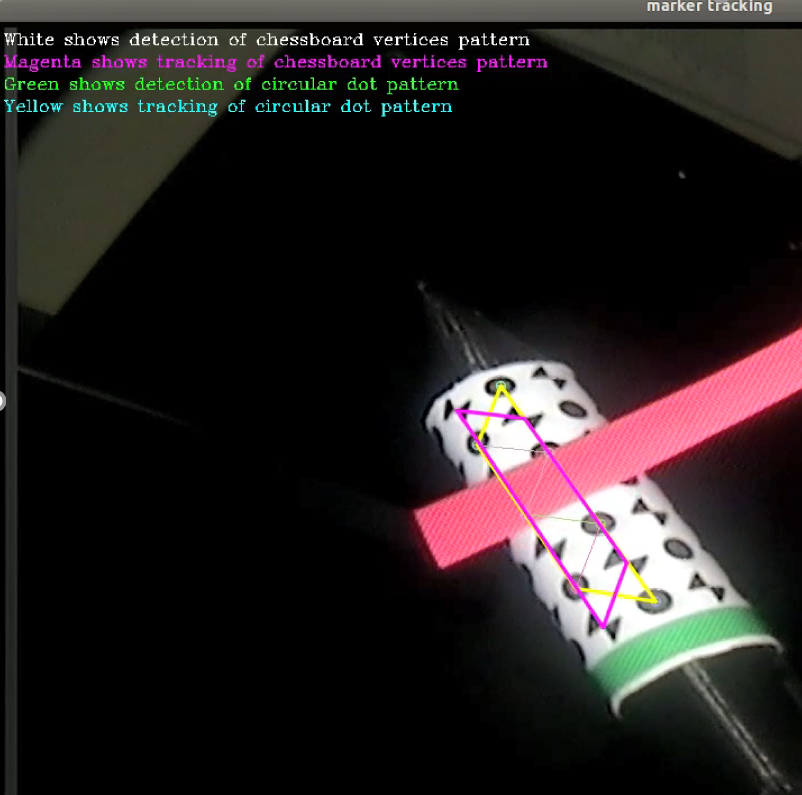}}
\hspace{9pt}
\subfloat[]{\includegraphics[height=0.46\textwidth]{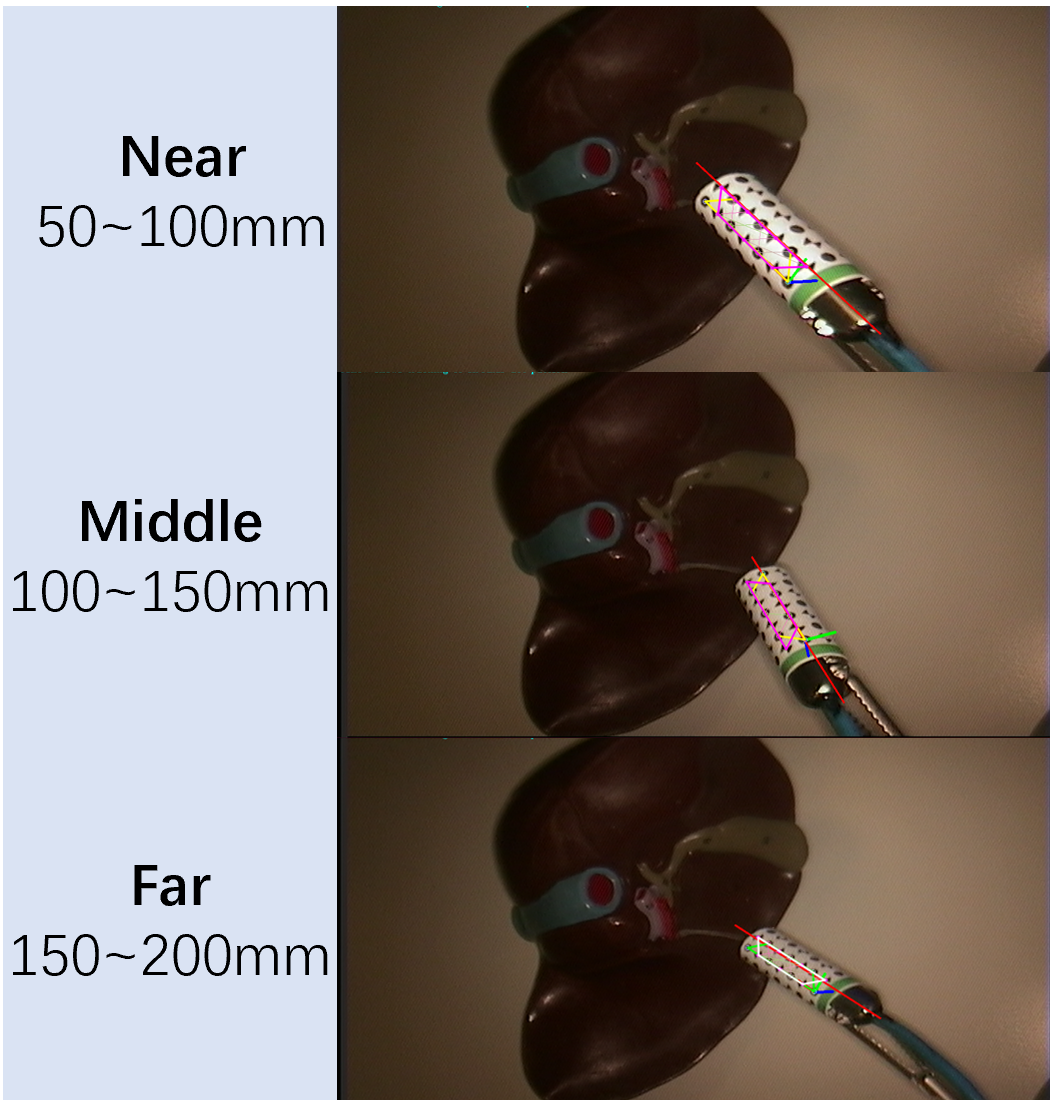}}
\caption{(a) Successful marker tracking in the case of occlusion; (b) The experimental visualisation results for different testing distances (Near: 50-100mm, Middle: 100-150mm, far: 150-200mm) between the probe and camera.}
\label{fig:distance_occlu}}
\end{figure*}

\subsection{Projection error}
Given the geometric parameters of the probe and the transformation matrix from the camera to the marker, the 3D position of the cone tip simulating a 2 cm working distance could be estimated. The probe was rotated with a fixed tip position. However, because of the pose estimation error, the calculated 3D tip position was found to vary from frame to frame, with the distance between the tips in every two frames calculated as the projection error. The results were compared to the previous hybrid marker \cite{Zhang2017Real-timeMarker}, although in this case it could not be tracked during probe axial rotation around its own axis, resulting in large errors. Hence, the projection errors presented below for \cite{Zhang2017Real-timeMarker} were recorded with and without the failed frames. 

\subsection{Detection limit and detection rate analysis}  

For further validation, the detection limits and detection rates were calculated by recording the maximal experimentally detectable distance and rotation angle of the probe. The distance was recorded from the camera to the probe and the limits of rotation were defined by the probe's local coordinate axes (roll, pitch, and yaw). When testing the distance limits, the probe was translated along the axis of the laparoscope until detection failed. To identify the rotational motion limits, the probe was placed 100mm from the laparoscope, a typical distance for practical tissue scanning. 

Since the detection of chessboard vertices relies on the intersection of edges, it was affected by image degrading effects like smudging and blooming. However, the circular dots detection algorithm was more robust because it did not rely on well-defined edge crossings. Regarding the dual-pattern marker detection, a frame was a success if either the chessboard vertices or circular dots pattern was detected because they were independent of each other. In the experiments, the focus was set at the phantom surface and the probe was placed at different distances to the camera Fig~\ref{fig:distance_occlu} (b): near (50mm $\sim$ 100mm), middle (100mm $\sim$ 150mm), far (150mm $\sim$ 200mm).

\section{Results and Discussion}
\label{sec:3-results}

\subsection{Pose estimation error}
Table~\ref{table:1} shows the validation results obtained from the dual pattern marker, which have a smaller mean error and a lower standard deviation than with the previous pattern. In addition, the pose estimation errors from the circular dots and the chessboard vertices patterns were quite similar and less than 2mm, which means that both patterns worked well. Specially, the mean translation error decreased from 2.53mm to 1.78mm and 1.81mm for circular dots and chessboard vertices separately, while mean rotation error decreased from 0.69$^{\circ}$ to 0.05$^{\circ}$ and 0.06$^{\circ}$ for circular dots and chessboard vertices separately, compared with the previous hybrid marker \cite{Zhang2017Real-timeMarker}. Note that all of these results were obtained from measurements on the same 3D-printed probe model, for both the proposed dual pattern marker and the previous hybrid marker. Given the position of the model points defined in the local coordinate frame on the marker and the correspondence-tracked projections on the image, the pose of the marker was estimated by using the \acrshort{ippe} method. Specifically, the \acrshort{ippe} will give two affine poses for each pattern and will compare the results to select the one with the smallest reprojection error as the first output. This is why the newly designed pattern and new pose estimation algorithm can lead to a smaller mean error and increase the tracking accuracy.

\begin{table}[h]
\caption{Summary of pose estimation error}
\resizebox{\textwidth}{!}{
\begin{tabular}{|c|c|c|c|c|}
\hline

\hline
\begin{tabular}[c]{@{}c@{}}\textbf{Different} \\ \textbf{marker}\end{tabular}                   & \multicolumn{2}{c|}{\begin{tabular}[c]{@{}c@{}}\textbf{Translation mean error} \\ $\pm$ STD (mm)\end{tabular}} & \multicolumn{2}{c|}{\begin{tabular}[c]{@{}c@{}}\textbf{Rotation mean error} \\ $\pm$ STD ($^{\circ}$)\end{tabular}} \\ \hline
\multirow{2}{*}{\textbf{\makecell[c]{Our hybrid\\ marker}}  } & Circular dots        & 
\begin{tabular}[c]{@{}c@{}}Chessboard \\
vertices\end{tabular}        & Circular dots       & \begin{tabular}[c]{@{}c@{}}Chessboard \\ vertices\end{tabular}      \\ \cline{2-5} 
& 1.78 $\pm$  0.81          & 1.81 $\pm$  0.80   & 0.05 $\pm$  0.02        & 0.06 $\pm$  0.02   \\ \hline
\begin{tabular}[c]{@{}c@{}}Previous\\ hybrid marker \cite{Zhang2017Real-timeMarker}\end{tabular}              & \multicolumn{2}{c|}{2.53 $\pm$  1.40}  & \multicolumn{2}{c|}{0.69 $\pm$  0.33}  \\ \hline

\hline
\end{tabular}}
\label{table:1}
\end{table}





\subsection{Projection error}

It can be seen from Table~\ref{table:3d2d} that for \cite{Zhang2017Real-timeMarker}, the failure frames cause large projection errors, unless the motion remains delicate. The errors calculated from our marker are lower due to pose estimation for every frame using two patterns. 

\begin{table}
\caption{3D tip distance when the cone tip is fixed}
\resizebox{\textwidth}{!}{
\begin{tabular}{cccc}
\toprule
\multicolumn{4}{c}{\textbf{3D Projection Error}}
\\ \midrule
\textbf{Different Marker}  & \begin{tabular}[c]{@{}c@{}}\textbf{Mean error $\pm$ STD}\\ (mm)\end{tabular} & \begin{tabular}[c]{@{}c@{}}\textbf{Maximum error}\\ (mm)\end{tabular} & \begin{tabular}[c]{@{}c@{}}\textbf{Minimum error}\\ (mm)\end{tabular} \\ \hline
\begin{tabular}[c]{@{}c@{}}Previous hybrid marker\cite{Zhang2017Real-timeMarker}\\ with the failed frames\end{tabular}    & 17.17 $\pm$ 16.33  & 137.72 & 0.00  \\ \midrule
\begin{tabular}[c]{@{}c@{}}Previous hybrid marker\cite{Zhang2017Real-timeMarker}\\ without the failed frames\end{tabular} & 1.73 $\pm$ 1.19  & 5.41  & 0.00   \\ \midrule
\rowcolor[RGB]{230,230,230} \textbf{Our hybrid marker}  & \textbf{0.22 $\pm$ 0.19}  & \textbf{1.90} & \textbf{0.00}  \\ \bottomrule
\end{tabular}}
\label{table:3d2d}
\end{table}
\subsection{Detection and tracking analysis}
The results of the detectable distance limits are shown in Table~\ref{table:2}. The farthest distance at which the probe could be detected was 220mm and the marker works well between 50mm and 150mm, which is a reasonable working range for \acrshort{mis}. The maximum detectable angles are displayed in Table~\ref{table:2}. Since the marker covered the entire probe surface circumferentially, detection results of the rotation around the roll axis are greatly improved. As the features in the marker are dense, the results when rotating around the pitch axis are also improved. As shown in Table~\ref{table:2}, rotation around both roll and pitch axes can reach $360^\circ$. It is worth noting that the detectable angle range around the yaw axis is not $360^\circ$ since the axis of the probe was aligned with the axis of the laparoscope and the marker becomes invisible due to occlusion. From experiments, the marker detection and tracking have an angular range of about $16^\circ$ within which it is undetectable.

The detection rates for the near and middle distance ranges were 100\%, which reduced to 99.7\% when the probe was in the long-distance range.

\begin{figure*}[]
\centering
{
\subfloat[]{\includegraphics[width=0.19\textwidth, height=5.4cm]{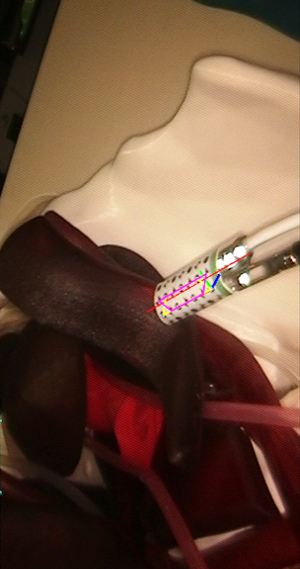}}
\hspace{0.2mm}
\subfloat[]{\includegraphics[width=0.189\textwidth, height=5.4cm]{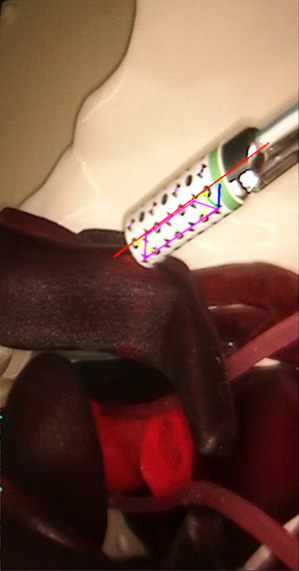}}
\hspace{0.2mm}
\subfloat[]{\includegraphics[width=0.189\textwidth, height=5.4cm]{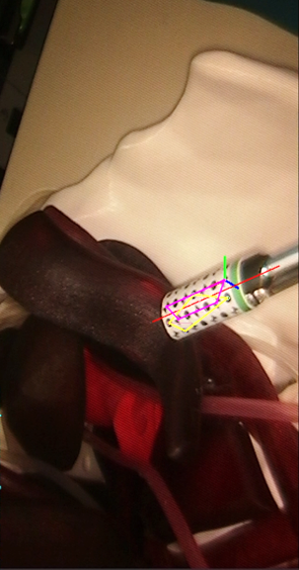}}
\hspace{0.2mm}
\subfloat[]{\includegraphics[width=0.19\textwidth, height=5.4cm]{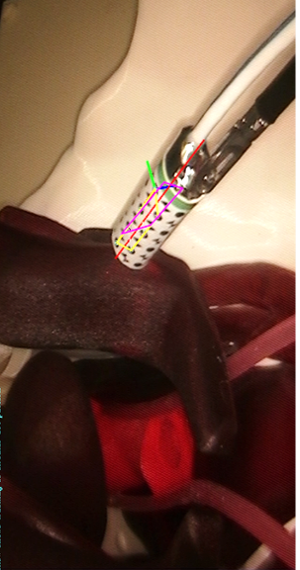}}
\hspace{0.12mm}
\subfloat[]{\includegraphics[width=0.1905\textwidth, height=5.4cm]{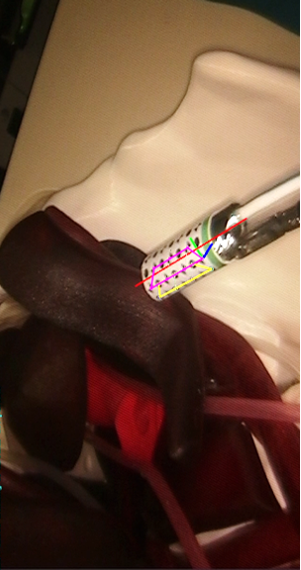}}
\caption{Examples where the pose estimation is more accurate by using (a) the circular dots pattern; (b) the chessboard vertices; Example where tracking failed for (c) the circular dots pattern and (d) the chessboard vertices. In (e) both vertices and dots pattern are detected in adjacent three marker lines}
\label{fig:diff_track_result}}
\end{figure*}

Since the pose estimations from chessboard vertices and circular dots are independent, if both are detected, the one with the smallest reprojection error will be selected. If identification of either fails, the system will rely on the other to get the probe pose. We list several different tracking scenarios in Fig~\ref{fig:diff_track_result}. Fig~\ref{fig:diff_track_result} (a) shows a case where the pose estimation result from the circular dots pattern is more accurate than that from chessboard vertices while Fig~\ref{fig:diff_track_result} (b) shows the opposite. In Fig~\ref{fig:diff_track_result} (c), the circular dots pattern tracking failed so the probe pose is estimated from the vertices, while the opposite situation is presented in Fig~\ref{fig:diff_track_result} (d). In Fig~\ref{fig:diff_track_result} (e), both vertices and dot patterns are detected for three adjacent marker lines with the vertices pattern providing a more accurate pose estimation result.

\begin{table}[h]
\centering
\caption{Maximum detectable distance and rotation angle around different axes. Distance to the camera is in mm.}
\begin{tabular}{ c c c } 
 \toprule
 \textbf{Rotation axis} & \textbf{Previous work}\cite{Zhang2017Real-timeMarker} & \textbf{Dual-pattern marker(ours)} \\ 
 \hline
 Roll ($^\circ$) & $\pm 85^\circ$ & $360^ \circ$ \\ 
 Pitch ($^\circ$) & $\pm 78^\circ$ & $ 360^\circ$ \\
 Yaw ($^\circ$) & $\pm 83^\circ$ & $8^{\circ}\sim82^{\circ}\cup188^{\circ}\sim352^{\circ}$ \\
 Distance to camera & $60 \sim 200$ & $50 \sim 220$ \\
 \bottomrule
\end{tabular}
\label{table:2}
\end{table}

\begin{figure*}[]
\centering
{
\subfloat[]{\includegraphics[width=0.49\textwidth, height=4.2cm]{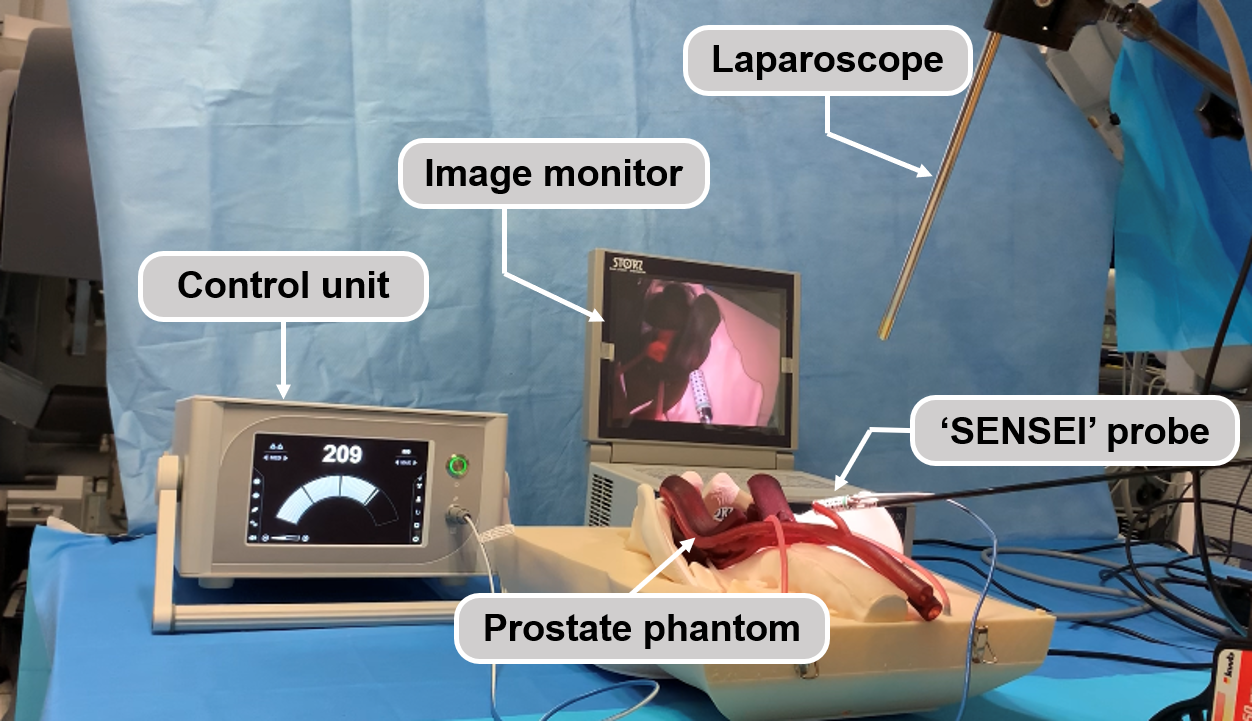}}
\hspace{1mm}
\subfloat[]{\includegraphics[width=0.49\textwidth, height=4.2cm]{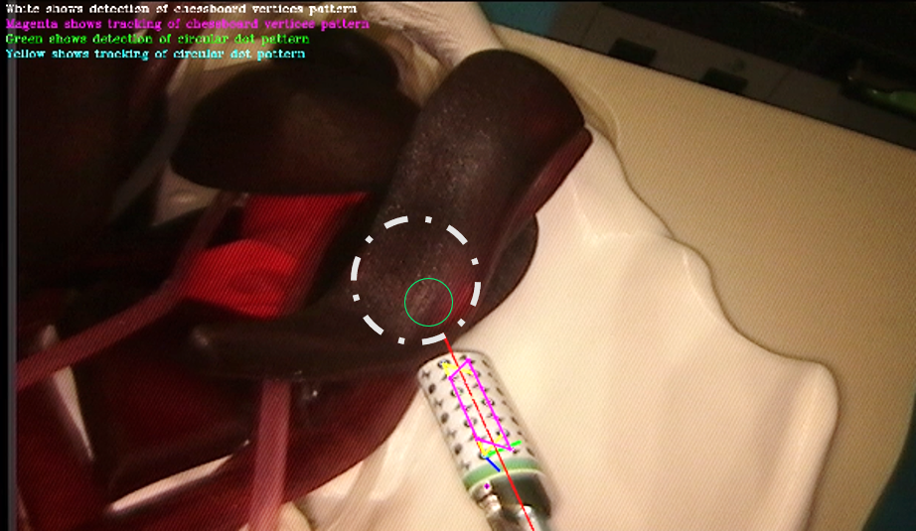}}
\vspace{1mm}
\subfloat[]{\includegraphics[width=0.49\textwidth, height=4.2cm]{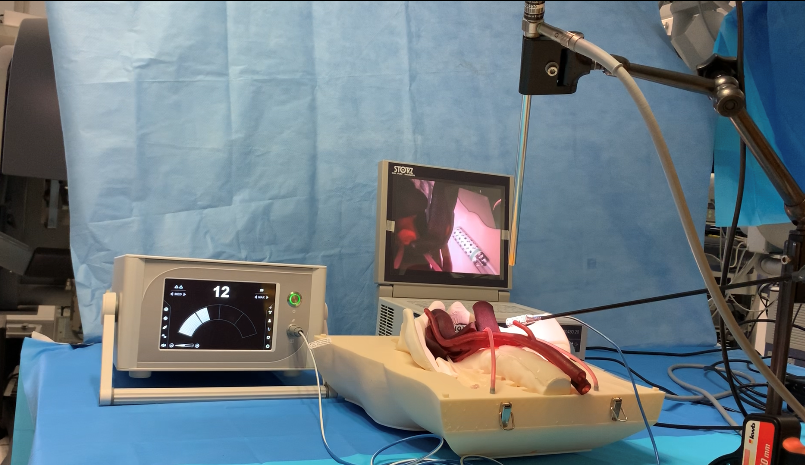}}
\hspace{1mm}
\subfloat[]{\includegraphics[width=0.49\textwidth, height=4.2cm]{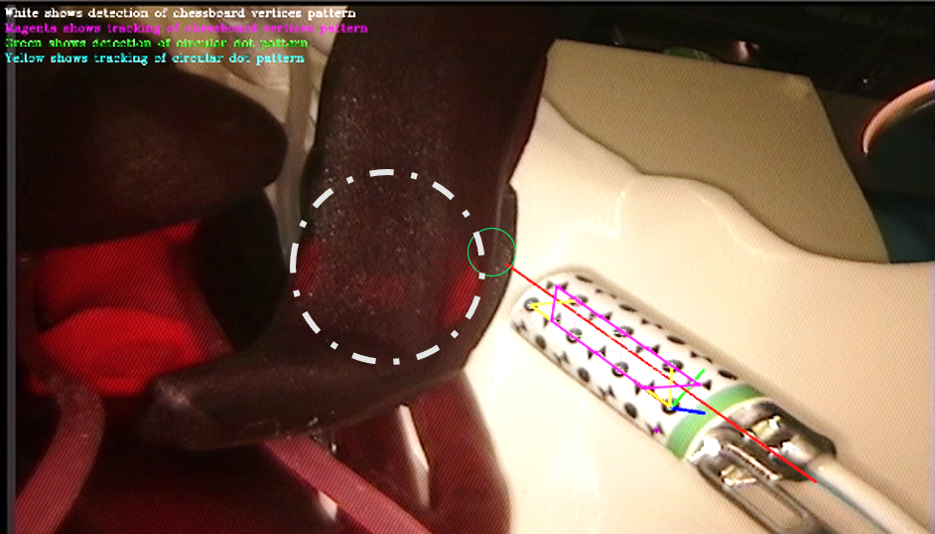}}
\caption{The hardware setup including laparoscope, image monitor, prostate phantom, \acrshort{sensei} probe, and control unit showing (a, b) a higher radiation level when the probe was pointing to and placed closer to the radioactive source; and (c, d) a lower radiation level when the probe was pointing to the edge of the source. The grey dashed circles in (b, d) show the position of radioactive source Cobalt-57 while the green circles represent the intersection area of the gamma probe axis and the tissue.}
\label{fig:AR_gamma}}
\end{figure*}


\subsection{Tracking results for simulated occlusions}
Fig~\ref{fig:distance_occlu} (a) shows an example of occlusion using a red stripe to block the markers. Although the number of remaining features was not enough to directly estimate the pose of the probe, they could still be used to calculate the homography. The position of the points that were occluded could then be inferred from the correspondence information between the coordinate reference frame and the current camera image frame with the help of the homography. Therefore, the marker tracking enhanced the robustness of the entire system to occlusions.

\subsection{Augmented reality}
Given a 3D point cloud representing the tissue surface and the equation of the probe axis, the intersection point was estimated and the results are shown in Fig~\ref{fig:AR_gamma}. The red line indicates the axis of the probe, the grey dashed circle shows the position of radioactive source Cobalt-57 and the green circle represents the intersection area of the Gamma probe axis and the tissue. In Fig~\ref{fig:AR_gamma} (a) and (b), the \acrshort{sensei} probe was close to and pointing towards the radioactive source, the probe recorded stronger gamma radiation of 209 gamma counts per second. Fig~\ref{fig:AR_gamma} (c) and (d) show the opposite where the \acrshort{sensei} probe was pointing at the edge of the buried source and the radiation was weak (12 counts per second). The \acrshort{ar} system with the green circle and the grey dashed circle (Fig~\ref{fig:AR_gamma}) can therefore allow the surgeon to know which part of the tissue the radiation is coming from so that they can do accurate node identification or tissue excision with this visual feedback. 
\section{Conclusion}
\label{sec:3-conclusion}

In this work, we proposed a new hybrid marker that incorporated both circular dots and chessboard vertices to increase the detection rate. The additional green stripe was included to introduce asymmetry and resolve direction ambiguity. The marker was designed such that it fully covered the tethered laparoscopic gamma probe using dense features. The experimental results show that the detection workspace, robustness pose estimation efficiency, and accuracy of the design outperformed previous works. We have therefore shown the feasibility and the potential of using the proposed framework to track the \acrshort{sensei} probe. In addition to the design of the new marker, we have also proposed a solution to provide clear visual feedback to indicate the tracer location on the tissue surface.

The work could be further extended to increase the registration accuracy by fusing the vision-based 3D pose estimation with the kinematic data of the instrument (robot) controlling the probe. Successive transformations from the probe to the instrument and endoscope coordinate frames will provide a robust initial viewpoint estimate and registration. The framework could also be used to track other types of probes, such as \acrshort{drs} probe.

\section{Chapter Transitions}

Although the \acrshort{sfm} can be used for 3D reconstruction in this setup, there are many constraints for the application of \acrshort{sfm}. First, \acrshort{sfm} has limited viewpoint coverage. \acrshort{sfm} relies on the availability of multiple images of the same scene taken from different viewpoints. If there are significant occlusions or limited viewpoint coverage, it can be challenging to accurately reconstruct the entire 3D structure. In laparoscopic surgery, moving the laparoscope around before the operation could bring more workload to surgeons and goes against the efficiency principle of surgery. Second, the accuracy of \acrshort{sfm} heavily depends on the ability to match features across multiple images. In \acrshort{mis} cases where images have few distinctive features or contain repetitive patterns, it is difficult to find reliable correspondences, leading to a less accurate reconstruction.

Third, it is difficult for \acrshort{sfm} to handle dynamic scenes. \acrshort{sfm} assumes a static scene, and moving objects can introduce errors in the reconstruction. The presence of dynamic elements, such as the moving surgical tools and the tissue movement because of the interaction with the tools, would lead to inaccurate 3D models. Additionally, the computational requirements for \acrshort{sfm} can be intensive, especially for high-resolution laparoscopic images. Processing time and memory usage can be limiting factors, particularly in real-time \acrshort{mis} applications.

Hence, we explored the learning-based depth estimation algorithms with stereo laparoscopic images as input and the corresponding depth maps as output while leveraging the geometrical relationship between the stereo laparoscopic images.








\setcounter{chapter}{3}

\chapter{Self-Supervised Generative Adversarial Network for Depth Estimation in Laparoscopic Images}
\chaptermark{Self-Supervised GAN for Depth Estimation}
\glsresetall
\label{chap:4-contribution_2-Depth2D}

\begin{cabstract}

Dense depth estimation and 3D reconstruction of a surgical scene are crucial steps in computer assisted surgery. Recent work has shown that depth estimation from a stereo image pair could be solved with convolutional neural networks. However, most recent depth estimation models were trained on datasets with per-pixel ground truth. Such data is especially rare for laparoscopic imaging, making it hard to apply supervised depth estimation to real surgical applications. To overcome this limitation, we propose SADepth, a new self-supervised depth estimation method based on Generative Adversarial Networks. It consists of an encoder-decoder generator and a discriminator to incorporate geometry constraints during training. Multi-scale outputs from the generator help to solve the local minima caused by the photometric reprojection loss, while adversarial learning improves the framework generation quality. Extensive experiments on two public datasets show that SADepth outperforms recent state-of-the-art unsupervised methods by a large margin, and reduces the gap between supervised and unsupervised depth estimation in laparoscopic images.

\end{cabstract}
This chapter’s research has been previously published at the International Conference on Medical Image Computing and Computer Assisted Intervention (MICCAI) in 2021~\cite{huang2021self}. 

\section{Introduction}
\label{sec:4-introduction}

Robot-assisted minimally invasive surgery with stereo laparoscopic vision has become popular due to the advantages of enhanced movement range, precision, vision, and proficiency \cite{mack2001minimally,nguyen2020end}. Surgical scene depth estimation is a fundamental problem in image-guided intervention and has received substantial prior interest due to its promise for robot navigation, 3D registration between pre- and intra-operative organ models, and augmented reality \cite{ye2017self}. Obtaining depth maps is not trivial due to the inherent problems such as tissue deformation, specular reflections, and lack of photometric constancy across frames \cite{liu2019dense}. 

Several traditional methods used multi-view stereo algorithms such as \acrfull{slam} and \acrshort{sfm} \cite{leonard2018evaluation}, but these struggle with less textured tissues. More recently deep learning-based depth estimation has used RGB images as the training data and \acrfull{cnns} for supervised learning  \cite{eigen2014depth,do2021multiple}. To produce accurate results in less than a second of GPU time, Luo \textit{et al.} \cite{luo2016efficient} treated the problem as a multi-class classification indicating all possible disparities, and exploited a product layer to simplify the representations of a Siamese architecture. Chang \textit{et al.} \cite{chang2018pyramid} proposed PSMNet, where the capacity of global context information at different scales and locations could be extracted by a spatial pyramid pooling module to form a cost volume. Duggal \textit{et al.} \cite{duggal2019deeppruner} sped up the runtime of stereo matching and developed a differentiable PatchMatch module that could discard most disparities without the need of full cost volume evaluation. 

The methods above are fully supervised and require ground truth depth during training. However, acquiring per-pixel ground truth depth data is challenging for real-world settings \cite{joung2019unsupervised} and especially for laparoscopic vision where port space is limited, the working distance is short and sterilisation is required \cite{huang2020tracking}. One alternative is self-supervised training of depth estimation models using image reconstruction as the supervisory signal \cite{garg2016unsupervised}. The input is usually a set of images in the form of monocular or stereo images \cite{zhou2017unsupervised}. Godard \textit{et al.} \cite{godard2017unsupervised} proposed a training loss that included a left-right depth consistency term and a reconstruction term for single image depth estimation, despite the absence of ground truth depth. This was extended by \cite{godard2019digging} with full-resolution multi-scale sampling to reduce visual artifacts, and a minimum reprojection loss to robustly handle occlusions. Johnston \textit{et al.} \cite{johnston2020self} further closed the gap with fully-supervised methods by including a self-attention mechanism and making use of contextual information. Ye \textit{et al.} \cite{ye2017self} proposed a deep learning framework for surgical scene depth estimation in self-supervised mode for scalable data acquisition by adopting a differentiable spatial transformer and an autoencoder. 

In this work, we present a new method for self-supervised adversarial depth estimation: SADepth. A U-Net architecture \cite{ronneberger2015u} was adopted as a generative structure and fed with stereo pairs as inputs to benefit from complementary information. To cope with local minima caused by classic photometric reprojection loss, we applied the disparity smoothness loss and formed the network across multiple scales. The use of a \acrshort{gan} allowed us to improve the reconstructed image quality, which formed a supervisory signal for training while keeping the overall end-to-end optimisation objective.  
\section{Methodology}
\label{sec:4-methods}

\subsection{Overview}
\begin{figure*}[t]
\centering
\includegraphics[width=\textwidth]{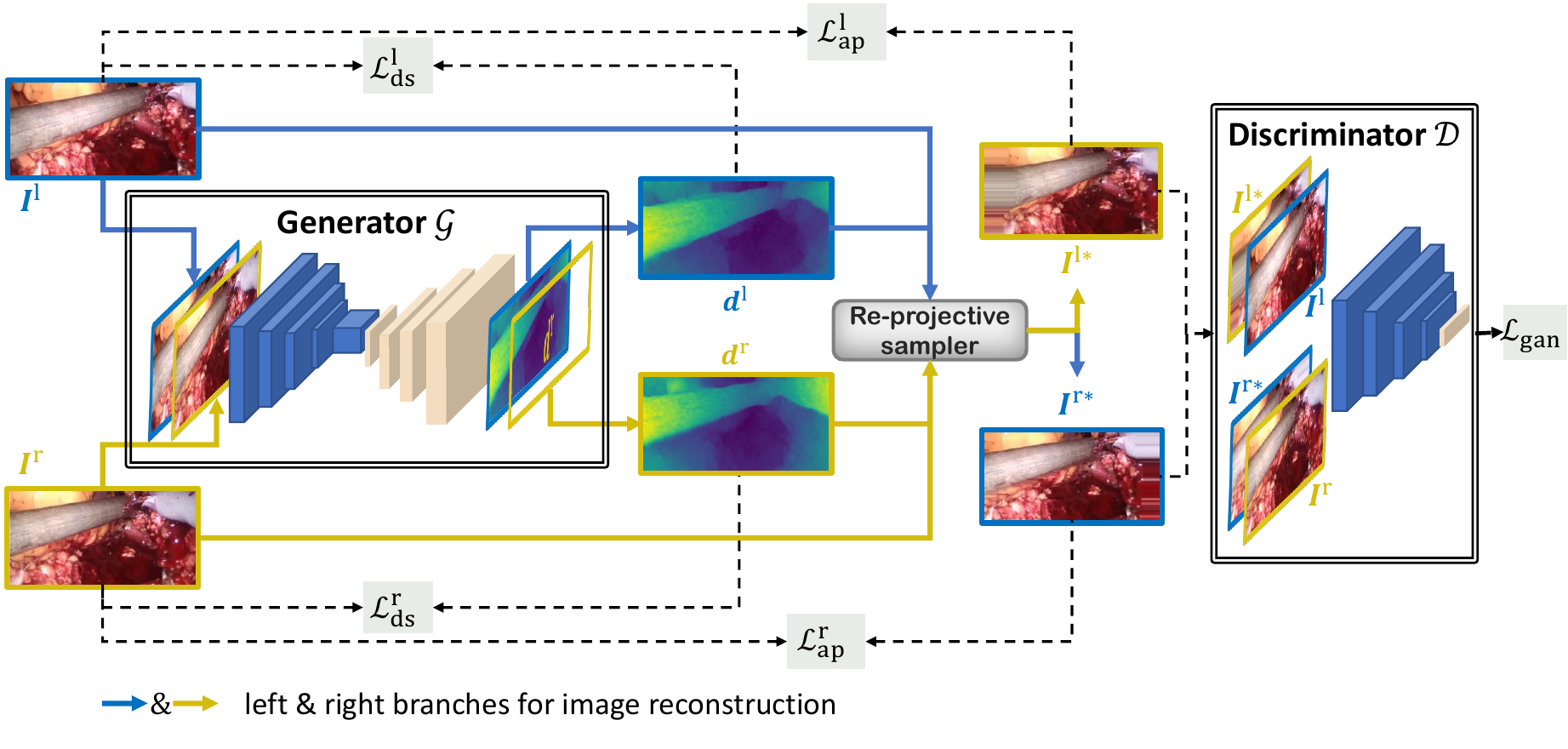}
\caption{Overview of the self-supervised adversarial depth estimation network, SADepth. 
} \label{fig:overview-depth2D}
\end{figure*}

Here we describe the proposed self-supervised adversarial depth estimation framework (see Fig~\ref{fig:overview-depth2D}), SADepth. Stereo depth estimation predicts depth maps $\textbf{\textit{D}}^{\rm l},\textbf{\textit{D}}^{\rm r}\in\mathbb{R}_{+}^{h\times w}$ based on the stereo RGB images ${\textbf{\textit{I}}^{\rm l}},{\textbf{\textit{I}}^{\rm r}}\in\mathbb{R}_{+}^{h\times w\times 3}$ of height and width $h,w$. A generative network $\mathcal{G}$ with stereo image pairs ${\textbf{\textit{I}}^{\rm l}}$ and ${\textbf{\textit{I}}^{\rm r}}$ as inputs, was used to produce two distinct left and right disparity maps ${\textbf{\textit{d}}^{\rm l}}$ and ${\textbf{\textit{d}}^{\rm r}}$, \textit{i.e.} ${\textbf{\textit{d}}^{\rm l}}$, ${\textbf{\textit{d}}^{\rm r}}$ = ${\mathcal{G}(\textbf{\textit{I}}^{\rm l}, \textbf{\textit{I}}^{\rm r})}$. As the two disparity maps were generated from different input images, a `reprojection sampler' \cite{jaderberg2015spatial} could be used for photometric reprojection loss computation of mutual counter-parts, \textit{i.e.} reconstructed left and right images ${\textbf{\textit{I}}^{\rm l*}}$ and ${\textbf{\textit{I}}^{\rm r*}}$. The discriminator $\mathcal{D}$ was exploited to indicate if the reconstructed images were real or fake (original input images were regarded as real). By forcing the reconstructed image to be consistent with the original input, we could derive accurate disparity maps for depth inference, as shown in the following sections.

\subsection{Network Architecture}
\begin{figure*}[t]
\includegraphics[width=\textwidth]{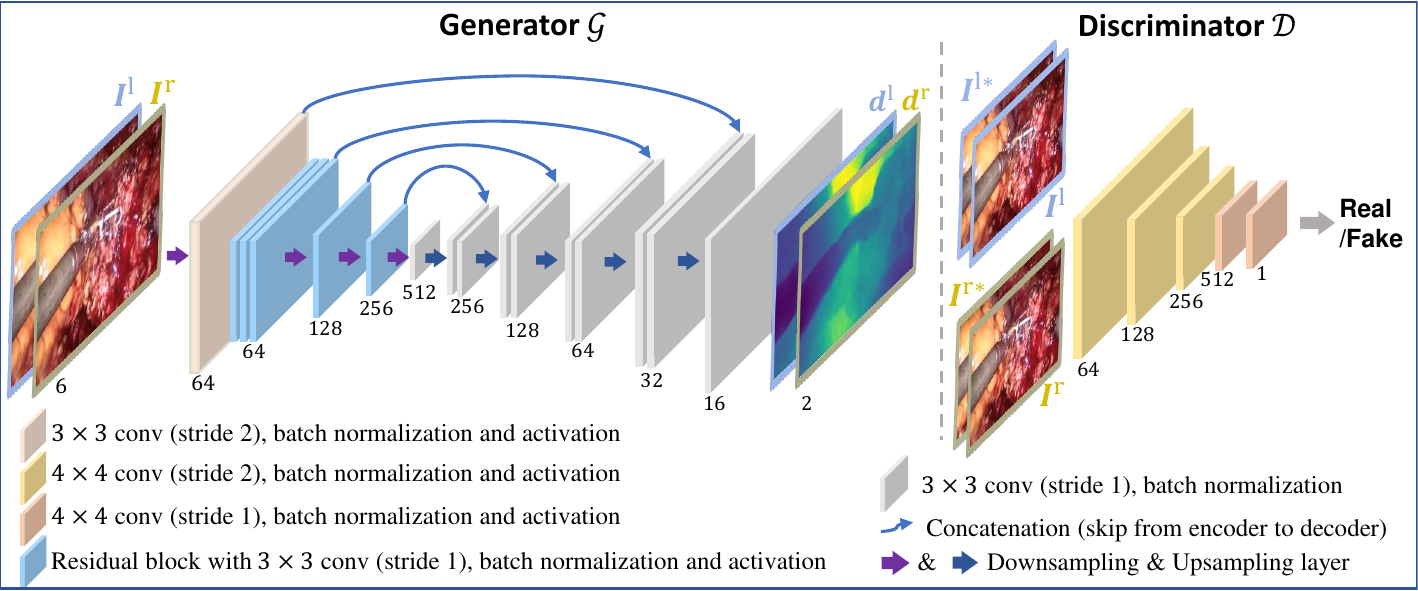}
\caption{The detailed architecture of the SADepth generator and discriminator. The generator was an autoencoder architecture with concatenated stereo image pairs as inputs and left and right disparity maps as outputs using a sigmoid function. These outputs were then transformed to reconstruct the counter-part camera input images using a `reprojection sampler', and these reconstructed images were fed into the discriminator together with the original input image pair. The discriminator outputs a scalar indicating whether the reconstructed images generated from the `reprojection sampler' were real or fake.} 
\label{fig:network-Depth2D}
\end{figure*}

\paragraph{Generator.}
The generator followed the general U-Net \cite{ronneberger2015u} architecture consisting of an encoder-decoder network, where the encoder was designed to obtain compact image representations and the decoder produced disparity maps for left and right input images, recovering them at the original scale (illustrated in Figure \ref{fig:network-Depth2D}). Encoder-decoder skip connections were applied to represent deep abstract features while preserving local information. To make the model compact - and different from less streamlined previous approaches which had two branches or two sub-networks for the encoder \cite{chang2018pyramid,pilzer2018unsupervised} - we first concatenated the left and right images into a 6-channel tensor and then fed it to a ResNet18 model \cite{he2016deep}. The input size was \(\emph{\#channels}\times h\times w=6 \times 192 \times 384\). Similar to \cite{godard2017unsupervised}, our decoder was formed of five cascaded blocks where each block had four parts: the first convolutional layer, an upsampling layer, a concatenation manipulation, and the second convolutional layer. In the upsampling layer, features were interpolated to twice the input size and both convolutional layers were followed by an \textit{ELU} activation function \cite{clevert2015fast}. In particular, sigmoids were applied at the output to generate a 2-channel tensor representing the left and right disparity ${\textbf{d}^{\rm l}}$ and ${\textbf{d}^{\rm r}}$. Finally the sigmoid outputs were converted to depth by \(\textbf{D}^{l(r)} = 1/(a \textbf{d}^{l(r)} + b)\), where parameters \(a\) and \(b\) were selected to constrain the depth \(\textbf{D}^{l(r)}\) between 0.1 and 100 units. The depth maps were then back-projected into point clouds by applying the intrinsic parameters and using the counter-part camera's extrinsic parameters to form reconstructed stereo images. The structural similarity between the original and reconstructed images was regarded as a supervisory signal to train the generator (see section 4.2.3 for the generator loss).

\paragraph{Discriminator.}
Goodfellow \textit{et al.} \cite{goodfellow2014generative} introduced a generative adversarial learning strategy and presented impressive results for image generation tasks. GANs have been widely exploited in different tasks with different \acrshort{gan} models including \textit{e.g.} DualGAN \cite{yi2017dualgan} and CycleGAN \cite{zhu2017unpaired}. To improve the generation quality of the reconstructed images ${\textbf{\textit{I}}^{\rm l*}}$ and ${\textbf{\textit{I}}^{\rm r*}}$, and following the work in \cite{pilzer2018unsupervised} for natural scenes, we applied an adversarial learning strategy for laparoscopic images to include geometry constraints during training and force the network to make a consistent depth map prediction. The original input stereo image pairs and reconstructed images \(\textbf{\textit{I}}^{\rm r*}\) and \(\textbf{\textit{I}}^{\rm l*}\) generated from the `reprojection sampler' were fed into the discriminator \(\mathcal{D}\), which consisted of convolutional, batch normalisation and activation function layers and classified the input and reconstructed images as real or fake. As training progressed, the reconstructed images became more similar to the original inputs, while the discriminator also became better at distinguishing between the input and reconstructed images, resulting in an overall improvement of the associated disparity maps. 


\subsection{Training Losses}
\paragraph{Generator Loss.}  

In the depth estimation generator network \(\mathcal{G}\), the loss \(\mathcal{L}_{\rm rec}^{\rm r}\) was formed from the appearance matching loss \(\mathcal{L}_{\rm ap}^{\rm r}\) and disparity smoothness loss \(\mathcal{L}_{\rm ds}^{\rm r}\) 
\begin{equation}
\mathcal{L}_{\rm rec}^{\rm r} = \mathcal{L}_{\rm ap}^{\rm r} + \alpha_{\rm ds} \mathcal{L}_{\rm ds}^{\rm r}
\end{equation}
where \(\alpha_{\rm ds}\) balanced the loss magnitude of the two parts to stabilise the training and was set to 0.001.

\paragraph{Appearance-Matching Loss.}
Self-supervised training in laparoscopic images typically assumes that the appearance and material properties (\textit{e.g.} brightness and Lambertian) of object surfaces are consistent between left and right frames. A local structure-based appearance loss \cite{godard2017unsupervised} can effectively improve the depth estimation performance compared with simple pairwise pixel differences \cite{zhou2017unsupervised}. Following \cite{godard2019digging}, we exploited the appearance-matching loss as part of the generator loss which forced the reconstructed image to be similar to the corresponding training inputs. 
During the training, the right disparity map ${\textbf{d}^{\rm r}}$ generated by the autoencoder was then transformed to produce ${\textbf{\textit{I}}^{\rm r*}}$ -- a reconstruction of the original right input image -- using RGB intensity information from the counter-part camera image ${\textbf{\textit{I}}^{\rm l}}$ (see Fig~\ref{fig:overview-depth2D}). This was achieved by first converting the disparity map ${\textbf{d}^{\rm r}}$ to a depth map ${\textbf{D}^{\rm r}}$, from which a point cloud of the surgical scene could be generated. Then the point cloud was transferred into the other camera's coordinate system and projected onto its image plane. The reconstructed input image  ${\textbf{\textit{I}}^{\rm r*}}$ was generated with bilinear interpolation for each output pixel using the weighted sum of the four neighbouring intensities. In contrast to \cite{garg2016unsupervised}, this bilinear sampling was locally fully differentiable, which allowed it to be integrated into the fully convolutional architecture without requiring simplification or approximation of the cost function. To compare the reconstructed image ${\textbf{\textit{I}}^{\rm r*}}$ and the original input image ${\textbf{\textit{I}}^{\rm r}}$, a combination of \acrfull{ssim}~\cite{wang2004image} and ${\mathcal{L}_1}$ loss were applied as the photometric image reconstruction cost ${\mathcal{L}_{ap}^{\rm r}}$: 
\begin{equation} \label{eq:appearance_matching_loss-Depth2D}
\mathcal{L}_{ap}^{\rm r} = \frac{1}{N}\sum_{i, j} \frac{\gamma}{2} (1-{\rm SSIM}(\textbf{\textit{I}}_{ij}^{\rm r}, \textbf{\textit{I}}_{ij}^{r*})) + (1-\gamma) {\|\textbf{\textit{I}}_{ij}^{\rm r}-\textbf{\textit{I}}_{ij}^{r*}\|}_1
\end{equation}
where \(N\) denotes the number of pixels and \(\gamma\) represents the weighting for L1-norm loss term, which was set to 0.85. Similar to \cite{godard2017unsupervised}, the calculation of \acrshort{ssim} here was simplified to a \(3\times 3\) block filter instead of a Gaussian. The training of the depth estimation generator then involved minimising the reconstruction loss between input and reconstructed images.

\paragraph{Disparity Smoothness Loss.}
Since disparities should be locally smooth and discontinuities usually occur at image gradients, we applied the disparity smoothness loss to penalise unexpected discontinuities in the disparity maps. Following \cite{heise2013pm}, this cost was an edge-aware term weighted with the input image gradients \(\partial \textbf{I}\):

\begin{equation}
\mathcal{L}_{\rm ds}^{\rm r} = \frac{1}{N}\sum_{ij}|\partial_x (\textbf{d}_{ij}^{\rm r})|e^{-|\partial_x \textbf{\textit{I}}_{ij}^{\rm r}|} + |\partial_y (\textbf{d}_{ij}^{\rm r})| e^{-|\partial_y \textbf{\textit{I}}_{ij}^{\rm r}|}
\end{equation}
where \(\textbf{d}^{\rm r}\) represents the generated disparity map and \(\textbf{\textit{I}}^{\rm r}\) is the original input right image. 

\paragraph{Discriminator Loss.}  
The adversarial objective of the generative network can be expressed as follows:
\begin{equation}
\label{GAN}
\mathcal{L}_{\rm gan}^{\rm r}(\textbf{\textit{I}}^{\rm r}, \textbf{\textit{I}}^{\rm r*};\mathcal{G}, \mathcal{D}) = \mathbb{E}_{\textbf{\textit{I}}^{\rm r}\sim P(\textbf{\textit{I}}^{\rm r})} [{\rm log} (\mathcal{D}(\textbf{\textit{I}}^{\rm r}))] + \mathbb{E}_{\textbf{\textit{I}}^{\rm r*}\sim P(\textbf{\textit{I}}^{\rm r*})} [{\rm log}(1-\mathcal{D}(\textbf{\textit{I}}^{\rm r*}))]
\end{equation}
where a cross-entropy loss measured the expectation of the reconstructed image \(\textbf{\textit{I}}^{\rm r*}\) against the distribution of the input image \(\textbf{\textit{I}}^{\rm r}\). 
Note that both generator and discriminator losses included losses for left and right images but only the right image equations are shown.

\paragraph{Multi-Scale Loss.}
One remaining issue with the above learning pipeline was that the training objective risked becoming stuck in local minima due to the application of a photometric reprojection loss \cite{watson2019self}.
The strategy introduced in~\cite{zhou2017unsupervised} indicated that combining the individual losses across multiple scales in the decoder was effective, which could improve the depth estimation performance and reduce sensitivity to architectural choices. Hence, the lower resolution depth maps (from the intermediate layers) were first upsampled to the input image resolution and then reprojected and resampled, with the errors computed at the higher input resolution. This manipulation is similar to matching patches, which enables low-resolution disparity maps to warp an entire patch of pixels in a high resolution image while promoting the depth maps at every scale to reconstruct the high resolution input image as accurately as possible~\cite{godard2019digging}.

\paragraph{Joint Optimization Loss} Finally, the joint optimization loss was a combination of generator loss and adversarial loss, written as:
\begin{equation} \label{eq:total_depth_loss}
\mathcal{L}_{\rm total} =\frac{1}{m}\sum_{s=1}^m \frac{\mathcal{L}_s^{\rm l} + \mathcal{L}_s^{\rm r}}{2}=  \frac{1}{m} \sum_{s=1}^m \big(\alpha(\mathcal{L}_{\rm rec}^{\rm l} + \mathcal{L}_{\rm rec}^{\rm r}) + \beta (\mathcal{L}_{\rm gan}^{\rm l} + \mathcal{L}_{\rm gan}^{\rm r})\big)
\end{equation}

\textbf{Training}
The depth estimation procedure was trained based on the reconstruction supervision signal and no per-pixel depth ground truth labels were needed. The augmentation of input data was performed on the fly by flipping 50 \% of the input images horizontally and reorienting the stereo pairs. Parameter $m$ was set to 4, which means that there were 4 output scales with resolutions \(\frac{1}{2^0}\), \(\frac{1}{2^1}\), \(\frac{1}{2^2}\) and \(\frac{1}{2^3}\) of the input resolution. \(\alpha\) and \(\beta\) were set to 0.5.

\begin{figure*}[t]
\includegraphics[width=\textwidth]{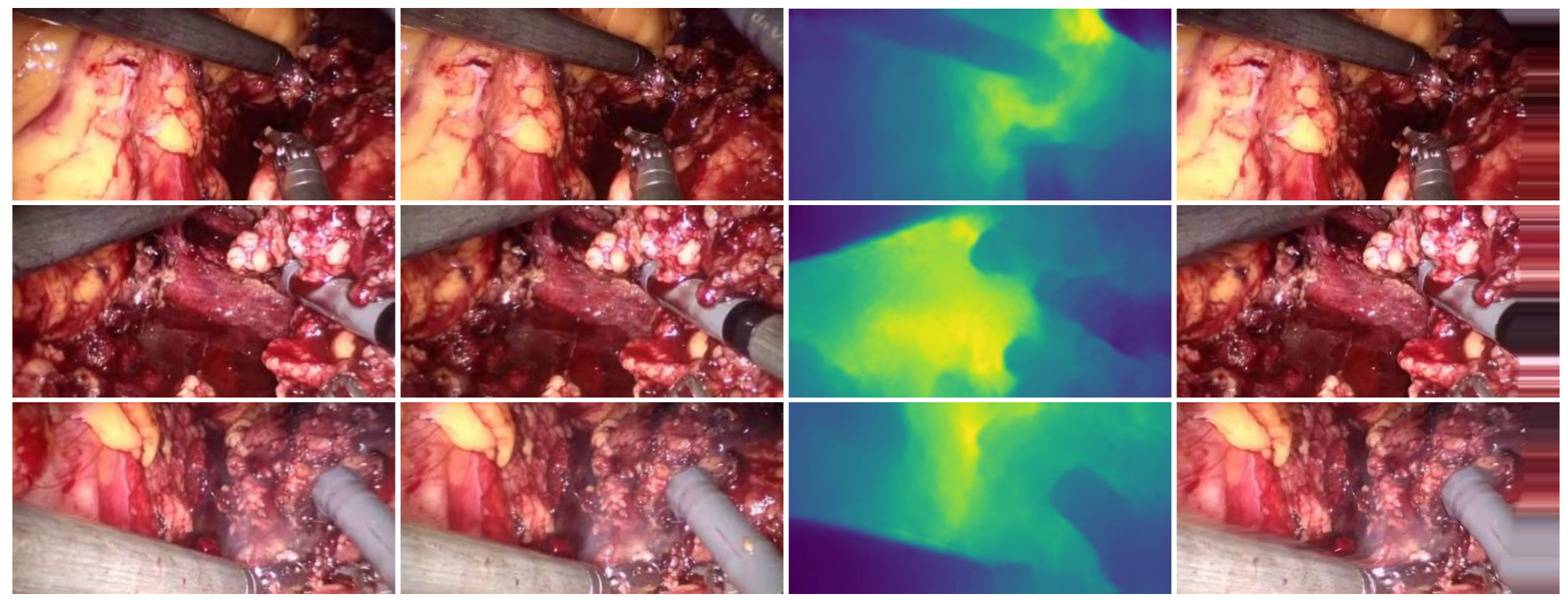}
\caption{Qualitative results on dVPN dataset. From left to right, they are left image, right image, right depth map, and reconstructed right image.} 
\label{fig:qualitative-Depth2D}
\end{figure*}

\begin{table}[h]
\centering
\caption{\acrshort{ssim} score for dVPN test set (higher is better).}\label{davinci_result}
\begin{tabular}{cccc}
\toprule
\textbf{Method} & \textbf{Training} &\textbf{Mean SSIM} & \textbf{Std. SSIM} \\
\midrule
ELAS \cite{geiger2010efficient}  &No training & 47.3  &0.079\\
SPS \cite{yamaguchi2014efficient}  &No training & 54.7 &0.092\\
V-Basic \cite{ye2017self} &Unsupervised  & 55.5 &0.106\\
V-Siamese \cite{ye2017self} &Unsupervised  & 60.4  &0.066\\
Monodepth \cite{godard2017unsupervised} &Unsupervised  &54.9   &0.087  \\
Monodepth2 \cite{godard2019digging}   &Unsupervised &71.2  & 0.075\\
\rowcolor[RGB]{230,230,230} \textbf{SADepth (ours)} &\textbf{Unsupervised} &\textbf{79.6} &\textbf{0.049}  \\ 
\bottomrule
\end{tabular}
\end{table} 

\section{Experiments and results}
\label{sec:4-results}

\subsection{Dataset}
We evaluated SADepth on two datasets. The first was the dVPN dataset, collected from da Vinci partial nephrectomy, with 34320 pairs of rectified stereo images for training and 14382 pairs for testing \cite{ye2017self}. The second was the SCARED dataset \cite{allan2021stereo} released during the Endovis challenge at MICCAI 2019, with 17206 pairs (dataset 1, 2, 3, 6, and 7) of rectified stereo images for training and 5637 pairs for testing. To verify the generalization of our framework, we only trained on the dVPN dataset but tested on both the dVPN and SCARED datasets.

\subsection{Evaluation Metrics, Baseline, and Implementation}
\textbf{Evaluation Metrics.} As the ground truth depth labels were not available for the \textit{in vivo} surgical data in the dVPN dataset, we adopted the \acrshort{ssim} index to evaluate the similarity between the reconstructed image and the original input image (\textit{i.e.} \(\textbf{\textit{I}}^{r*}\) and \(\textbf{\textit{I}}^{r}\)) as the evaluation metric. For the SCARED dataset, the team at Intuitive Surgical collected the ground truth by using structured light, thus we used the absolute error to assess our SADepth model.

\textbf{Baseline.}
We compared SADepth with several recent works. For the dVPN dataset, we compared our method with stereo matching-based methods: ELAS \cite{geiger2010efficient} and SPS \cite{yamaguchi2014efficient}; Siamese-based networks: V-Basic \cite{ye2017self} and V-Siamese \cite{ye2017self}; and recent deep learning methods: Monodepth \cite{godard2017unsupervised} and the stereo mode of Monodepth2 \cite{godard2019digging}. For the SCARED dataset, we compared our results with the methods summarized by the recent MICCAI sub-challenge paper \cite{allan2021stereo}. Note that they are all laparoscopic images and align well with the application environment of the gamma probe. Hence, the algorithms and experimental exploration on these scenes could lay a good foundation for further integrating the gamma probe into the environment.

\textbf{Implementation Details.}
The SADepth model was implemented in PyTorch \cite{paszke2017automatic}, with a batch size of 16 and input/output resolution of \(192\times384\). The learning rate was set to \(10^{-4}\) for the first 15 epochs and then dropped to \(10^{-5}\) for the remainder. The model was trained for 20 epochs using the Adam optimizer which took about 22 hours on a single NVIDIA 2080 Ti GPU.

\begin{table}
\centering
\renewcommand\tabcolsep{9pt}
\caption{The mean absolute depth error for the SCARED test set 1 and 2 (unit: mm) (lower is better).}\label{Endovis_test}
\begin{tabular}{cccc}
\toprule
\textbf{Method}  & \textbf{Training}  & \makecell[c]{\textbf{Test Set 1} \\ \textbf{Average}} & \makecell[c]{\textbf{Test Set 2} \\ \textbf{Average}}\\ \midrule
Lalith Sharan \cite{allan2021stereo} &Supervised  &43.03 &48.72\\
Xiaohong Li \cite{allan2021stereo} &Supervised  &22.77 &20.52\\
Huoling Luo \cite{allan2021stereo} &Supervised  &19.52 &18.21\\
Zhu Zhanshi \cite{allan2021stereo} &Supervised  &9.60  &21.20\\
Wenyao Xia \cite{allan2021stereo} &Supervised  &6.73 &9.44\\

Congcong Wang \cite{allan2021stereo} &Supervised  &4.10 &4.28\\
Trevor Zeffiro \cite{allan2021stereo} &Supervised  &3.60 &3.47\\
J.C. Rosenthal \cite{allan2021stereo} &Supervised  &3.44  &4.05\\

Dimitris Psychogyios 1 \cite{allan2021stereo} &Supervised  &3.00  &1.67  \\
Dimitris Psychogyios 2 \cite{allan2021stereo} &Supervised &2.95  &2.30  \\
\midrule
KeXue Fu \cite{allan2021stereo} &Unsupervised &20.94 &17.22  \\
Monodepth \cite{godard2017unsupervised} &Unsupervised  &23.56 &21.62  \\
Monodepth2 \cite{godard2019digging} &Unsupervised &21.92  &15.25  \\
\rowcolor[RGB]{230,230,230} \textbf{SADepth (ours)}  &\textbf{Unsupervised}   &\textbf{17.42}  &\textbf{11.23}  \\
\bottomrule
\end{tabular}
\end{table}

\subsection{Results}
Qualitative results generated by our method are presented in Fig~\ref{fig:qualitative-Depth2D} using the dVPN dataset, which includes left and right images, right estimated depth maps, and reconstructed right images. Note that in the reconstructed right image, the far right border is a repeat of the last column pixel due to existence of the unseen area of the right image in the left image. The SADepth and other state-of-the-art quantitative results for the dVPN dataset are summarized in Table~\ref{davinci_result} using the mean and \acrfull{std} of the \acrshort{ssim} index. The SADepth model effectively outperformed other methods with an \acrshort{ssim} of 79.6, \textit{i.e.} 24.7 units higher than Monodepth \cite{godard2017unsupervised},  8.4 units higher than Monodepth2 \cite{godard2019digging}, and 19.2 units higher than the Siamese architecture~\cite{ye2017self}.

Table~\ref{Endovis_test} presents the results of SADepth on test set 1 and test set 2 (as defined in the SCARED dataset), together with the performance reported in the MICCAI sub-challenge summary paper \cite{allan2021stereo}. The results show an improvement over the unsupervised methods from the summary paper and recent baselines, while it is also competitive with some supervised approaches. It is expectable that the supervised methods outperformed unsupervised methods due to the extra ground truth information. The results from different experimental settings in Table~\ref{davinci_result} and Table~\ref{Endovis_test} also confirm that SADepth generalises well across different datasets collected from different laparoscopes and subjects, while still producing superior performance compared with the state-of-the-art unsupervised approaches.


\section{Conclusion}
\label{sec:4-conclusions}

We have presented a new self-supervised adversarial depth estimation framework SADepth with an encoder-decoder generator and a concatenated stereo image pair as the input. The adversarial learning strategy improved the generation quality of the framework and led to the state-of-the-art performance on two public datasets. Furthermore, SADepth did not require any per-pixel depth labels and generalized well across different laparoscopes, suggesting excellent applicability to scalable data acquisition when accurate ground truth depth cannot be collected.

\section{Chapter Transitions}

In this study, our focus was primarily on investigating left-right consistency in 2D, along with exploring properties of the disparity map and appearance matching in 2D. However, it is crucial to acknowledge that the essence of depth recovery from images lies in the comprehension of 3D space. Furthermore, the input data employed in this research comprises stereo laparoscopic images, which demand increased computational resources and memory usage, rendering them incompatible with a single-image input stream and \acrshort{ar} processing. 

Consequently, we introduced an alternative depth estimation module that capitalizes on the 3D geometrical consistency inherent in the image data. This module is designed to be adaptable to both monocular and stereo image input networks, enhancing the versatility of our approach. 

\setcounter{chapter}{4}

\chapter{Self-Supervised Depth Estimation in Laparoscopic Image using 3D Geometric Consistency}
\chaptermark{3D Geometric Consistency for Depth Estimation}
\glsresetall
\label{chap:5-contribution_3-Depth3D}

\begin{cabstract}

Depth estimation is a crucial step for image-guided intervention in robotic surgery and laparoscopic imaging systems. Since per-pixel depth ground truth is difficult to acquire for laparoscopic image data, it is rarely possible to apply supervised depth estimation to surgical applications. As an alternative, self-supervised methods have been introduced to train depth estimators using only synchronised stereo image pairs. However, most recent work focused on the left-right consistency in 2D and ignored valuable inherent 3D information on the object in real world coordinates, meaning that the left-right 3D geometric structural consistency is not fully utilised. To overcome this limitation, we present M3Depth, a self-supervised depth estimator to leverage 3D geometric structural information hidden in stereo pairs while keeping monocular inference. The method also removes the influence of border regions unseen in at least one of the stereo images via masking, to enhance the correspondences between left and right images in overlapping areas. Intensive experiments show that our method outperforms previous self-supervised approaches on both a public dataset and a newly acquired dataset by a large margin, indicating a good generalisation across different samples and laparoscopes. 

\end{cabstract}
This chapter’s research was previously published at the International Conference on Medical Image Computing and Computer Assisted Intervention (MICCAI) in 2022~\cite{huang2021self}. 

\section{Introduction}
\label{sec:5-introduction}

Perception of 3D surgical scenes is a fundamental problem in computer assisted surgery. Accurate perception, tissue tracking, 3D registration between intra- and pre-operative organ models, target localisation and augmented reality \cite{liu2019dense,huang2020tracking} are predicated on having access to correct depth information. Range-finding sensors such as multi-camera systems or LiDAR that are often employed in autonomous systems and robotics are not convenient for robot-assisted minimally invasive surgery because of the limited port size and requirement of sterilisation. Furthermore, strong `dappled' specular reflections as well as less textured tissues hinder the application of traditional methods \cite{luo2016efficient}. This has led to the exploration of learning-based methods, among which fully convolutional neural networks are particularly successful \cite{allan2021stereo,tran2022light}.   

Since it is challenging to obtain per-pixel ground truth depth for laparoscopic images, there are far fewer datasets in the surgical domain compared with mainstream computer vision applications~\cite{geiger2013vision,huang2022h}. It is also not a trivial task to transfer approaches that are based on supervised learning to laparoscopic applications due to the domain gap. To overcome these limitations, view-synthesis methods are proposed to provide self-supervised learning for depth estimation \cite{vuong2024scene,huang2021self,liu2019dense,}, with no supervision via per-pixel depth data. Strong depth prediction baselines have been established in \cite{godard2017unsupervised,godard2019digging,johnston2020self}. However, all of these methodologies employed left-right consistency and smoothness constraints in 2D, \textit{e.g.} \cite{godard2017unsupervised,guizilini20203d,huang2022simultaneous}, 
and ignored the important 3D geometric structural consistency of the stereo images.

Recently, a self-supervised semantically-guided depth estimation method was proposed to deal with moving objects \cite{klingner2020self}, which made use of mutually beneficial cross-domain training of semantic segmentation. Jung \textit{et al.} \cite{jung2021fine} extended this work by incorporating semantics-guided local geometry into intermediate depth representations for geometric representation enhancement. However, semantic labels are not common in laparoscopic applications except for surgical tool masks, impeding the extension of this work. Mahjourian \textit{et al.} \cite{mahjourian2018unsupervised} presented an approach for unsupervised learning of depth by enforcing consistency of ego-motion across consecutive frames to infer 3D geometry of the whole scene. However, in laparoscopic applications, the interaction between the surgical tools and tissue creates a dynamic scene, leading to the failure of local photometric and geometric consistency across consecutive frames in both 2D and 3D. Nevertheless, the 3D geometry inferred from left and right synchronised images can be assumed identical, allowing the adoption of 3D- as well as 2D-loss.   

In this work, we propose a new framework for self-supervised laparoscopic image depth estimation called M3Depth, leveraging not only the left-right consistency in 2D but also the inherent geometric structural consistency of real-world objects in 3D (see Section \ref{3dloss} for the 3D geometric consistency loss), while enhancing the mutual information between stereo pairs. A U-Net architecture \cite{ronneberger2015u} was employed as the backbone and the network was fed with only the left image as inputs but was trained with the punitive loss formed by stereo image pairs. To cope with the unseen areas at the image edges that were not visible in both cameras, blind masking was applied to suppress and eliminate outliers and give more emphasis to feature correspondences that lay on the shared vision field. 
Extensive experiments on both a public dataset and a new experimental dataset demonstrated the effectiveness of this approach and a detailed ablation study indicated the respective positive influence of each proposed novel module on the overall performance.  
\section{Methodology}
\label{sec:5-methods}

\subsection{Network Architecture}

\begin{figure*}[t]
\includegraphics[width=\textwidth]{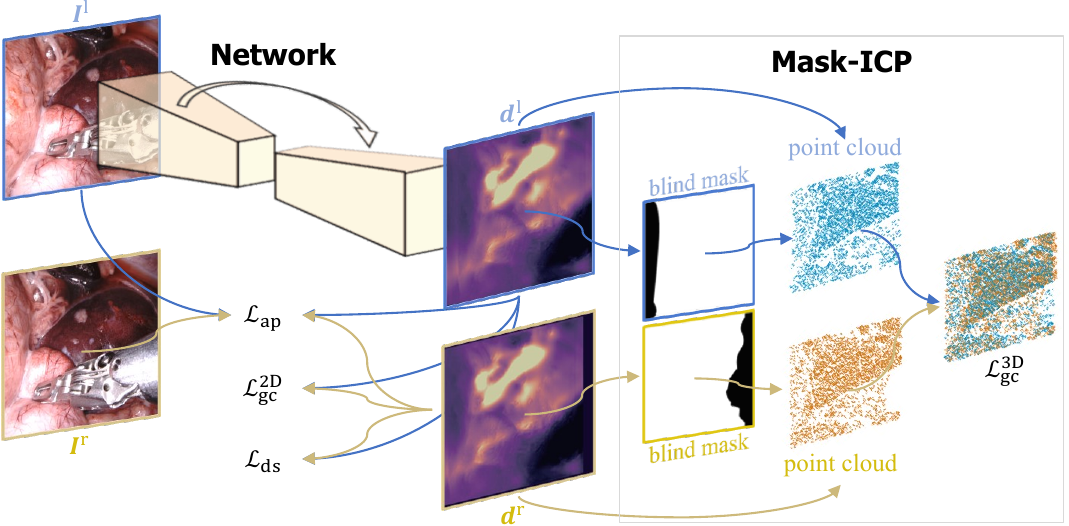}
\caption{Overview of the proposed self-supervised depth estimation network. ResNet18 was adopted as the backbone and received a left image from a stereo image pair as the input. Left and right disparity maps were produced simultaneously and formed 2D losses with the original stereo pair. 3D point clouds were generated by applying the intrinsic parameters of the camera and iterative closest point loss was calculated between them. Blind masks were applied to the 2D disparity maps to remove outliers from areas not visible in both cameras. }
\label{fig:overview-depth3D}
\end{figure*}

\textbf{Network Architecture}
The backbone of the M3Depth followed the general U-Net \cite{ronneberger2015u} architecture, \textit{i.e.} an encoder-decoder network, in which an encoder was employed to extract image representations while a decoder with convolutional layer and upsampling manipulation was designed to recover disparity maps at the original scale. Skip connections were applied to obtain both deep abstract features and local information. To keep a lightweight network, a ResNet18 \cite{he2016deep} was employed as the encoder with only 11 million parameters. To improve the regression ability of the network from intermediate high-dimensional features maps to disparity maps, one more ReLU \cite{nair2010rectified} activation function and a convolutional layer with decreased last latent feature map dimension were added before the final sigmoid disparity prediction. Similar to Monodepth1 \cite{godard2017unsupervised}, in M3Depth, the left image ${\textbf{\textit{I}}^{\rm l}}$ of a stereo image pair (${\textbf{\textit{I}}^{\rm l}},{\textbf{\textit{I}}^{\rm r}}\in\mathbb{R}_{+}^{h\times w\times 3}$) was always the input and the framework generated two distinct left and right disparity maps ${\textbf{\textit{d}}^{\rm l}}$, ${\textbf{\textit{d}}^{\rm r}} \in\mathbb{R}_{+}^{h\times w}$ simultaneously, \textit{i.e.}  ${\mathcal{Z}: \textbf{\textit{I}}^{\rm l}\mapsto({\textbf{\textit{d}}^{\rm l}},{\textbf{\textit{d}}^{\rm r}})}$. Given the camera focal length $\textbf{\textit{f}}$ and the baseline distance $\textbf{\textit{b}}$ between the cameras, left and right depth maps $\textbf{\textit{D}}^{\rm l},\textbf{\textit{D}}^{\rm r}\in\mathbb{R}_{+}^{h\times w}$ could then be trivially recovered from the predicted disparity, ($\textbf{\textit{D}}^{\rm l},\textbf{\textit{D}}^{\rm r}$)=$\textbf{\textit{bf}}$/(${\textbf{\textit{d}}^{\rm l}}$, ${\textbf{\textit{d}}^{\rm r}}$). $h$ and $w$ denote image height and width. The overview of the architecture is presented in Fig~\ref{fig:overview-depth3D}.

\textbf{Image Reconstruction Loss in 2D}
With the predicted disparity maps and the original stereo image pair, left and right images could then be reconstructed by warping the counter-part RGB image with the disparity map mimicking optical flow \cite{godard2017unsupervised} \cite{lipson2021raft}. Similar to Monodepth1 \cite{godard2017unsupervised}, an appearance matching loss \(\mathcal{L}_{\rm ap}\), disparity smoothness loss \(\mathcal{L}_{\rm ds}\) and left-right disparity consistency loss \(\mathcal{L}_{\rm lr}^{2D}\) were used to encourage coherence between the original input and reconstructed images (\(\textbf{\textit{I}}^{\rm l*}, \textbf{\textit{I}}^{\rm r*}\)) as well as consistency between left and right disparities while forcing disparity maps to be locally smooth.

\begin{equation} \label{eq:appearance_matching_loss}
\mathcal{L}_{\rm ap}^{\rm r} = \frac{1}{N}\sum_{i, j} \frac{\gamma}{2} (1-{\rm SSIM}(\textbf{\textit{I}}_{ij}^{\rm r}, \textbf{\textit{I}}_{ij}^{r*})) + (1-\gamma) {\|\textbf{\textit{I}}_{ij}^{\rm r}-\textbf{\textit{I}}_{ij}^{r*}\|}_1
\end{equation}

\begin{equation}
\mathcal{L}_{\rm lr}^{\rm 2D(r)} = \frac{1}{N}\sum_{ij}| \textbf{d}_{ij}^{\rm r} +  \textbf{d}_{ij+\textbf{d}_{ij}^{\rm r}}^{\rm l}|
\end{equation}

\begin{equation}
\mathcal{L}_{\rm ds}^{\rm r} = \frac{1}{N}\sum_{ij}|\partial_x (\textbf{d}_{ij}^{\rm r})|e^{-|\partial_x \textbf{\textit{I}}_{ij}^{\rm r}|} + |\partial_y (\textbf{d}_{ij}^{\rm r})| e^{-|\partial_y \textbf{\textit{I}}_{ij}^{\rm r}|}
\end{equation}
where \(N\) is the number of pixels and \(\gamma\) was set to 0.85. Note that 2D losses were applied on both left and right images but only equations for the right image are presented here.

\subsection{Learning 3D Geometric Consistency}
\label{3dloss}
Instead of using the inferred left and right disparities only to establish a mapping between stereo coordinates and generate reconstructed original RGB input images, a loss function was also constructed that registered and compared left and right point clouds directly to enforce the 3D geometric consistency of the whole scene. The disparity maps of the left and right images were first converted to depth maps and then backprojected to 3D coordinates to obtain left and right surgical scene point clouds ($\textbf{\textit{P}}^{\rm l},\textbf{\textit{P}}^{\rm r}\in\mathbb{R}^{hw\times3}$) by multiplying the depth maps with the intrinsic parameter matrix (\(\textbf{\textit{K}}\)). The 3D consistency loss employed \acrfull{icp} \cite{mahjourian2018unsupervised,rusinkiewicz2001efficient}, a classic rigid registration method that derives a transformation matrix between two point clouds by iteratively minimising point-to-point distances between correspondences. 

From an initial alignment, \acrshort{icp} alternately computed corresponding points between two input point clouds using the closest point heuristic and then recomputed a more accurate transformation based on the given correspondences. The final residual registration error after \acrshort{icp} minimisation was output as one of the returned values. More specifically, to explicitly explore global 3D loss, the \acrshort{icp} loss at the original input image scale was calculated with only 1000 randomly selected points to reduce the computational workload. 

\subsection{Blind Masking}
Some parts of the left scene were not visible in the right view and vice versa, leading to non-overlapping generated point clouds. These areas are mainly located at the left edge of the left image and the right edge of the right image after rectification. Depth and image pixels in those areas had no useful information for learning, either in 2D and 3D. Our experiments indicated that retaining the contribution to the loss functions for such pixels and voxels degraded the overall performance. Many previous approaches solved this problem by padding these areas with zeros \cite{godard2017unsupervised} or values from the border \cite{godard2019digging}, but this can lead to edge artifacts in depth images~\cite{mahjourian2018unsupervised}.

To tackle this problem, we present a blind masking module \(\mathcal{M}^{\rm l,r}\) that suppressed and eliminated these outliers and gave more emphasis to correspondences between the left and right views. First, a meshgrid was built with the original left image pixel coordinates in both horizontal \(\mathcal{X}_{\rm grid}\) and vertical \(\mathcal{Y}_{\rm grid}\) directions. Then the \(\mathcal{X}_{\rm grid}\) was shifted along the horizontal direction using the right disparity map ${\textbf{\textit{d}}^{\rm r}}$ to a get a new grid  \(\mathcal{X'}_{\rm grid}\), which was then stacked with \(\mathcal{Y}_{\rm grid}\) to form a new meshgrid. Finally, grid sampling was employed on the new meshgrid with the help of the original left image coordinates, from which the pixels that were not covered by the right view for the current synchronised image pair were obtained. By applying the blind masking on the depth maps for the stereo 3D point cloud generation, a 3D alignment loss was obtained as follows. 

\begin{equation}
\textbf{\textit{M}}_{ij}^{\rm l,r} =  \left\{
\begin{array}{cc}
1, & (\textbf{d}_{ij}^{\rm l,r}+\mathcal{X}_{ij})\in\{\mathcal{X}\}  \\
0, & (\textbf{d}_{ij}^{\rm l,r}+\mathcal{X}_{ij})\notin\{\mathcal{X}\}
\end{array}
\right.
\end{equation}
\begin{equation}
\textbf{\textit{P}}^{\rm l,r}={\rm backproj}(\textbf{d}^{\rm l,r},\textbf{\textit{K}},\textbf{\textit{M}}^{\rm l,r})
\end{equation}
\begin{equation}
\mathcal{L}_{\rm gc}^{\rm 3D} = 
{\rm ICP}(\textbf{\textit{P}}^{\rm l}, \textbf{\textit{P}}^{\rm r})
\end{equation}

\subsection{Training Loss}
Pixel-wise, gradient-based 2D losses and point cloud-based 3D losses were applied to force the reconstructed image to be identical to the original input while encouraging the left-right consistency in both 2D and 3D to derive accurate disparity maps for depth inference. Finally, an optimisation loss used a combination of these, written as:

\begin{equation}
\begin{split}
\mathcal{L}_{\rm total} & = (\mathcal{L}_{\rm 2D}^{\rm r} + \mathcal{L}_{\rm 2D}^{\rm l}) +  \mathcal{L}_{\rm lr}^{\rm 3D} \\
& = \alpha_{\rm ap}(\mathcal{L}_{\rm ap}^{\rm r} + \mathcal{L}_{\rm ap}^{\rm l}) + \alpha_{\rm ds}( \mathcal{L}_{\rm ds}^{\rm r} + \mathcal{L}_{\rm ds}^{\rm l}) + \alpha_{\rm lr}^{2D}(\mathcal{L}_{\rm lr}^{\rm 2D(r)} + \mathcal{L}_{\rm lr}^{\rm 2D(l)}) + \beta \mathcal{L}_{\rm gc}^{\rm 3D}
\end{split}
\end{equation}
where \(\alpha_{\rm *}\) and \(\beta\) balanced the loss magnitude of the 2D and 3D parts to stabilise the training. More specifically, \(\alpha_{\rm ap}\), \(\alpha_{\rm ds}\), \(\alpha_{\rm lr}^{\rm 2D}\) and \(\beta\) were experimentally set to 1.0, 0.5, 1.0 and 0.001. 
\section{Experiments}
\label{sec:5-Experiments}

\subsection{Dataset}


M3Depth was evaluated on two datasets. The first was the SCARED dataset \cite{allan2021stereo} released at the MICCAI Endovis Challenge 2019. As only the ground truth depth map of key frames in each dataset was provided (from structured light), the other depth maps were created by reprojection and interpolation of the key frame depth maps using the kinematic information from the \textit{da Vinci} robot, causing a misalignment between the ground truth and the RGB data. Hence, only key-frame ground truth depth maps were used in the test dataset while the remainder of the RGB data formed the training set but with the similar adjacent frames removed. 

\begin{figure*}[t]
\includegraphics[width=\textwidth]{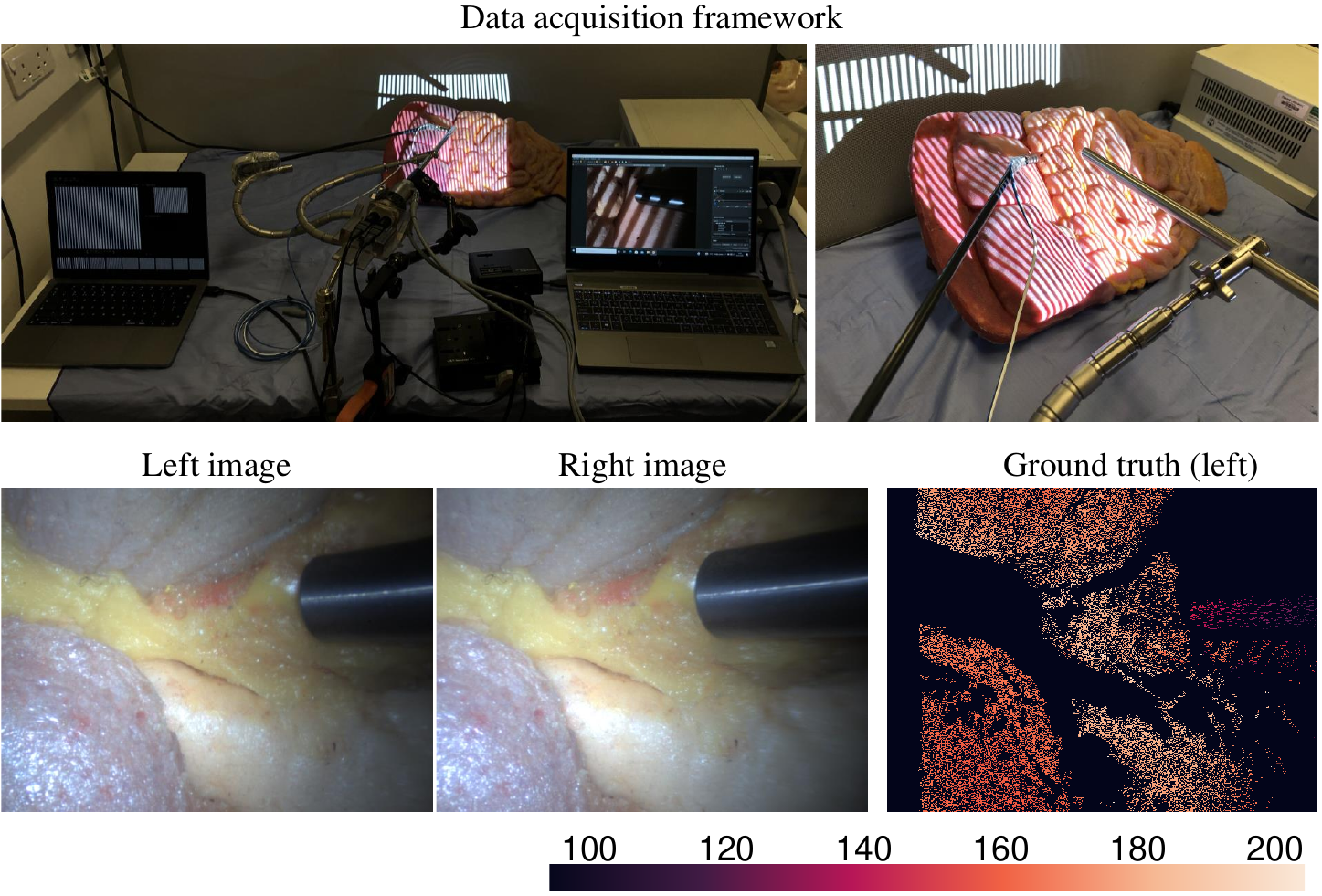}
\caption{
The data acquisition setting and examples of the LATTE dataset.}

\label{fig:data-acqu-depth2D}
\end{figure*}

To overcome the SCARED dataset misalignment and improve the validation, an additional laparoscopic image dataset (namely LATTE) was experimentally collected, including RGB laparoscopic images and corresponding ground truth depth maps calculated from a custom-built structured lighting pattern projection. More specifically, the grey-code detection and decoder algorithm \cite{xu2007robust} were used with both original and inverse patterns. To remove the uncertainty brought by occlusions and uneven illumination conditions, we used a more advanced 3-phase detection module, in which sine waves were shifted by $\pi/3$ and $2\pi/3$ and the modulation depth $\mathcal{T}$ was calculated for every pixel. Pixels with modulation depth under $\mathcal{T}$ were defined as uncertain pixels, and the equation for calculating the modulation is written as Eq.~\ref{eq:modulation}. Because of the limitations of using animal tissue in the lab, we conducted the experiments on a tissue phantom that is realistic and similar to real tissue. This provided 739 extra image pairs for training and 100 pairs for validation and testing.

\begin{equation}
\mathcal{T} = \frac{2\sqrt{2 }}{3}\times\sqrt{(\mathcal{I}_{1} - \mathcal{I}_{2})^2 + (\mathcal{I}_{2} - \mathcal{I}_{3})^2 + (\mathcal{I}_{1} - \mathcal{I}_{3})^2}
\label{eq:modulation}
\end{equation}
where $\mathcal{I}_{1}, \mathcal{I}_{2}, \mathcal{I}_{3}$ denotes original modulation and modulations after shifts.

\begin{figure*}[!t]
  \centering
  \includegraphics[width=\textwidth]{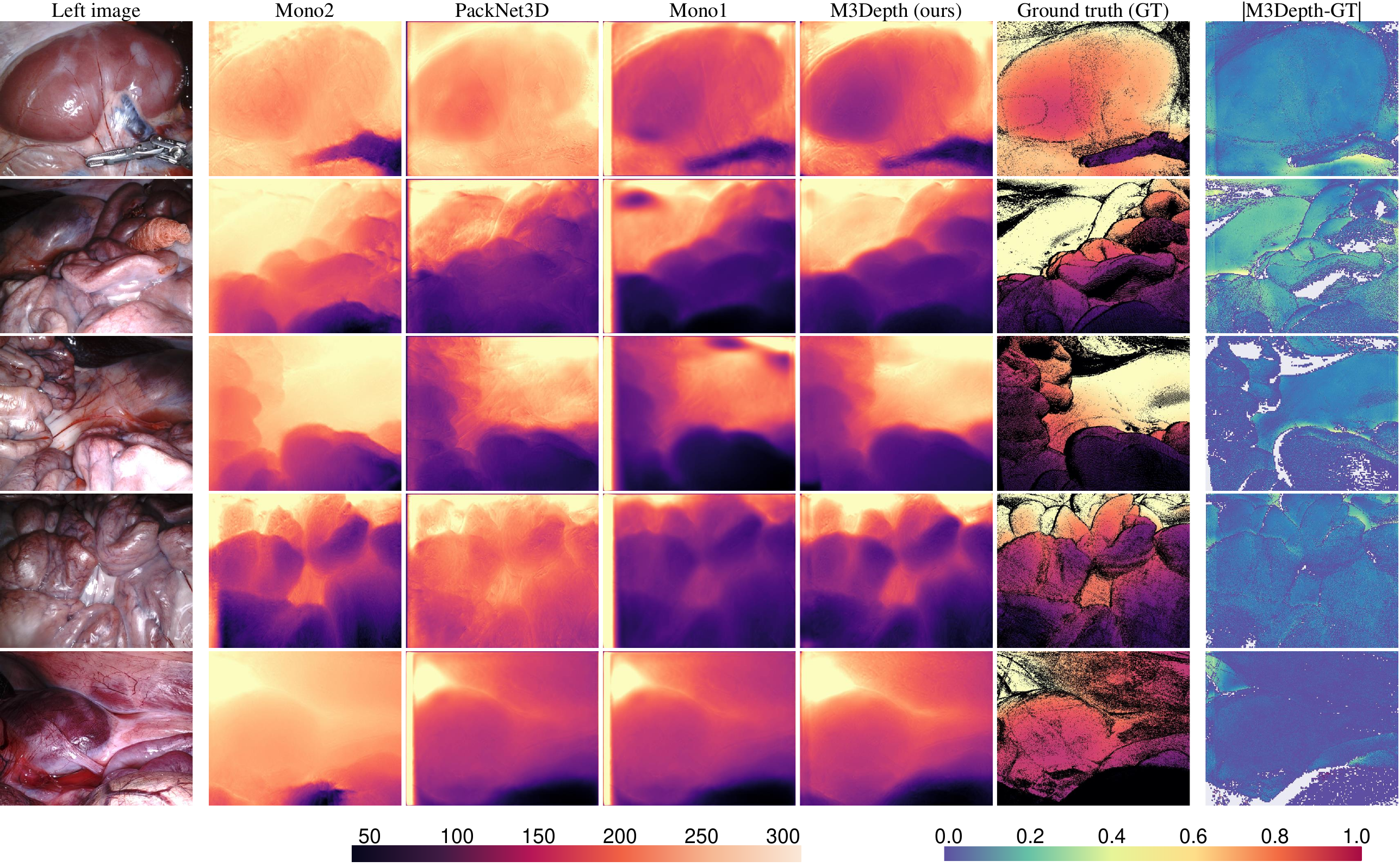}
  
  \caption{Qualitative results on the SCARED dataset with error map of M3Depth. The depth predictions are all for the left input image. M3Depth generated depth maps with high contrast between the foreground and background and performed better at distinguishing different parts of the scene, reflecting the superior quantitative results in Table \ref{quantitive results SCARED}.}
  \label{Fig:visualization}
\end{figure*}

\begin{table}[b]
\caption{Quantitative results on the SCARED dataset. 
Metrics labelled with blue headings mean \textit{lower is better} while those labelled with red mean \textit{higher is better}.
}
\resizebox{\textwidth}{!}{
\begin{tabular}{|c|c|c|c|c|c|c|c|c|}
\hline

\hline
\textbf{Method}   & \cellcolor[RGB]{193,218,243}\textbf{Abs Rel} & \cellcolor[RGB]{193,218,243}\textbf{Sq Rel} & \cellcolor[RGB]{193,218,243}\textbf{RMSE}  & \cellcolor[RGB]{193,218,243}\textbf{RMSE log} & \cellcolor[RGB]{249,196,196}${\delta<1.25 }$ &\cellcolor[RGB]{249,196,196} $\delta<1.25^2$  &\cellcolor[RGB]{249,196,196}${\delta<1.25^3}$ \\ 
\hline

Mono2 \cite{godard2019digging}  &1.100 &74.408 &56.548 &0.717 &0.102 &0.284 &0.476\\

PackNet3D \cite{guizilini20203d} &   0.733  &  37.690  &  32.579  &   0.649  &   0.288  &   0.538  &   0.722  \\

Mono1 \cite{godard2017unsupervised} &0.257  &20.649  &33.796  &0.404  &0.696  &0.837  &0.877 \\

\rowcolor[RGB]{230,230,230}
\textbf{M3Depth}  
&   \textbf{0.116}  &   \textbf{1.822}  &   \textbf{9.274}  &   \textbf{0.139}  &   \textbf{0.865}  &   \textbf{0.983}  &   \textbf{0.997} \\

\hline

\hline
\end{tabular}}
\label{quantitive results SCARED}
\end{table}


\begin{table}[b]
\caption{Quantitative results on LATTE  dataset. 
Metrics labelled with blue headings mean \textit{lower is better} while labelled by red mean \textit{higher is better}.
}
\resizebox{\textwidth}{!}{
\begin{tabular}{|c|c|c|c|c|c|c|c|c|}
\hline

\hline
\textbf{Method}   & \cellcolor[RGB]{193,218,243}\textbf{Abs Rel} & \cellcolor[RGB]{193,218,243}\textbf{Sq Rel} & \cellcolor[RGB]{193,218,243}\textbf{RMSE}  & \cellcolor[RGB]{193,218,243}\textbf{RMSE log} & \cellcolor[RGB]{249,196,196}${\delta<1.25 }$ &\cellcolor[RGB]{249,196,196} $\delta<1.25^2$  &\cellcolor[RGB]{249,196,196}${\delta<1.25^3}$ \\ 
\hline

Mono2 \cite{godard2019digging}     &   1.601  & 306.823  &  87.694  &   0.913  &   0.169  &   0.384  &   0.620   \\

PackNet3D \cite{guizilini20203d} &   0.960  & 357.023  & 259.627  &   0.669  &   0.135  &   0.383  &   0.624  \\

Mono1 \cite{godard2017unsupervised} 
&   0.389  &  57.513  &  99.020  &   0.424  &   0.268  &   0.709  &   0.934   \\

\rowcolor[RGB]{230,230,230}
\textbf{M3Depth}    &   \textbf{0.236}  &  \textbf{21.839}  &  \textbf{57.739}  &   \textbf{0.245}  &   \textbf{0.665}  &   \textbf{0.893}  &   \textbf{0.969} \\

\hline

\hline
\end{tabular}}
\label{quantitive results on LATTE}
\end{table}

\subsection{Evaluation Metrics, Baseline, and Implementation}
\textbf{Evaluation Metrics} To evaluate depth errors, seven criteria were adopted that are commonly used for monocular depth estimation tasks \cite{godard2017unsupervised}\cite{godard2019digging}: \acrfull{absrel}, \acrfull{sqrel}, \acrfull{rmse}, \acrfull{rmselog}, and the ratio between ground truth and prediction values, for which the threshold was denoted as ${\delta}$.

\textbf{Baseline}
The M3Depth model was compared with several recent deep learning methods including Monodepth \cite{godard2017unsupervised}, Monodepth2 \cite{godard2019digging}, and PackNet \cite{guizilini20203d}, and both quantitative and qualitative results were generated and reported for comparison. To further study the importance of each M3Depth component, the various components of M3Depth were removed in turn.

\textbf{Implementation Details}
M3Depth was implemented in PyTorch \cite{paszke2017automatic}, with an input/output resolution of \(256\times320\) and a batch size of 18. The learning rate was initially set to \(10^{-5}\) for the first 30 epochs and was then halved until the end. The model was trained for 50 epochs using the Adam optimiser which took about 65 hours on two NVIDIA 2080 Ti GPUs. 
\section{Results and Discussion}
\label{sec:5-Results}

\begin{table}[h]
\caption{Ablation study results on the SCARED dataset. 
Metrics labelled with blue headings mean \textit{lower is better} while those labelled with red mean \textit{higher is better}.
}
\resizebox{\textwidth}{!}{
\begin{tabular}{|c|c|c|c|c|c|c|c|c|c|c|}
\hline

\hline
\textbf{Method} &\textbf{\acrshort{3gc}} &\textbf{Blind masking}  & \cellcolor[RGB]{193,218,243}\textbf{Abs Rel} & \cellcolor[RGB]{193,218,243}\textbf{Sq Rel} & \cellcolor[RGB]{193,218,243}\textbf{RMSE}  & \cellcolor[RGB]{193,218,243}\textbf{RMSE log} & \cellcolor[RGB]{249,196,196}${\delta<1.25 }$ &\cellcolor[RGB]{249,196,196} $\delta<1.25^2$  &\cellcolor[RGB]{249,196,196}${\delta<1.25^3}$ \\ 
\hline

Mono1 \cite{godard2017unsupervised}  &\ding{55} &\ding{55}     &0.257  &20.649  &33.796  &0.404  &0.696  &0.837  &0.877 \\
M3Depth w/ mask &\checkmark &\ding{55}   &   0.150  &   3.069  &  13.671  &   0.249  &   0.754  &   0.910  &   0.956  \\
\rowcolor[RGB]{230,230,230}
\textbf{M3Depth}  &\checkmark &\checkmark 
&   \textbf{0.116}  &   \textbf{1.822}  &   \textbf{9.274}  &   \textbf{0.139}  &   \textbf{0.865}  &   \textbf{0.983}  &   \textbf{0.997} \\
\hline

\hline
\end{tabular}}
\label{ablation results on SCARED dataset}
\end{table}

The M3Depth and other results on the SCARED and LATTE datasets are shown in Table \ref{quantitive results SCARED} and Table~\ref{quantitive results on LATTE} using the seven criteria evaluation metrics. M3Depth outperformed all other methods by a large margin on all seven criteria, which shows that considering the 3D structure benefited the overall performance of the depth estimation task. In particular, the M3Depth model had 0.141, 18.827, 24.522, and 0.265 units error lower than Monodepth1~\cite{godard2017unsupervised} in \acrshort{absrel}, \acrshort{sqrel}, \acrshort{rmse}, and \acrshort{rmselog}, and 0.169, 0.146, and 0.12 units higher than Monodepth1 in three different threshold criteria. Furthermore, the average inference time of M3Depth was 105 frames per second, satisfying the real-time depth map generation requirements.

An ablation study is an important experimental technique used in machine learning and computer vision to understand the contribution of different components of a model to its overall performance. Detailed ablation study results on the SCARED dataset are shown in Table~\ref{ablation results on SCARED dataset} and the impact of the proposed modules, \acrfull{3gc} and blind masking were evaluated. The evaluation measures steadily improved when the various components were added. More specifically, the addition of the blind masking further boosted the \acrshort{3gc} term, which shows the importance and necessity of removing invalid information from areas that are not visible to both cameras.

Qualitative results comparing our depth estimation results against prior work using the SCARED dataset are presented in Fig \ref{Fig:visualization}.
As the sample images show, the application of temporal consistency encouraged by the 3D geometric consistency loss can reduce the errors caused by subsurface features, and better recover the real 3D surface shape of the tissue.  
\section{Conclusion}
\label{sec:5-conclusions}

A novel framework for self-supervised monocular laparoscopic image depth estimation was presented. By combining the 2D image-based losses and 3D geometry-based losses from an inferred 3D point cloud of the whole scene, the global consistency and small local neighbourhoods were both explicitly taken into consideration. The incorporation of blind masking avoided penalising areas where no useful information exists. The modules proposed can easily be plugged into any similar depth estimation network, monocular and stereo, while the use of a lightweight ResNet18 backbone enabled real time depth map generation in laparoscopic applications. Extensive experiments on both public and newly acquired datasets, together with the qualitative and quantitative results in Table \ref{quantitive results SCARED}, Table~\ref{quantitive results on LATTE} and Fig \ref{Fig:visualization}, demonstrated good generalisation across different laparoscopes, illumination conditions, and samples, indicating the capability to large scale data acquisition where precise ground truth depth cannot be easily collected.

\section{Chapter Transitions}

The two depth estimation tasks mentioned above are limited to individual depth estimation work. However, numerous studies indicate that concurrent execution of multiple tasks can assist each other, sharing weights and saving inference time. Therefore, we considered another task in the field of minimally invasive surgical vision—tool segmentation. As the appearance of surgical tools inevitably disrupts the original depth of the laparoscopic scene, the recovery of depth is closely related to outlining the contours of the surgical tools. In the subsequent work, we attempted to merge these two tasks, creating a joint-task learning approach. This approach aims to enhance the accuracy of both tasks in situations where a reasonable amount of extra time can be allocated. 


\setcounter{chapter}{5}

\chapter{Simultaneous Depth Estimation and Surgical Tool Segmentation in Laparoscopic Images}
\chaptermark{Simultaneous Depth Estimation and Surgical Tool Segmentation}
\glsresetall
\label{chap:6-contribution_4-DepthSeg}

\begin{cabstract}

Surgical instrument segmentation and depth estimation are crucial steps to improve autonomy in robotic surgery as they provide the working distance and segmented regions. Most recent works treat these problems separately, making the deployment challenging. In this work, we propose a unified framework for depth estimation and surgical tool segmentation in laparoscopic images. The network has an encoder-decoder architecture and comprises two branches for simultaneously performing depth estimation and segmentation. To train the network end to end, we propose a new multi-task loss function that effectively learns to estimate depth in an unsupervised manner, while requiring only semi-ground truth for surgical tool segmentation. We conducted extensive experiments on different datasets to validate these findings. The results showed that the end-to-end network successfully improved the state-of-the-art for both tasks while reducing the complexity during their deployment.

\end{cabstract}
This chapter’s research was previously published in the IEEE Transactions on Medical Robotics and Bionics (T-MRB) in 2022~\cite{huang2022simultaneous}.

\section{Introduction}
\label{sec:6-introduction}

\acrshort{mis}, including robot-assisted procedures, provides significant advantages such as reducing operative trauma and the risk of infection. Advanced robotic surgery systems such as the da Vinci surgical platform~\cite{zhang2018self} allow multiple types of information to be integrated with effective feedback to the surgeon. However, interpreting visual surgical data is complex and involves many tasks such as tissue deformation modeling~\cite{giannarou2016vision}, tool tracking~\cite{huang2020tracking}, and scene depth estimation~\cite{ye2017self,alexandridis2022long,huang2021self}.

In recent years there has been much work on depth estimation and surgical tool segmentation. Notably, learning-based algorithms have shown excellent prediction capability of the relationship between colour images and depth, as well as image segmentation into meaningful regions. These depth-predicting algorithms may use monocular or stereo input data, with either supervised, self-supervised, or unsupervised~\cite{hoyer2021three} training approaches depending on the availability of ground truth labels. Instrument segmentation may also use supervised or unsupervised methods~\cite{liu2020unsupervised}. Knowing the tissue depth and the instrument masks could facilitate tissue scanning~\cite{zhan2020autonomous} or dynamic image overlays~\cite{zevallos2018surgical}, which are useful for laparoscopic and endoscopic surgery.

To date, depth estimation and surgical tool segmentation have been mainly treated as separate challenges, requiring time-consuming sequential task completion \cite{nekrasov2019real}. In this work, we propose a novel unified framework that can perform simultaneous depth estimation and surgical tool segmentation: \textsc{SDSNet}. Our method does not require manually labelled ground truth, and achieves the state-of-the-art performance for both tasks, as well as reducing the deployment complexity.

\textbf{Depth Estimation} 
Most existing methods treat depth estimation as a supervised regression problem \cite{hu2019revisiting}, however, collecting per-pixel ground truth for laparoscopic imaging is challenging. To overcome this limitation, Liu \textit{et al.} \cite{liu2019dense} introduced a self-supervised algorithm for dense depth estimation in stereo endoscopy. The authors in \cite{vijayanarasimhan2017sfm} proposed a geometry-aware network for motion estimation. By enforcing consistency between left and right RGB images, Godard \textit{et al.} ~\cite{godard2017unsupervised} produced results that outperformed contemporary supervised methods.  

\textbf{Surgical Tool Segmentation}
Semantic segmentation of robotic instruments has also attracted a lot of attention in robot-assisted surgery research. Some discriminative models such as Naive Bayesian classifiers \cite{speidel2006tracking} and maximum likelihood Gaussian Mixture Models \cite{pezzementi2009articulated} can be trained on color features. More recently, the state of the art has increasingly focused on fully convolutional neural networks. Zhao \textit{et al.} \cite{islam2019real} proposed a dual motion-based method that enhanced segmentation by motion flows leveraging temporal dynamics.

\textbf{Simultaneous Depth Estimation and Segmentation}
The task of depth estimation and segmentation is usually tackled separately, with few works unifying both tasks, especially for laparoscopy. Most recent methods used RGB images as the training data, for instance in EdgeStereo \cite{song2019edgestereo} the authors incorporated edge detection to accurately estimate depth changes across object boundaries. In medical imaging, self-supervised depth estimation was used to regularise the semantic segmentation during knee arthroscopy \cite{liu2020self}.

\section{Methodology}
\label{sec:6-methods}

\begin{figure*}[t]
    \centering
    \includegraphics[width=\textwidth]{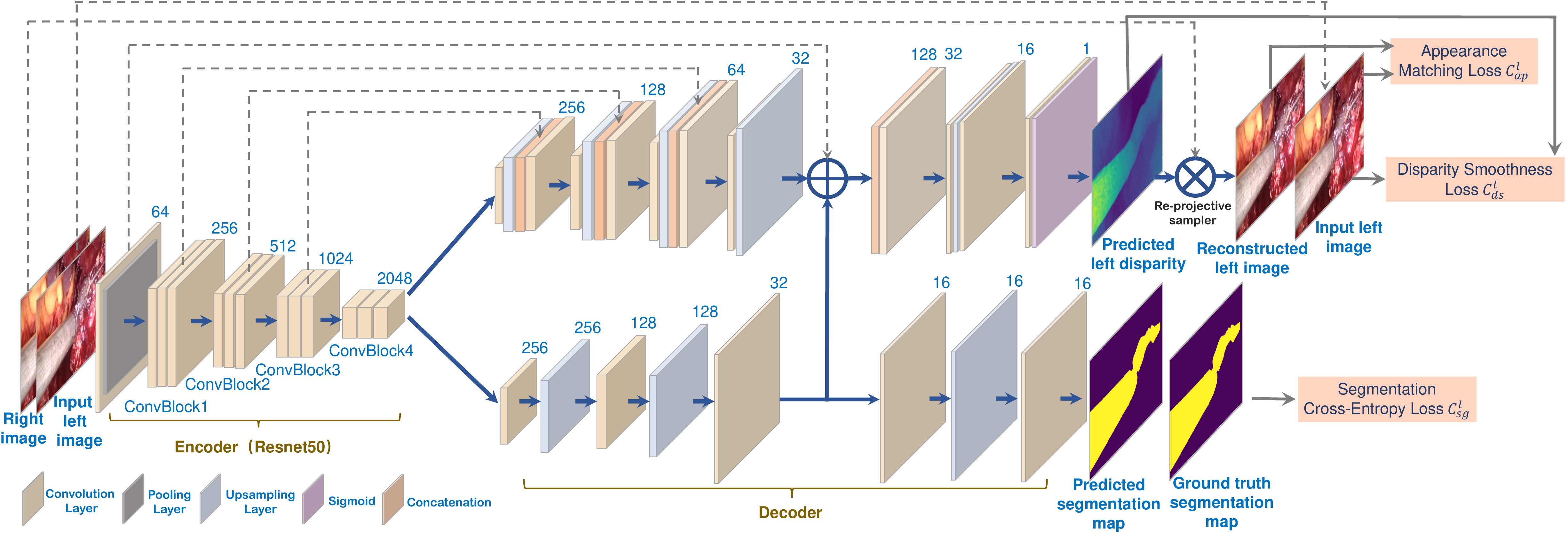} 
    \caption{The detailed architecture of \textsc{SDSNet}. The depth branch and the segmentation branch share the same encoder network. The features of the third convolutional layer in the segmentation branch decoder are fused with the features from the fourth block in the depth branch decoder. We use only one ConvBlock to represent all repeated blocks for better visualisation}.
    \label{Fig:6-network} 
\end{figure*}

An overview of our proposed \textsc{SDSNet} can be found in {Fig.~\ref{Fig:6-network}}. We first combined the depth estimation and tool segmentation tasks by sharing an encoder network, where essential geometric features from the input images were extracted. After the encoder, the features flowed separately to two branches (segmentation and depth estimation). By forcing the disparity map to generate a reconstructed input image that is consistent with the original, we could derive an accurate disparity map for depth inference.

\subsection{Depth Estimation Branch} 
The depth estimation branch was based on the general U-Net architecture \cite{ronneberger2015u}, \emph{i.e.} an encoder-decoder network with skip connections, which represents local information as well as deep abstract features. The size of the input batch was \(b \times 3 \times 192 \times 384\), where b was the batch size, 3 was the number of channels and \(192 \times 384\) was the size of the input image. A Resnet50 was adopted as the encoder to extract features from the input colour image. 

The decoder consisted of five cascade blocks of multiple scales. Previously, multi-scale depth predictions and image reconstruction used gradient locality of a bilinear sampler \cite{jaderberg2015spatial}, which was prone to create `holes' and texture-copy artifacts in large low-texture regions. In our work, similar to \cite{godard2017unsupervised}, this problem is tackled by decoupling the resolutions of the disparity maps and corresponding colour images used to compute the reprojection error. The lower resolution depth maps were first upsampled to the input image resolution and then reprojected and resampled. 
From the second block, the output of each block was taken by the convolutional layer and followed by a sigmoid activation function, which generated the disparity map at each scale. In total, 4 scales were used with output sizes \(b \times 1 \times 24 \times 48\), \(b \times 1 \times 48 \times 96\), \(b \times 1 \times 96 \times 192\), \(b \times 1 \times 192 \times 384\). The largest of these was the final disparity map which was the same size as the input image.

The final disparity map (the sigmoid output) \(\hat D\) was converted to a depth map by \(D = 1/(p\hat D + q)\), in which \(p\) and \(q\) constrained \(D\) between 0.1 and 80 units.
 
\subsection{Segmentation Branch}
The shared encoder features were also fed into the segmentation map decoder, which consisted of five convolutional layers and three upsampling layers to interpolate the features to full image resolution. The first layer was \(b \times 256 \times 6 \times 12\) and took the \(b \times 2048 \times 6 \times 12\) input followed by an ELU activation function. The upsampling layer interpolated the features to four times the input size. After the third convolutional layer, the features were concatenated into the corresponding layer from the depth estimation decoder to perform feature fusion between the two branches. The size of the segmentation subnetwork output was \(b \times k \times 192 \times 384\), where $k=2$ was the number of classes.  In practice, the depth decoder features were concatenated with the segmentation branch decoder in the fourth block. We generated the surgical instrument segmentation semi-ground truth by applying the network from \cite{shvets2018automatic} pretrained on the EndoVis dataset \cite{allan2021stereo}.

\subsection{Multi-Task Loss}
The network was trained end-to-end using a multi-task loss function \(\mathcal{C}_{\rm t}^{\rm l}\), which was formed as 
\begin{equation} \label{eq:mean}
\begin{split}
\mathcal{C}_{\rm t}^{\rm l} = \alpha_{\rm dp} \mathcal{C}_{\rm dp}^{\rm l} + \alpha_{\rm sg} \mathcal{C}_{\rm sg}^{\rm l}
\end{split}
\end{equation}
where \(C_{dp}^{\rm l}\) is the loss from the depth estimation branch and \(C_{sg}^{\rm l}\) is from semantic segmentation, as described below.

\textbf{Depth Loss}
In the depth estimation branch, the depth loss \(\mathcal{C}_{\rm dp}^{\rm l}\) consisted of the appearance matching loss \(\mathcal{C}_{\rm ap}^{\rm l}\) and disparity smoothness loss \(\mathcal{C}_{\rm ds}^{\rm l}\) as:
 \begin{equation} \label{eq:total_depth_loss}
\begin{split}
\mathcal{C}_{dp}^{\rm l} = \sum_{s=1}^4 \mathcal{C}_s^{\rm l} = \sum_{s=1}^4 (\mathcal{C}_{ap}^{\rm l} + \alpha_{ds} \mathcal{C}_{ds}^{\rm l})
\end{split}
\end{equation}
where $\alpha_{ds}$ was set to 0.001.

\emph{Appearance Matching Loss}
The appearance matching loss \(\mathcal{C}_{\rm ap}^{\rm l}\) forced the reconstructed image to be similar to the corresponding training input and was computed for the higher input resolution. During training, the autoencoder in the depth estimation branch generated a disparity map \(\hat D_{t}\) from the input left colour image \(I_{t}^l\). This map was then transformed using an image sampler from the \acrfull{stn} \cite{jaderberg2015spatial}, along with the right input image \(I_{t}^r\) (the counter-part of \(I_{t}^l\)), to reconstruct the left image \(I_{t}^{{\rm l}*}\). This sampler model used bilinear interpolation and the output pixel was the weighted sum of four input pixels. This bilinear sampler was locally fully differentiable and could be seamlessly integrated into the fully convolutional architecture, in contrast to \cite{garg2016unsupervised}. Hence, there was no need to simplify or approximate the cost function. As in \cite{zhao2015loss}, we applied a combination of \(L_1\) loss and structural similarity (SSIM) index as the photometric image reconstruction cost \(\mathcal{C}_{\rm ap}^{\rm l}\). Training the depth estimation network then required minimising the reconstruction loss between the reconstructed image \(I^{{\rm l}*}\) and the corresponding training input \(I^{\rm l}\), where \(N\) denotes the number of pixels. 


\begin{equation} \label{eq:appearance_matching_loss}
\mathcal{C}_{\rm ap}^{\rm l} = \frac{1}{N}\sum_{i, j} \frac{\gamma}{2} (1-{\rm SSIM}(I_{ij}^{\rm l}, I_{ij}^{{\rm l}*})) + (1-\gamma) {\|I_{ij}^{\rm l}-I_{ij}^{{\rm l}*}\|}_1
\end{equation}

Similar to \cite{godard2017unsupervised}, the \acrshort{ssim} was simplified to a \(3 \times 3\) block filter rather than a Gaussian, and \(\gamma\) was set to 0.85.

\emph{Disparity Smoothness Loss}
Smooth disparities were favoured by this loss, and since discontinuities usually occur at image gradients \cite{godard2017unsupervised}, this cost was weighted by an edge-aware term based on the image gradients \(\partial I\).
\begin{equation} \label{eq:disparity_smoothness_loss}
\begin{split}
\mathcal{C}_{\rm ds}^{\rm l} = \frac{1}{N}\sum_{ij}\mid \partial_x \hat D_{ij}^{\rm l}\mid e^{-\mid \partial_x I_{ij}^{\rm l}\mid} + \mid \partial_y \hat D_{ij}^{\rm l} \mid e^{-\mid \partial_y I_{ij}^{\rm l}\mid}
\end{split}
\end{equation}

\textbf{Segmentation Loss} The segmentation branch only considers the full resolution image to reduce the computational complexity. 
For the segmentation subnetwork, given a sequence of input images and the corresponding sequence of semi-ground truth segmentation annotations, we performed end-to-end training by minimising the normalised pixel-wise cross-entropy loss \cite{long2015fully}, which is denoted as \(\mathcal{C}_{\rm sg}^{\rm l}\).
\begin{equation} \label{eq:segmentation_loss}
\begin{split}
\mathcal{C}_{\rm sg}^{\rm l} = -\sum_{i=1}^N \hat{y_i}*\log(y_i)
\end{split}
\end{equation}
where \(y_i\), \(\hat{y_i}\) are the predicted value, and semi-ground truth.

\subsection{Training}
As there was no per-pixel depth ground truth label available, the depth estimation relied on the image reconstruction similarity, trained in self-supervised mode. For depth estimation, the data augmentation was performed by flipping \(50\%\) the input images horizontally. For segmentation, the semi-ground truth was provided for supervised training. 
The whole \textsc{SDSNet} was trained end-to-end with the combination of losses from each branch that involved the generation of a depth map and segmentation map.

\section{Experiments}
\label{sec:6-Experiments}

\textbf{Experimental Setup} We evaluated our \textsc{SDSNet} on two datasets. The first one was the \( \mathcal{D} \)\(^{sia}\) dataset from \cite{ye2017self}, which has 34320 pairs of rectified stereo images for training and 14382 for testing, collected during da Vinci partial nephrectomy. For the depth estimation branch, similar to~\cite{godard2019digging}, we used the SSIM index to evaluate the unsupervised depth estimation. To evaluate the result of the segmentation branch, we manually labelled $400$ images with the surgical tool ground truth, and the segmentation performance was assessed by the Jaccard index and the Dice Score~\cite{ronneberger2015u}.  
 Apart from the \( \mathcal{D} \)\(^{sia}\) dataset, we also used the \( \mathcal{D} \)\(^{por}\) dataset~\cite{mountney2010three} to validate the generalisation of the network.

\textbf{Baseline} For depth estimation, we compared the results to those from the Basic and Siamese architectures \cite{ye2017self}, Monodepth2 \cite{godard2019digging}, and two non-learning methods, SPS~\cite{yamaguchi2014efficient} and ELAS~\cite{geiger2010efficient}. For surgical instrument segmentation, results from the \textsc{SDSNet} were compared with the popular U-Net \cite{ronneberger2015u} architecture and the dual motion-based method proposed by Zhao \textit{et al.} \cite{zhao2020learning}.

\textbf{Implementation} The \textsc{SDSNet} model was implemented in PyTorch \cite{paszke2017automatic}, with a batch size of 16 and an input/output resolution of 192 x 384. The learning rate was set to \(10^{-4}\) for the first 15 epochs and dropped to \(10^{-5}\) for the remainder. The hyperparameters $\alpha_{dp}$ and $\alpha_{sg}$ in Equation (\ref{eq:mean}) were empirically set to $10$ and $1$, respectively. The network was trained for 20 epochs using Adam optimiser~\cite{kingma2014adam} and the training took about 8 hours on a single NVIDIA 2080 Ti GPU. 
 
\begin{figure*}[]
    \centering
    \includegraphics[width=\textwidth]{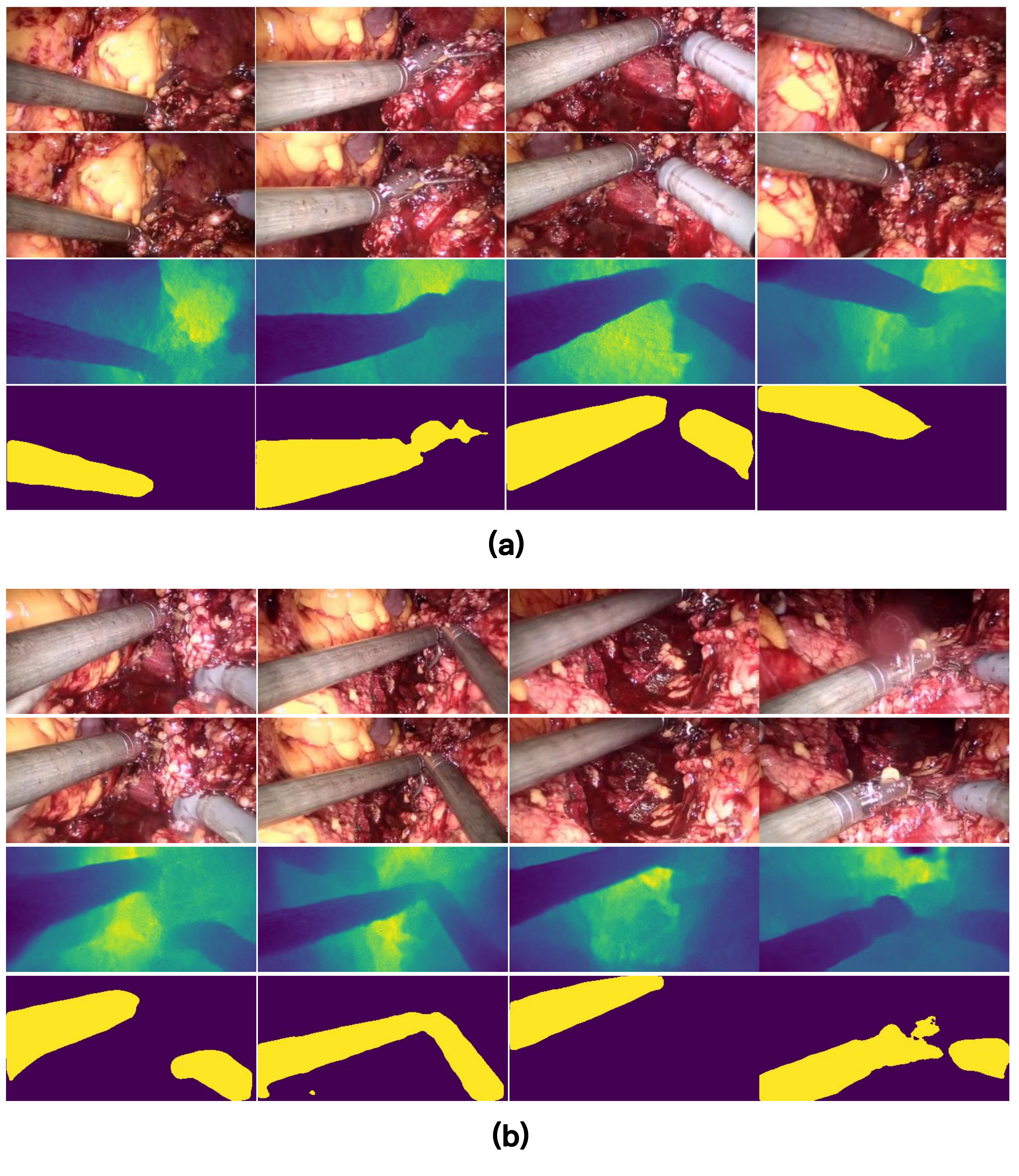} 
    \caption{Qualitative results on stereo pairs. (a) and (b) from top to bottom: the left input image, the corresponding right image in the stereo pair, the predicted depth map, and the segmentation result from \textsc{SDSNet}.}
    \label{visualisation} 
\end{figure*}

\section{Results}
\label{sec:6-Results}

\begin{table}[]
\centering
\caption{SSIM Scores on the \( \mathcal{D} \)\(^{sia}\) Test Set}
\label{SSIM_dataset1}
\renewcommand\tabcolsep{10pt}
\begin{tabular}{@{}lccc@{}}
\toprule
&\textbf{Mean SSIM} & \textbf{Std.SSIM}\\
\midrule
ELAS~\cite{geiger2010efficient} & 47.3  &0.079  \\
SPS~\cite{yamaguchi2014efficient} & 54.7 &0.092 \\
V-Basic~\cite{ye2017self} & 55.5  &0.106  \\
V-Siamese~\cite{ye2017self} & 60.4  &\textbf{0.066} \\
Monodepth~\cite{godard2017unsupervised} &58.4  & 0.114 \\
Monodepth2~\cite{godard2019digging} &71.2  & 0.075  \\
\hline

\textsc{SDSNet} without fusion (ours)   &71.9  &0.079 \\
\rowcolor[RGB]{230,230,230}\textbf{\textsc{SDSNet} with fusion (ours)}   & \textbf{72.8}  &0.073 \\
\bottomrule
\end{tabular}
\end{table}


\begin{table}[]
\centering
\caption{Segmentation Results on the \( \mathcal{D} \)\(^{sia}\) Test Set}
\label{seg_dataset1}
\renewcommand\tabcolsep{9pt}

\begin{tabular}{@{}lccc@{}}
\toprule
&\textbf{IoU}   &\textbf{Dice} \\
\midrule
U-Net~\cite{ronneberger2015u} & 71.16 & 80.90 \\

Zhao \textit{et al.} \cite{zhao2020learning} &72.70  &82.70\\

\hline
\textsc{SDSNet} (segmentation only) (ours) & 73.34 & 84.13 \\

\textsc{SDSNet} without fusion (ours)  & 73.44  & 84.59 \\
\rowcolor[RGB]{230,230,230}\textbf{\textsc{SDSNet} with fusion (ours)} & \textbf{74.92} & \textbf{85.63}\\
\bottomrule

\end{tabular}
\end{table}

\begin{table}[]
\centering
\caption{SSIM Scores on the \( \mathcal{D} \)\(^{por}\) Test Set}
\label{SSIM_dataset2}
\renewcommand\tabcolsep{10pt}
\begin{tabular}{@{}lccc@{}}
\toprule
&\textbf{Mean SSIM}   &\textbf{Std.SSIM}\\
\midrule
Monodepth2~\cite{godard2019digging} & 76.67 & 0.047\\
\rowcolor[RGB]{230,230,230}\textbf{\textsc{SDSNet} with fusion (ours)} &\textbf{77.53} & \textbf{0.041}\\
\bottomrule
\end{tabular}
\end{table}

Table \ref{SSIM_dataset1} summaries the \textsc{SDSNet} results as well as other depth estimation methods, using the mean and \acrshort{std} of the \acrshort{ssim} index. The \textsc{SDSNet} outperforms the other methods. More specifically, it is \(1.6\%\) higher than Monodepth2~\cite{godard2019digging} and \(12.4\%\) higher than the Siamese architecture~\cite{ye2017self}. This is a significant improvement and interestingly, from Table \ref{SSIM_dataset1} we can also see that we achieve the best result when both the depth estimation branch and the segmentation branch were fused together, with the added benefit of surgical instrument segmentation included.

Table \ref{seg_dataset1} summaries the segmentation results using the \acrfull{iou} and \acrshort{dice} index. It can be seen that \textsc{SDSNet} produces superior segmentation results \(5.28\%\) higher than U-Net \cite{ronneberger2015u} for IoU and \(5.85\%\) for Dice index while \(3.05\%\) higher than Zhao \textit{et al.} \cite{zhao2020learning} for IoU and \(3.54\%\) for Dice index. To evaluate if the \textsc{SDSNet} works well on video processing, we calculated the inference time of the \textsc{SDSNet} with the batch size set as 1. The inference time of the \textsc{SDSNet} was 0.0058s per frame and 172 frames per second, which satisfied the requirements of real-time video processing. Table~\ref{seg_dataset1} also confirms that the use of a fusion operation when performing depth estimation and segmentation simultaneously can improve the segmentation result. Example qualitative results are presented in Fig~\ref{visualisation}, showing that \textsc{SDSNet} provides consistent depth estimation and accurate segmentation simultaneously.

\textbf{Generalisation}
To validate the generalisation of our network, an additional experiment used the model trained on the \( \mathcal{D} \)\(^{sia}\) dataset but tested directly on the \( \mathcal{D} \)\(^{por}\) dataset, without retraining the whole network. Table~\ref{SSIM_dataset2} represents the results of \textsc{SDSNet} and Monodepth2 in this experiment using the SSIM index. Overall, \textsc{SDSNet} with fusion from both segmentation and depth estimation branch achieved a higher SSIM index, confirming that the \textsc{SDSNet} generalises well across different datasets, while still achieving competitive performance compared to the recent state-of-the-art methods.

\section{Conclusions}
\label{sec:6-conclusions}

\textsc{SDSNet} is presented in this work, a joint learning network that can simultaneously segment surgical tools and estimate the depth for each pixel. The proposed fusion network achieved state-of-the-art performance in both tasks. Besides, the framework does not require any depth labels and segmentation ground truth, and thus allows superior applicability on large-scale \emph{in vivo} video processing where ground truth for per-pixel depth maps and manual segmentation labels are not easy to obtain.

\section{Chapter Transitions}

In the aforementioned chapters, when speculating the intersection points of the probe and tissue -- the detection points of the probe, we take into consideration various prior information. This includes the 6D pose estimation of the probe, depth information of the surgical scene, and the segmented contours of surgical tools. However, these pieces of information either necessitate the introduction of additional hardware devices, such as the pose ground truth acquirement for pose estimation or rely on accurate depth map information for evaluation. Both approaches introduce uncertainties and difficulties in predicting the intersection points. Consequently, we contemplated whether it is possible to deduce the position of the intersection point -- sensing area where the probe is pointed to solely from a single laparoscopic RGB image, without introducing any additional depth information or prior knowledge about the probe's pose. In the subsequent phase of our work, we made alterations to the gamma probe to more intuitively indicate the position of the detection point. Our neural network was also enhanced into an end-to-end intersection point prediction network, resulting in a further improvement in inference speed. 

\setcounter{chapter}{6}
\chapter{Detecting the Sensing Area of A Laparoscopic Probe in Minimally Invasive Cancer Surgery}
\chaptermark{Detecting Laparoscopic Sensing Area}
\glsresetall
\label{chap:7-contribution_-5PtRegress}

\begin{cabstract}
Endoscopic radio-guided cancer detection and resection can enhance endoscopic imaging and complement preoperative nuclear imaging data. However, currently the gamma signal is recorded as a number indicated on the control unit, therefore, it is challenging to know where the radioactive source originates on the tissue surface. To tackle this challenge, initial failed attempts used segmentation or geometric methods, but led to the discovery that it could be resolved by leveraging high-dimensional image features and probe position information. To demonstrate the effectiveness of this solution, we designed and implemented a simple regression network that successfully addressed the problem. To further validate the proposed solution, we acquired and publicly released two datasets captured using a custom-designed, portable stereo laparoscope system. Through intensive experimentation, we demonstrated that our method can successfully and effectively detect the sensing area, establishing a new performance benchmark. 
Code and data are available at \href{https://github.com/br0202/Sensing_area_detection.git}{this link}.

\end{cabstract}
This chapter’s research was previously published in MICCAI 2023~\cite{huang2022simultaneous}. 

\section{Introduction}


\begin{figure*}[h]
\centering
\subfloat[]
{\includegraphics[width=0.49\linewidth, height=0.38\linewidth]{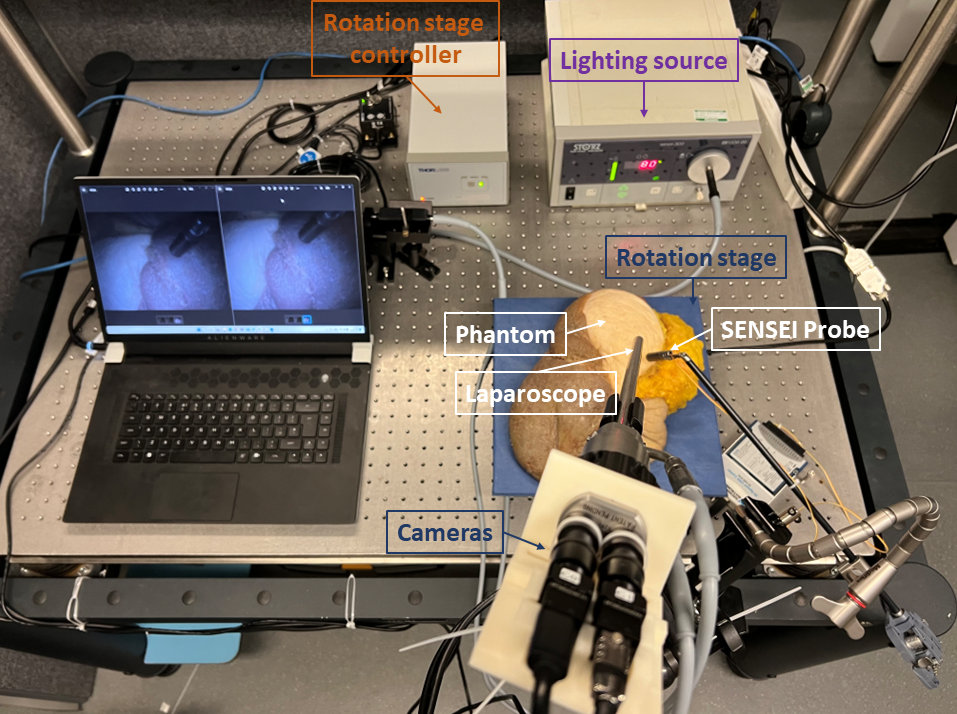}}
\hspace{1pt}
\subfloat[]
{\includegraphics[width=0.49\linewidth, height=0.38\linewidth]{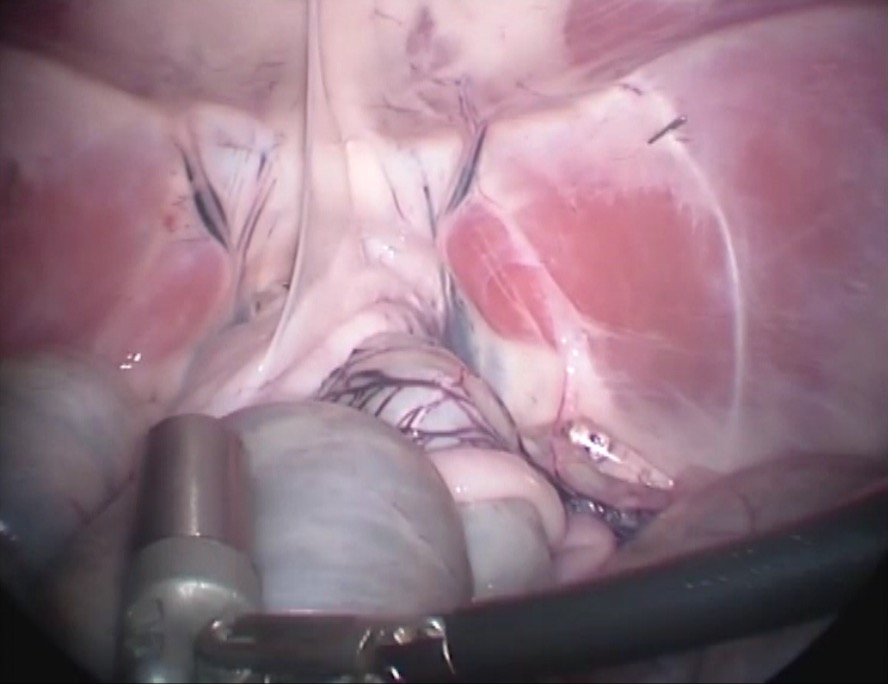}}
\caption{(a) Hardware set-up for experiments, including a customised portable stereo laparoscope system and the \acrshort{sensei} probe, a rotation stage, a laparoscopic lighting source, and a phantom; (b) An example of the use of the \acrshort{sensei} probe in \acrshort{mis}.}
\label{chaper7_fig:intro}
\end{figure*}

The cancer detection probe \acrshort{sensei} developed by Lightpoint Medical Ltd. leverages the cancer-targeting ability of nuclear agents typically used in nuclear imaging to more accurately identify cancer intra-operatively from the emitted gamma signal while presenting the visualisation challenge as the probe is non-imaging and is air-gapped from the tissue, making it challenging for the surgeon to locate the probe-sensing area on the tissue surface. Geometrically, the sensing area is defined as the intersection point between the gamma probe axis and the tissue surface in 3D space but projected onto the 2D laparoscopic image. However, it is not trivial to determine this using traditional methods due to poor textural definition of tissues and lack of per-pixel ground truth depth data. Similarly, it is also challenging to acquire the probe pose during the surgery. 

\textbf{Problem redefinition.} In this study, in order to provide sensing area visualisation ground truth, we modified a non-functional \acrshort{sensei} probe by adding a miniaturised laser module to clearly optically indicate the sensing area on the laparoscopic images - i.e. the `probe axis-surface intersection'. Our system consists of four main components: a customised stereo laparoscope system for capturing stereo images, a rotation stage for automatic phantom movement, a shutter for illumination control, and a DAQ-controlled switchable laser module (see Fig.~\ref{chaper7_fig:intro}). 
With this setup, we aim to transform the sensing area localisation problem from a geometrical issue to a high-level content inference problem in 2D. It is noteworthy that this remains a challenging task, as ultimately in the real surgery we need to infer the probe axis-surface intersection without the aid of the laser module to realistically simulate the use of the \acrshort{sensei} probe.

\section{Related Work}

Laparoscopic images play an important role in computer-assisted surgery and have been used in several problems such as object detection~\cite{jo2019robust}, image segmentation~\cite{yoon2022surgical}, depth estimation ~\cite{tukra2022stereo} or 3D reconstruction~\cite{liu2022sage}. Recently, supervised or unsupervised depth estimation methods have been introduced~\cite{liu2019dense}. Ye
\textit{et al.}~\cite{ye2017self} proposed a deep learning framework for surgical scene depth estimation
in self-supervised mode and achieved scalable data acquisition by incorporating a differentiable spatial transformer and an autoencoder into their framework. A 3D displacement module was explored in~\cite{xu2022self} and 3D geometric consistency was utilised in~\cite{huang2022self} for self-supervised monocular depth estimation. Tao~\textit{et al.} \cite{tao2023svt} presented a spatiotemporal vision transformer-based method and a self-supervised generative adversarial network was introduced in~\cite{huang2021self} for depth estimation of stereo laparoscopic images. Recently, fully supervised methods were summarised in~\cite{allan2021stereo} for depth estimation. 
However, acquiring per-pixel ground truth depth data is challenging, especially for laparoscopic images, which makes it difficult for large-scale supervised training~\cite{huang2022self}. 

Laparoscopic segmentation is another important task in computer-assisted surgery as it allows for accurate and efficient identification of instrument position, anatomical structures, and pathological tissue. For instance, a unified framework for depth estimation and surgical tool segmentation in laparoscopic images was proposed in~\cite{huang2022simultaneous}, with simultaneous depth estimation and segmentation map generation. In~\cite{liu2020self}, self-supervised depth estimation was utilised to regularise the semantic segmentation in knee arthroscopy. Marullo~\textit{et al.}~\cite{marullo2023multi} introduced a multi-task convolutional neural network for event detection and semantic segmentation in laparoscopic surgery. The dual swin transformer U-Net was proposed in~\cite{lin2022ds} to enhance the medical image segmentation performance, which leveraged the hierarchical swin transformer into both the encoder and the decoder of the standard U-shaped architecture, benefiting from the self-attention computation in swin transformer as well as the dual-scale encoding design. 

Although the intermediate depth information was not our final aim and can be bypassed, the 3D surface information was necessary in the intersection point inference. \acrshort{resnet}~\cite{he2016deep} has been commonly used as the encoder to extract the image features and geometric information of the scene. In particular, in~\cite{xu2022self}, concatenated stereo image pairs were used as inputs to achieve better results, and such stereo image types are also typical in robot-assisted minimally invasive surgery with stereo laparoscopes. Hence, stereo image data were also adopted in this paper.

If the problem of inferring the intersection point is treated as a geometric problem, both data collection and intra-operative registration would be difficult, which inspired us to approach this problem differently. In practice, we utilise the laser module to collect the ground truth of the intersection points when the laser is on. We note that the standard illumination image from the laparoscopic probe is also captured with the same setup when the laser module is on. Therefore, we can establish a dataset with an image pair (RGB image and laser image) that shares the same intersection point ground truth from the laser image (see Fig.~\ref{fig:dataset}a and Fig.~\ref{fig:dataset}b).
The assumptions made are that the probe's 3D pose when projected into the two 2D images is the observed 2D pose, and that the intersection point is located on its axis. Hence, we input these axes to the network as another branch and randomly sampled points along them to represent the probe.

\section{Dataset}
\begin{figure*}[h]
\centering
\includegraphics[width=0.99\linewidth]{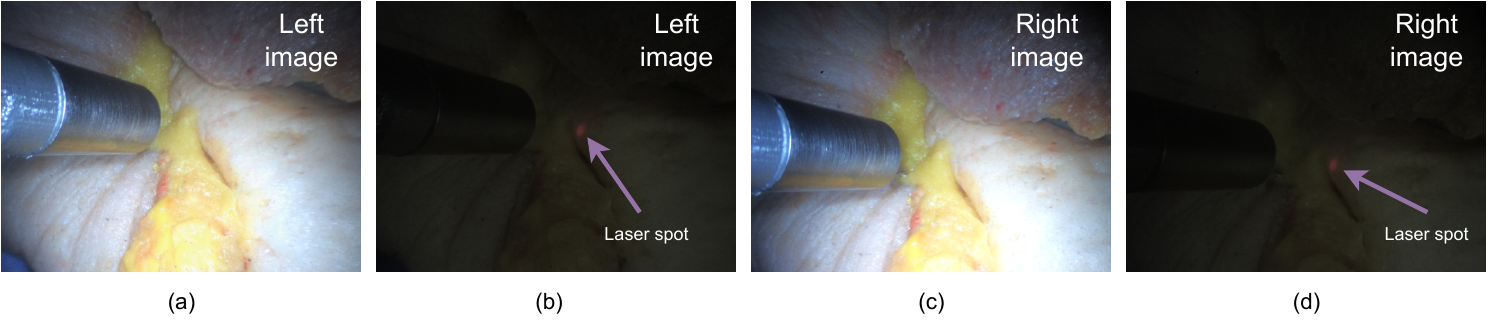}
\caption{Example data. (a) Standard illumination left RGB image; (b) left image with laser on and laparoscopic light off; same for (c) and (d) but for right images.}
\label{fig:dataset}
\end{figure*}

To validate our proposed solution for the newly formulated problem, we acquired and publicly released two new datasets. In this section, we introduce the hardware and software design that was used to achieve our final goal, while Fig.~\ref{fig:dataset} shows a sample from our dataset.

\textbf{Data Collection.}
Two miniaturised, high-resolution cameras were coupled onto a stereo laparoscope using a custom-designed connector. The accompanying API allowed for automatic image acquisition, exposure time adjustment, and white balancing. An electrically controllable shutter was incorporated into the standard laparoscopic illumination path. To indicate the probe axis-surface intersection, we incorporated a DAQ controlled cylindrical miniature laser module into a \acrshort{sensei} probe shell so that the adapted tool was visually identical to the real probe. The laser module emitted a red laser beam (wavelength 650 \textit{nm}) that was visible as a red spot on the tissue surface. This portable setup (illustrated in Fig.~\ref{chaper7_fig:intro}) allowed us to acquire the data in the lab, which can be easily extended to a standard laparoscopic system.

We acquired the dataset on a silicone tissue phantom which was \(30 \times 21 \times 8 \) \textit{cm} and was rendered with tissue colour manually by hand to be visually realistic. Hence, it will be easy to extend the experiments to \textit{in-vivo} data and conduct further experiments in the hospital for clinical trial validation. The phantom was placed on a rotation stage that stepped $10$ times per revolution to provide views separated by a 36-degree angle. At each position, stereo RGB images were captured \textit{i)} under normal laparoscopic illumination with the laser off; \textit{ii)} with the laparoscopic light blocked and the laser on; and \textit{iii)} with the laparoscopic light blocked and the laser off. Subtraction of the images with laser on and off readily allowed segmentation of the laser area and calculation of its central point, i.e. the ground truth probe axis-surface intersection.

\begin{figure*}[h]
\centering
\includegraphics[width=0.99\linewidth]{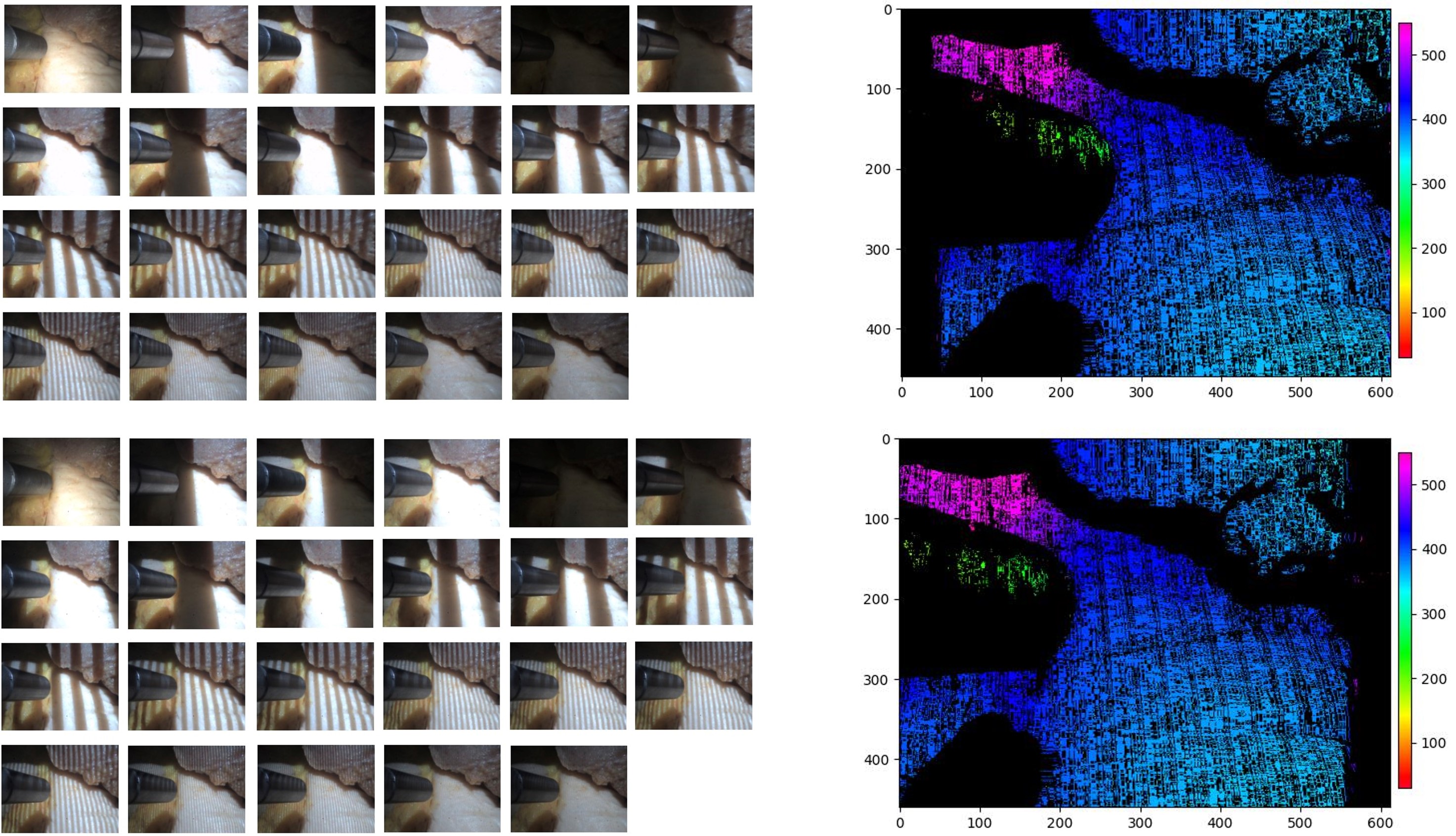}
\caption{The structure lighting patterns and example depth maps. The top row shows the left camera. The bottom row shows the right camera.}
\label{fig:sl+depmap}
\end{figure*}

All data acquisition and devices were controlled by Python and LABVIEW programs, and complete data sets of the above images were collected on visually realistic phantoms for multiple probe and laparoscope positions. This provided 10 tissue surface profiles for a specific camera-probe pose, repeated for 120 different camera-probe poses, mimicking how the probe may be used in practice. Therefore, our first newly acquired dataset, namely Jerry, contains 1200 sets of images. Since it is important to report errors in 3D and in millimetres, we recorded another dataset similar to Jerry but also including ground truth depth map for all frames by using structured-lighting system~\cite{huang2022self} --- namely the Coffbee dataset. Fig.~\ref{fig:sl+depmap} shows the structure lighting patterns and example depth maps in our dataset.

Our datasets have multiple uses such as:
\begin{itemize}
    \item Intersection point detection: detecting intersection points is an important problem that can bring accurate surgical cancer visualisation. We believe this is an under-investigated problem in surgical vision.
    \item Depth estimation and surgical tool segmentation. 
\end{itemize}

\section{Probe Axis-Surface Intersection Detection}

\subsection{Overview}

The problem of detecting the intersection point is trivial when the laser is on and can be solved by training a deep segmentation network. However, segmentation requires images with a laser spot as input, while the real gamma probe produces no visible mark and therefore this approach produces inferior results.

An alternative approach to detect the intersection point is to reconstruct the 3D tissue surface and estimate the pose of the probe in real time. The tracking and pose estimation method for the gamma probe in Chapter \ref{chap:3-contribution_1} involved attaching a dual-pattern marker to the probe to improve detection accuracy. This enabled the derivation of a 6D pose, comprising a rotation matrix and translation matrix with respect to the laparoscope camera coordinate. To obtain the intersection point, the authors used \acrshort{sfm} to compute the 3D tissue surface, combining it with the estimated pose of the probe, all within the laparoscope coordinate system. However, marker-based tracking and pose estimation methods have sterilisation implications for the instrument, and the \acrshort{sfm} method requires the surgeon to constantly move the laparoscope, reducing the practicality of these methods for surgery.

In this work, we propose a simple, yet effective regression approach to address this problem. Our approach relies solely on the 2D information and works well without the need for the laser module after training. Furthermore, this simple methodology facilitated an average inference time of $50$ frames per second, enabling real-time sensing area map generation for intraoperative surgery. 

\subsection{Intersection Detection as Segmentation} 
We utilised different deep segmentation networks (Table \ref{chapter7_seg}) as a first attempt to address our problem. The central point of the segmented area was calculated from the segmentation map as the intersection point. The problem of the segmentation solution is the inferior success rate. We observed that when we do not use images with the laser, the network was not able to make any good predictions. Only 66\% and 78\% frames on the test dataset had a valid segmented area on the segmentation map and the rest were labelled entirely as background. This is understandable as the red laser spot provides the key information for the segmentation. Therefore the network does not have any visual information to make predictions from images of the gamma probe. We note that to enable real-world applications, we need to estimate the intersection point using the images when the laser module is turned off.

\begin{table}[!h]
\caption{Results using segmentation solutions. For the invalid segmentation maps (labelled entirely as background), there was no central point of the segmented area, and the error was `NAN'. Hence, the accuracy was only calculated for valid segmentation maps.}
\centering
\resizebox{0.82\columnwidth}{!}{%
\begin{tabular}{|c|c|ccc|}
\hline
\small
\multirow{2}{*}{Methods} & \multirow{2}{*}{Failure rate} & \multicolumn{3}{c|}{Accuracy on valid frames (pixels)}                                   \\ \cline{3-5} 
                         &                               & \multicolumn{1}{c|}{Mean error} & \multicolumn{1}{c|}{Std.} & Median \\ \hline
Swin-unet~\cite{cao2023swin}   & 34\%                          & \multicolumn{1}{c|}{41.1}          & \multicolumn{1}{c|}{47.0}    & 27.8      \\ \hline
Transunet~\cite{chen2021transunet} & 22\%                          & \multicolumn{1}{c|}{34.5}          & \multicolumn{1}{c|}{40.6}    & 22.2      \\ \hline
\end{tabular}}
\vspace{3ex}

\label{chapter7_seg}
\end{table}

\subsection{Intersection Detection as Regression}

\begin{figure*}[t]
\centering
\includegraphics[width=0.99\linewidth]{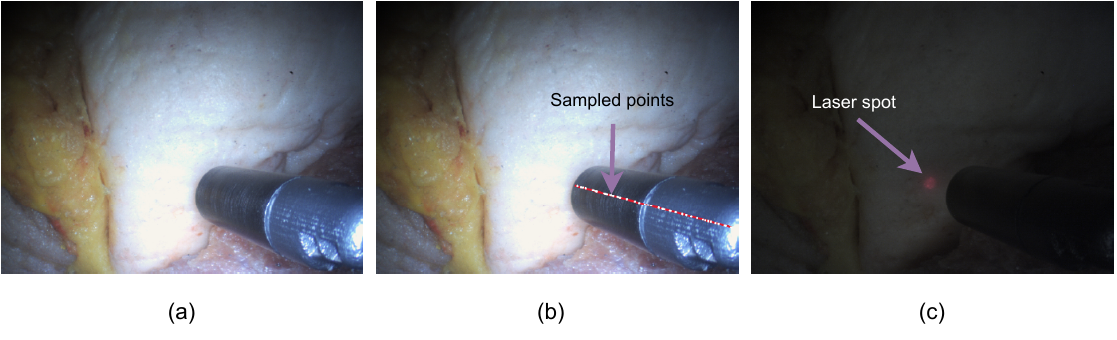}
\caption{Sensing area detection. (a) The input RGB image, (b) The estimated line using PCA for obtaining principal points, (c) The image with laser on that we used to detect the intersection ground truth.}
\label{fig_problem_define}
\end{figure*}

\begin{figure*}[t]
\centering
\includegraphics[width=0.99\linewidth]{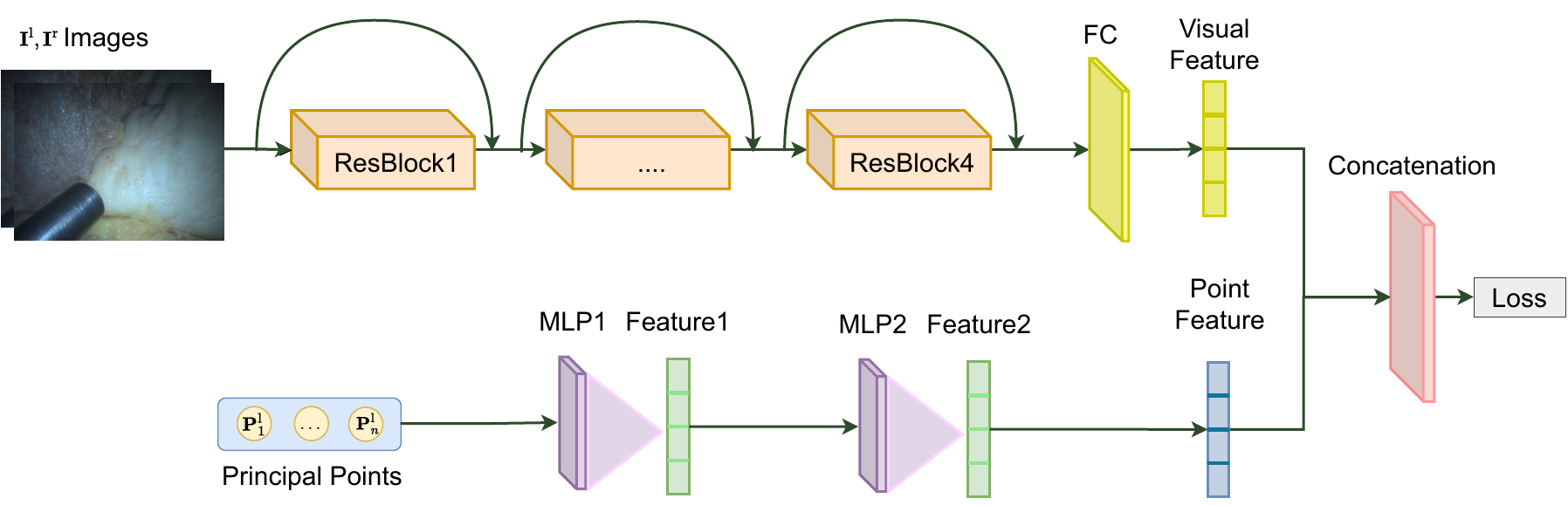}
\caption{An overview of our approach using \acrshort{resnet} and \acrshort{mlp}. 
}
\label{fig_network}
\end{figure*}

\textbf{Problem Formulation.} Formally, given a pair of stereo images $\mathbf{I}^{\rm l},\mathbf{I}^{\rm r}$,  $n$ points $\{\mathbf{P}_1^{\rm l}, \mathbf{P}_2^{\rm l}, ...,\mathbf{P}_n^{\rm l}\}$ were sampled along the principal axis of the probe, $\mathbf{P}_i^{\rm l} \in {R}^{2}$ from the left image. The same process was repeated for the right image. The goal was to predict the intersection point $\mathbf{P}_{\rm intersect}$ on the surface of the tissue. During the training, the ground truth intersection point position was provided by the laser source, while during testing the intersection was estimated solely based on visual information without laser guidance (see Fig.~\ref{fig_problem_define}).

\textbf{Network Architecture.} Unlike the segmentation approach, the intersection point was directly predicted using a regression network. The images fed to the network were `laser off' stereo RGB, but crucially, the intersection point for these images was known \textit{a priori} from the paired `laser on' images. The raw image resolution was $4896\times3680$ but these were binned to $896\times896$. \acrfull{pca}~\cite{mackiewicz1993principal} was used to extract the central axis of the probe and $50$ points were sampled along this axis as an extra input dimension. A network was designed with two branches, one branch for extracting visual features from the image and one branch for learning the features from the sequence of principal points using \acrshort{resnet}~\cite{he2016deep} and \acrfull{vit}~\cite{dosovitskiy2020image} as two backbones. The principal points were learned through a \acrfull{mlp} or a \acrfull{lstm} network~\cite{hochreiter1997long}. The features from both branches were concatenated and used for regressing the intersection point (see Fig.~\ref{fig_network}). Finally, the whole network is trained end-to-end using the mean square error loss.

\subsection{Implementation}
\textbf{Evaluation Metrics.} To evaluate sensing area location errors, Euclidean distance was adopted to measure the error between the predicted intersection points and the ground truth laser points. We reported the mean absolute error, the standard derivation, and the median in pixel units.

\textbf{Implementation Details.}
The networks were implemented in PyTorch \cite{paszke2017automatic}, with an input resolution of \(896\times896\) and a batch size of $12$. We partitioned the Jerry dataset into three subsets, the training, validation, and test set, consisting of 800, 200, and 200 images, respectively, and the same for the Coffbee dataset. The learning rate was set to \(10^{-5}\) for the first $300$ epochs, then halved until epoch $400$, and quartered until the end of the training. The model was trained for $700$ epochs using the Adam optimiser on two NVIDIA 2080 Ti GPUs, taking approximately $4$ hours to complete. 

\begin{figure*}[t]
\centering
\includegraphics[width=0.99\linewidth]{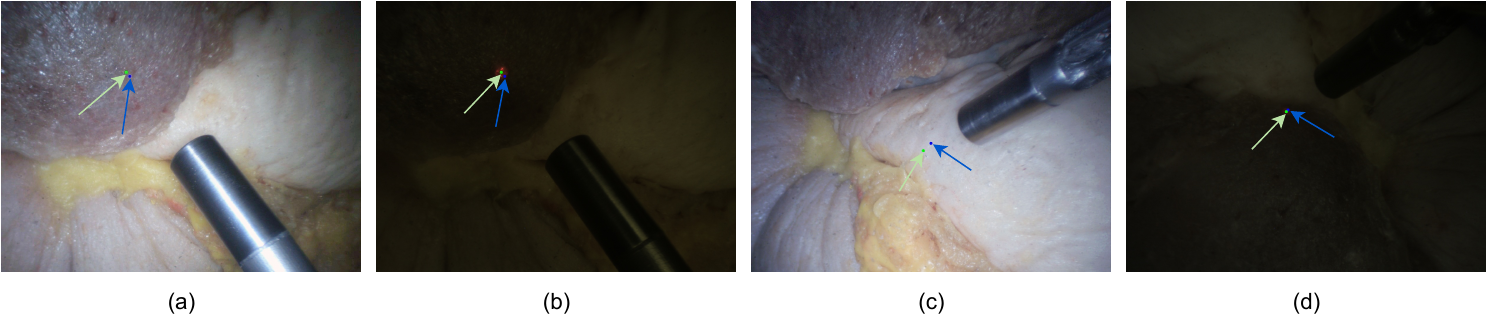}
\caption{Qualitative results. (a) and (c) are standard illumination images and (b) and (d) are images with laser on and laparoscopic light off. The predicted intersection point is shown in blue and the green point indicates the ground truth, which are further indicated by arrows for clarity.}
\label{vis}
\end{figure*}

\begin{table}[t]
\centering
\footnotesize
\caption{Results using ResNet50. Grey colour denotes the Jerry dataset and Blue colour is for Coffbee dataset (2D errors are in pixels and 3D errors are in mm).}
\renewcommand\tabcolsep{11pt}
{\renewcommand{\arraystretch}{1.2}
\begin{tabular}{c|c|c|c|c|c|c}
\toprule
\multicolumn{2}{c|}{\textbf{ResNet}}     & \checkmark & \checkmark  & \checkmark  & \checkmark  & \checkmark \\ \hline
\multicolumn{2}{c|}{\textbf{MLP}} &      & \checkmark   & & & \checkmark    \\ \hline
\multicolumn{2}{c|}{\textbf{LSTM}}   &  &   & \checkmark   & &     \\ \hline
\multicolumn{2}{c|}{\textbf{Stereo}} & \checkmark     &      \checkmark     &\checkmark    &  &   \\ \hline
\multicolumn{2}{c|}{\textbf{Mono}} & & &    & \checkmark  & \checkmark        \\ \hline
\hline

\rowcolor[RGB]{230,230,230}\multicolumn{2}{c|}{2D Mean E.} & $73.5$ & $70.5$ &  $73.7$  & $75.6$  & $76.7$        \\ 
\rowcolor[RGB]{230,230,230}\multicolumn{2}{c|}{2D Std.} & $65.1$ & $56.8$ &  $62.1$  & $62.9$  & $64.4$        \\ 
\rowcolor[RGB]{230,230,230}\multicolumn{2}{c|}{2D Median} & $57.5$ & $59.8$ &  $56.9$  & $58.8$  & $68.4$        \\ \hline \hline

\rowcolor[RGB]{193,218,243}\multicolumn{2}{c|}{2D Mean E.} & $63.2$ & $52.9$ &  $62.0$  & $55.8$  & $60.2$        \\ 
\rowcolor[RGB]{193,218,243}\multicolumn{2}{c|}{2D Std.} & $71.4$ & $42.9$ &  $63.4$  & $55.3$  & $42.1$        \\ 
\rowcolor[RGB]{193,218,243}\multicolumn{2}{c|}{2D Median} & $44.9$ & $44.6$ &  $43.4$  & $42.5$  & $52.3$        \\\hline
\rowcolor[RGB]{193,218,243}\multicolumn{2}{c|}{R2 Score} & $0.55$ & $0.82$ &  $0.63$  & $0.73$  & $0.78$        \\ \hline

\rowcolor[RGB]{193,218,243}\multicolumn{2}{c|}{3D Mean E.} & $8.5$ & $7.4$ &  $6.5$  & $6.4$  & $11.2$        \\ 
\rowcolor[RGB]{193,218,243}\multicolumn{2}{c|}{3D Std.} & $15.7$ & $6.7$ &  $6.8$  & $7.1$  & $18.2$        \\ 
\rowcolor[RGB]{193,218,243}\multicolumn{2}{c|}{3D Median} & $4.5$ & $4.6$ &  $4.0$  & $4.3$  & $5.4$        \\
\bottomrule
\end{tabular}
}

\label{tab_resnet}
\end{table}


\begin{table}[t]
\centering
\footnotesize
\caption{Results using \acrshort{vit}. Grey colour denotes the Jerry dataset and Blue colour is for Coffbee dataset (2D errors are in pixels and 3D errors are in mm).}
\renewcommand\tabcolsep{11pt}
{\renewcommand{\arraystretch}{1.3}
\begin{tabular}{c|c|c|c|c|c|c}
\hline
\multicolumn{2}{c|}{\textbf{ViTNet}}       & \checkmark & \checkmark   & \checkmark  & \checkmark   & \checkmark                                              \\ \hline
\multicolumn{2}{c|}{\textbf{MLP}}     &      & \checkmark     &  &  & \checkmark \\ \hline
\multicolumn{2}{c|}{\textbf{LSTM}}   &   &              & \checkmark   &              &    \\ \hline
\multicolumn{2}{c|}{\textbf{Stereo}} & \checkmark     &  \checkmark &  \checkmark     &  &  \\ \hline
\multicolumn{2}{c|}{\textbf{Mono}} &  &   &  & \checkmark            & \checkmark   \\ \hline
\hline

\rowcolor[RGB]{230,230,230}\multicolumn{2}{c|}{2D Mean E.} & $77.9$ & $92.3$ &  $80.9$  & $87.7$  & $112.1$        \\ 
\rowcolor[RGB]{230,230,230}\multicolumn{2}{c|}{2D Std.} & $69.1$  & $71.0$ &  $67.4$  & $68.6$  & $84.2$        \\ 
\rowcolor[RGB]{230,230,230}\multicolumn{2}{c|}{2D Median} & $59.0$ & $75.0$ &  $64.8$  & $74.9$ & $90.0$        \\ \hline \hline

\rowcolor[RGB]{193,218,243}\multicolumn{2}{c|}{2D Mean E.} & $76.3$ & $75.0$ &  $88.0$  & $56.5$  & $82.7$        \\ 
\rowcolor[RGB]{193,218,243}\multicolumn{2}{c|}{2D Std.} & $69.8$  & $60.6$ &  $83.3$  & $75.8$  & $63.9$        \\ 
\rowcolor[RGB]{193,218,243}\multicolumn{2}{c|}{2D Median} & $59.9$ & $59.6$ &  $68.3$  & $34.5 $ & $69.1$        \\ \hline

\rowcolor[RGB]{193,218,243}\multicolumn{2}{c|}{R2 Score} & $0.58$ & $0.66$ &  $0.33$  & $0.65$ & $0.60$        \\ \hline

\rowcolor[RGB]{193,218,243}\multicolumn{2}{c|}{3D Mean E.} & $7.9$ & $9.1$ &  $11.4$  & $11.6$  & $7.7$        \\ 
\rowcolor[RGB]{193,218,243}\multicolumn{2}{c|}{3D Std.} & $6.9$  & $8.2$ &  $16.7$  & $21.3$  & $7.0$        \\ 
\rowcolor[RGB]{193,218,243}\multicolumn{2}{c|}{3D Median} & $6.0$ & $5.9$ &  $7.1$  & $5.3$ & $6.2 $        \\ \hline

\end{tabular}
}
\label{tab_vit}
\end{table}

\section{Results}
Quantitative results on the released datasets are shown in Table \ref{tab_resnet} and Table \ref{tab_vit} with different backbones for extracting image features, \acrshort{resnet} and \acrshort{vit}. For the 2D error on two datasets, among the different settings, the combination of \acrshort{resnet} and \acrshort{mlp} gave the best performance with a mean error of $70.5$ pixels and a standard deviation of $56.8$. The median error of this setting was $59.8$ pixels while the R2 score was $0.82$ (higher is better for R2 score). Comparing the Table \ref{tab_resnet} and Table \ref{tab_vit}, we found that the \acrshort{resnet} backbone was better than the \acrshort{vit} backbone in the image processing task, while \acrshort{mlp} was better than \acrshort{lstm} in probe pose representation. 
\acrshort{resnet} processed the input images as a whole, which was better suited for utilising the global context of a unified scene composed of the tissue and the probe, compared to the \acrshort{vit} scheme, which treated the whole scene as several patches. 
Similarly, the sampled $50$ principal points on the probe axis were better processed using the simple \acrshort{mlp} rather than using a recurrent procedure \acrshort{lstm}. It is worth noting that the results from stereo inputs exceeded those from mono inputs, which can be attributed to the essential 3D information included in the stereo image pairs.

For the 3D error, the \acrshort{resnet} backbone still gave generally better performance than the \acrshort{vit} backbone while under the \acrshort{resnet} backbone, \acrshort{lstm} and \acrshort{mlp} gave competitive results and they are all in sub-millimetre level. We note that the 3D error subjected to the quality of the acquired ground truth depth maps, which had limited resolution and non-uniformly distributed valid data due to hardware constraints. Hence, we used the median depth value of a square area of 5 pixels around the points where the depth value was not available. 

Fig.~\ref{vis} shows visualisation results of our method using \acrshort{resnet} and \acrshort{mlp}. This figure illustrates that our proposed method can detect the intersection point using solely standard RGB laparoscopic images as the input. Furthermore, based on the simple design, our method achieved the inference time of $50$ frames per second, making it well-suitable for intraoperative surgery.

\section{Conclusion}

In this work, a new framework for using a laparoscopic drop-in gamma detector in manual or robotic-assisted minimally invasive cancer surgery was presented, where a laser module mock probe was utilised to provide training guidance, and the problem of detecting the probe axis-tissue intersection point was transformed to laser point position inference. Both the hardware and software design of the proposed solution were illustrated and two newly acquired datasets were publicly released. Extensive experiments were conducted on various backbones and the best results were achieved using a simple network design, enabling real time inference of the sensing area. We believe that our problem reformulation and dataset release, together with the initial experimental results, will establish a new benchmark for the surgical vision community.

\setcounter{chapter}{7}
\chapter{Conclusion}
\chaptermark{Conclusion and Future Work}
\glsresetall
\label{chap:8-conclusion}

As stated in Chapter~\ref{chap:1-introduction} and Chapter~\ref{chap:2-rw}, despite the increasing popularity and attention garnered by robot-assisted or manually operated \acrshort{mis} due to its potential for smaller incisions, reduced blood loss, and faster recovery times, there still exist numerous challenges in \acrshort{mis}. Particularly in the context of cancer surgery, accurate identification of cancerous tissue remains a significant hurdle. This thesis represents new methods that combine computer vision, machine learning, and laparoscopic imaging to address the challenges in \acrshort{mis} for cancer treatment. The exploration across the chapters addresses a series of classical laparoscopic intervention problems such as tool pose estimation, tracking, depth estimation, and more advanced problems in endoscopic radio-guided cancer resection field, with a novel tethered laparoscopic gamma detector applied, such as gamma probe sensing area detection and \acrshort{ar} for visual feedback. The outcomes of the thesis have been applied to real-world products with commercialisation potential, and further possibilities for applying the gamma probe to other organs are being explored.

Chapter~\ref{chapter_marker_probe_tracking} introduced a dual-pattern marker for probe tracking and 6D pose estimation. Different from the pure circular dots marker or single chessboard vertices based marker, the new dual-pattern marker made use of the advantages of both type of markers, helped with improving the robustness and reducing the detection failure rate, providing more accurate pose estimation results and a larger workspace. The system was validated by the ground truth collected from the OptiTrack system and a Cobalt-57 disk was put under the tissue phantom to show the augmented reality results. In this work, the visualisation challenge of locating the detection area was solved by combining the probe pose estimation results and the 3D tissue surface reconstruction, therefore showing the feasibility and the potentiality of using the gamma probe with visual feedback.

To improve the reliability and accuracy of 3D tissue surface reconstruction results acquired from algorithm \acrshort{sfm} used in Chapter~\ref{chapter_marker_probe_tracking}, subsequent chapters illustrated different depth estimation approaches, such as exploring self-supervised depth estimation techniques in laparoscopic imaging leveraging the power of \acrshort{gan} (Chapter~\ref{chap:4-contribution_2-Depth2D}) and 3D masked geometric consistency (Chapter~\ref{chap:5-contribution_3-Depth3D}). These contributions show the creation solutions of highly accurate depth maps that serve as indispensable tools for image-guided intervention in the \acrshort{mis} and a navigating foundation towards fully automatic robot-based surgery for intricate anatomy and pathology. The improved depth perception achieved by these methodologies not only provides better visualisation of the 3D structure of surgical scenes, but also holds the potential to redefine surgical strategies, for potentially minimising the likelihood of residual disease and maximising the precision of cancer resections.

The integrative approach of simultaneous depth estimation and surgical tool segmentation in Chapter~\ref{chap:6-contribution_4-DepthSeg} showed a leap in the fusion of methods for multi-purpose tasks in \acrshort{mis}. This holistic methodology not only refines depth perception, but also provides a nuanced understanding of the surgical environment, equipping surgeons with insights into the spatial relationships between the surgical tools and surrounding interactive tissue structures. The simultaneous segmentation and depth estimation of surgical tools further streamlined the surgical workflow, thereby enhancing the precision and efficacy of \acrshort{mis}.

Chapter~\ref{chap:7-contribution_-5PtRegress} illustrated a novel sensing area detection method of the laparoscopic probe, representing a pivotal step towards solving real-world problems in cancer tissue localisation by using a single RGB image. By utilising a simple yet effective technique, the proposed method ensured the application of laparoscopic gamma probes in the real-time detection and assessment of cancerous lesions. We note that while the laser images were necessary for collecting labels during training, they are not needed during real-world application because the trained model can be deployed without the laser images. This work also contributed new datasets that can serve as a new benchmark for the community in this task.

The outcomes of this thesis have the potential to make the impact beyond the academic inquiry, signalling new advancements in the detection and treatment of cancer. The convergence of computer vision, machine learning, surgical scene analysis and new surgical instrumentation emerges as a new direction in cancer surgery, providing possibility of enhanced precision and efficiency in real-time surgery applications. As this journey unfolds, these contributions become not just a proof of technological prowess but a profound testament to the potential of collaborative endeavours to positively impact the lives of individuals affected by cancer.

As the first work building the hardware and software system for the gamma probe in \acrshort{mis}, this thesis has certain limitations. Expanding upon the groundwork, the future work for this thesis could encompass several avenues:

\textbf{Refinement of Detection Algorithms.} Building upon the initial success of using a simple network design for real-time inference of the sensing area, further refinement and optimisation of detection algorithms can be pursued. This refinement may involve exploring advanced machine learning techniques, such as transformer architectures, which offer greater capacity for capturing complex patterns in data. Additionally, incorporating supplementary information, such as corresponding depth maps, can potentially enhance the accuracy and robustness of detecting the intersection point between the gamma probe axis and the tissue surface. Continuous algorithmic improvements, combined with extensive validation, will ensure that the system remains reliable and effective in diverse clinical scenarios, ultimately improving intraoperative decision-making and patient outcomes.

\textbf{Integration of Multi-Modal Data and Generative Models.} Incorporating additional modalities of data, such as pre-operative imaging data (\acrshort{ct}, \acrshort{mri}, \acrshort{pet}) and real-time intraoperative imaging (e.g., fluorescence imaging), can provide complementary information for improved localisation and characterisation of cancerous tissues and metastases. The fusion of multi-modal data streams has the potential to enhance the overall accuracy and reliability of intraoperative visualisation. Moreover, the use of foundation models is becoming a powerful tool in computer-assisted \acrshort{mis} particularly with the application of large language-vision model~\cite{vuong2024languagegrasppp,vuong2023grasp}.

\textbf{Development of Surgical Navigation Systems.} Integrating the proposed visualisation tools into surgical navigation systems can offer surgeons real-time guidance during minimally invasive procedures. This integration requires seamless compatibility with existing surgical platforms and interfaces, ensuring intuitive and user-friendly interactions. By minimising disruption to established surgical workflows, these systems can enhance precision and efficiency in the operating room. Additionally, the incorporation of \acrshort{ar} and extra display devices can further improve the accuracy of tissue differentiation and instrument tracking. Continuous feedback from these systems can support surgical decision-making, potentially reducing operative times and improving patient outcomes.

\textbf{Clinical Validation and Translation.} Conducting comprehensive clinical studies to validate the efficacy and safety of the developed intraoperative visualisation tools is paramount. This process involves rigorous testing in diverse clinical settings to ensure robustness and reliability. Collaborating closely with clinical partners and regulatory agencies to navigate the path towards regulatory approval and clinical adoption of the technology is essential for eventual translation of these advanced tools into routine clinical practice, ultimately enhancing surgical outcomes and patient care.

 %


\bibliographystyle{unsrt}
\bibliography{meta/0-references} 

\end{document}